\documentclass[a4paper,11pt]{article}
\PassOptionsToPackage{hyperfootnotes=false}{hyperref}
\usepackage{jheppub} 
\usepackage{lineno}
\usepackage{feynmp-auto}
\usepackage{subcaption}
\usepackage[section]{placeins}
\usepackage{flafter}
\usepackage{mathabx}
\usepackage{nicematrix}
\usepackage{fancyvrb}
\usepackage{snapshot}
\usepackage{slashed}
\VerbatimFootnotes
\newcommand{\beq}      {\begin{equation}}
\newcommand{\eeq}      {\end{equation}}
\newcommand{\beqn}     {\begin{eqnarray}}
\newcommand{\eeqn}     {\end{eqnarray}}
\newcommand{\slsh}     {\slashed}
\newcommand{\myfrac}[2]{\mbox{\small$\frac{#1}{#2}$}}
\newcommand{\half}     {\myfrac{1}{2}}
\newcommand{\ldot}[2]  {\:{#1}\!\cdot\!{#2}\:}
\newcommand{\Herwig}{\textsc{Herwig}}
\newcommand{\HerwigSeven}{\Herwig~7}

\newcommand{\POLDIS}{\textsc{POLDIS}}
\graphicspath{{./}{figures/}}

\title{\boldmath Polarized Deep-Inelastic Scattering with Spin Correlations in Herwig 7}

\author[1]{M.R.~Masouminia,}
\author[2]{A.~Papaefstathiou,}
\author[1]{P.~Richardson}
\affiliation[1]{Institute for Particle Physics Phenomenology, Durham University, Durham, UK}
\affiliation[2]{Department of Physics, Kennesaw State University, 830 Polytechnic Lane, Marietta, GA 30060, USA}

\emailAdd{mohammad.r.masouminia@durham.ac.uk}
\emailAdd{apapaefs@kennesaw.edu}
\emailAdd{peter.richardson@durham.ac.uk}

\abstract{We investigate polarized deep-inelastic scattering (DIS) in the context of fully exclusive Monte Carlo simulations for the Electron-Ion Collider (EIC). We present a next-to-leading-order (NLO) treatment of polarized DIS in \HerwigSeven\ using the POWHEG matching scheme, including neutral-current $\gamma/Z$ exchange and charged-current scattering. We also construct a spin-correlation treatment for NLO-matched events, first by initializing the shower from the spin-density matrix of the polarized DIS Born process and then by propagating the spin-density matrix of the accepted real-emission configuration without changing the POWHEG event weight or hardest-emission choice. We validate the integrated cross sections against fixed-order calculations, use parton-level comparisons without subsequent shower evolution to isolate the accepted POWHEG real-emission kinematics, and employ shower-level observables to test the numerical importance of the Born-level and real-emission spin information. We find that the Born-level spin-density initialization can have a visible impact on shower-sensitive observables, while the additional spin information from the accepted real-emission configuration is not resolved as a robust separate effect within the independent-sample Monte Carlo precision for the longitudinally polarized observables considered here.}

\begin{document}
\begin{flushright}
\footnotesize

IPPP-26-47\\
MCnet-26-17

\end{flushright}

\maketitle
\flushbottom

\section{Introduction}\label{sec:intro}

Polarized deep-inelastic scattering (DIS) remains one of the cleanest ways of probing how the spin of a hadron is assembled from its partonic degrees of freedom. The upcoming Electron-Ion Collider (EIC) will bring polarized lepton and hadron beams into a kinematic regime with both high luminosity and a broad reach in $x$ and $Q^2$, allowing helicity-dependent dynamics to be studied with differential final-state information. Such measurements are expected to sharpen constraints on polarized parton distributions~\cite{Accardi:2012qut,AbdulKhalek:2021gbh}. To make full use of this programme, theory predictions must go beyond inclusive structure functions and retain the correlations among kinematics, flavour, and spin that are present in the hadronic final state.

The theoretical description of DIS has a long history. Building on the parton model and the QCD evolution equations~\cite{Altarelli:1977zs}, complete next-to-leading-order (NLO) calculations for unpolarized DIS were already available by the late 1970s~\cite{Bardeen:1978yd}. Over the following decades this programme was extended to next-to-next-to-leading order (NNLO) with the computation of the exact three-loop splitting functions~\cite{Moch:2004pa,Vogt:2004mw}, establishing DIS as one of the primary benchmark processes for precision PDF determinations. The treatment of the polarized process followed a parallel development, though with a significant time lag: NLO helicity-dependent radiative corrections and splitting functions were worked out in the mid-1990s~\cite{Zijlstra:1993sh,Mertig:1995ny,Vogelsang:1996im}. At NNLO, the main three-loop helicity-dependent splitting functions for singlet evolution and helicity-sum non-singlet combinations were obtained in Ref.~\cite{Moch:2014sna}. The additional non-singlet quark-antiquark-difference anomalous dimension relevant for polarized charged-current combinations was obtained in Ref.~\cite{Blumlein:2021enk}. Its corrected expression was given in the erratum~\cite{Blumlein:2026erratum} to the polarized-singlet calculation of Ref.~\cite{Blumlein:2021ryt}, with an independent recalculation reported in Ref.~\cite{Behring:2025avs}. The massless three-loop Wilson coefficients for $F_2$, $F_L$, $xF_3$ and $g_1$ were calculated in Ref.~\cite{Blumlein:2022gpp}, and the full set of inclusive polarized DIS structure functions at NNLO was subsequently presented in Ref.~\cite{Borsa:2022irn}. At the differential fixed-order level, parton-level tools such as \textsc{DISENT} supplied NLO benchmarks for unpolarized DIS jet observables~\cite{Catani:1996vz}. The polarized \POLDIS\ calculations, used in the present article as comparison benchmarks, extend this class of predictions to NLO dijet and NNLO inclusive-jet production, including EIC kinematics and neutral- and charged-current channels~\cite{Borsa:2020polarizedjets,Borsa:2022nnlojets}.

The phenomenological emphasis has meanwhile moved from inclusive calculations towards fully differential event generation. Complete-event DIS simulations have long been available through process-specific Monte Carlo programs, including \textsc{PEPSI} for polarized leptoproduction~\cite{Mankiewicz:1991jc}, \textsc{LEPTO}~\cite{Ingelman:1996mq}, and \textsc{DJANGO6}/\textsc{DJANGOH}, which combine electroweak and QED radiative corrections with \textsc{LEPTO} hadronic final states and have also been used for polarized EIC studies~\cite{Charchula:1994kf,Aschenauer:2013iia}. The development of matching between NLO calculations and the parton shower (PS), conventionally denoted NLO+PS matching, made it possible to retain this event-level description while improving the perturbative accuracy of the hard process. The general MC@NLO and POWHEG formalisms provide standard frameworks for matching NLO QCD calculations to parton showers~\cite{Frixione:2002ik,Nason:2004rx,Frixione:2007vw}. In DIS, matched and merged developments include the \textsc{Sherpa} implementation with multi-leg matrix-element merging~\cite{Carli:2009cg}, the automated POWHEG implementation in \textsc{Sherpa}, for which deep-inelastic lepton-hadron scattering was one of the validation processes~\cite{Hoche:2010kg}, and the POWHEG treatment in \Herwig~\cite{DErrico:2011wfa}. Subsequent work reached NNLO+PS accuracy for hadronic final states in DIS using \textsc{Sherpa}~\cite{Hoche:2018suf} and provided NLO+PS \textsc{Sherpa} predictions in event-shape studies of neutral-current DIS~\cite{Knobbe:2023jth}. More recent developments include dedicated POWHEG BOX generators for neutral- and charged-current DIS~\cite{Banfi:2023dis}, for polarized DIS~\cite{Borsa:2024rmh}, for lepton-hadron DIS with mass effects~\cite{Buonocore:2024pdv}, and for neutrino-induced DIS~\cite{FerrarioRavasio:2024disnu}. Further work has provided MEPS@NLO merging for neutral- and charged-current DIS at the EIC~\cite{Meinzinger:2025dis}, multi-jet merging in \textsc{Pythia}~8/\textsc{Vincia}~\cite{Helenius:2024disjets}, and a multiplicative matching strategy for unpolarized neutral-current DIS in \textsc{Pythia}~8~\cite{Helenius:2026pym}.

Within this broader programme, general-purpose Monte Carlo event generators provide the framework in which perturbative calculations are combined with parton showers, hadronization, and decays to produce realistic event simulations. \Herwig\ brings these ingredients together within a unified simulation framework~\cite{Bahr:2008pv,Gieseke:2011na,Arnold:2012fq,Bellm:2013hwb,Bellm:2015jjp,Bellm:2017bvx,Bellm:2019zci,Bewick:2023tfi,Bellm:2025pcw}, and already incorporates matching to higher-order corrections as well as a dedicated spin-correlation algorithm for propagating polarization information through showering and decays~\cite{Richardson:2018pvo}. An NLO-matched treatment of unpolarized DIS was incorporated in an earlier version of \Herwig~\cite{DErrico:2011wfa} via the POWHEG method. The corresponding polarized treatment has not previously been available in \Herwig. This includes both the initialization of the shower with the spin-density matrix of the polarized DIS hard process and the further propagation of the spin information of an accepted real-emission configuration.

We present two related developments. First, we formulate polarized DIS in \Herwig\ at NLO accuracy using the POWHEG method, including the real-emission, virtual, and collinear-subtraction contributions required for matched event generation. This gives a spin-dependent description of the hard scattering inside a general-purpose event generator, rather than relying only on spin correlations generated by the shower. The same framework also supplies the ingredients needed for polarized hard scattering in \Herwig\ beyond the DIS process considered here. Extensions to other processes are left to future work.

The second contribution concerns the propagation of spin information from the hard event to the parton shower. The polarized DIS matrix element already provides a nontrivial Born-level spin-density matrix for the shower, giving a spin-aware baseline that is absent if shower spin correlations are switched off. Once the hardest POWHEG emission has been generated, the accepted configuration is a specific real-emission state whose final-state particles carry additional correlated spin information. To preserve this information in the event record, we instead initialize the parton shower with the spin-density information of this accepted state. We construct that density matrix from the exact real-emission helicity amplitudes and pass it to the shower without modifying the accepted hard-event normalization or the POWHEG hardest-emission choice. The resulting setup therefore separates two effects: the Born-level polarized DIS spin-density initialization, and the incremental information carried by the accepted POWHEG emission in longitudinally polarized DIS jet observables at EIC energies.

The remainder of the paper is organized as follows. In Section~\ref{sec:formalism} we review the spin-density-matrix formalism for polarized hadrons and the corresponding modifications to the parton shower. Section~\ref{sec:polarizeddis} presents the polarized DIS calculation, including the real and virtual contributions and their fixed-order validation, and then turns to the Born-level and real-emission spin-correlation framework for the shower and to the jet observables used to test the possible impact of this spin information at EIC energies. We summarize the results and outline possible extensions in Section~\ref{sec:conclusions}. Appendix~\ref{app:standalone-fixed-order-validation} presents an independent validation of the fixed-order components, and Appendix~\ref{app:charged-current-differential-validation} provides supplementary results for the charged-current process.

\section{Formalism and Definitions}\label{sec:formalism}
\subsection{Polarized Hadrons}

For unpolarized hadrons, the relevant non-perturbative input is the parton distribution function $f_i(x,Q^2)$, which specifies the density of partons of type $i$ carrying momentum fraction $x$ at factorization scale $Q^2$.\footnote{In this section and in the real-emission derivations we write the PDF scale argument as $Q^2$. When the factorization scale is distinguished from the DIS virtuality, as in Section~\ref{sec:polarizeddis}, it is denoted by $\mu_F$, with central choice $\mu_F^2=Q^2$.}
Following the notation of Ref.\,\cite{Barone:2001sp}, for longitudinally polarized hadrons
we define the helicity-dependent parton distributions $f_{i\pm}(x,Q^2)$ for partons of type $i$ with helicity $\pm$.
The helicity PDF $\Delta f_i(x,Q^2)$ is then introduced such that
\begin{subequations}
\begin{eqnarray}
  f_i(x,Q^2)        &=& f_{i+}(x,Q^2) + f_{i-}(x,Q^2),\\
  \Delta f_i(x,Q^2) &=& f_{i+}(x,Q^2) - f_{i-}(x,Q^2).
\end{eqnarray}
\end{subequations}
For transversely polarized spin-$1/2$ hadrons, the corresponding leading-twist input is the quark transversity distribution $\Delta_T f_q(x,Q^2)$. It gives the difference between quark densities with polarization parallel and antiparallel to that of the hadron. For compactness we keep the parton label $i$ in the density-matrix formulae below, with $\Delta_T f_i$ understood only where a transversity distribution exists. For details, see Sec.~1 of Ref.\,\cite{Barone:2001sp}.

In terms of these distributions and the components of the hadron polarization vector in the $x,y,z$ directions,\footnote{The hadron momentum is taken to define the $z$ direction.} $P_{x,y,z}$,
we define a spin-density matrix\footnote{This is Eq.~(4.3.7) of Ref.\,\cite{Barone:2001sp}, apart from a sign convention arising from our choice to order helicity states as $(-,+)$ rather than $(+,-)$.}
\begin{equation}
  H_{i\tau\tau'}(x,Q^2) = \frac12 \left( \begin{array}{cc} 1-\lambda_i(x,Q^2) & s_{i,x}(x,Q^2)+is_{i,y}(x,Q^2) \\ s_{i,x}(x,Q^2)-is_{i,y}(x,Q^2) & 1 + \lambda_i(x,Q^2) \end{array}\right)\;,
\end{equation}
where
\begin{subequations}
  \begin{eqnarray}
    \lambda_i(x,Q^2) f_i(x,Q^2) &=& P_z \Delta f_i(x,Q^2)\;, \\
    s_{i,\{x,y\}}(x,Q^2) f_i(x,Q^2) &=& P_{\{x,y\}} \Delta_T f_i(x,Q^2)\;.
\end{eqnarray}
\end{subequations}

Considering a process with one incoming hadron, as in DIS, in which a parton of type $i$ initiates the process and $n$ particles are produced, the partonic contribution to the differential cross section is given by
\begin{equation}
  \mathrm{d}\sigma_n = C \frac{(2\pi)^4}{2\hat{s}}\, \mathrm{d}x\, \mathrm{d}\Phi_n\, f_i(x,Q^2)\, H_{i\tau\tau'}(x,Q^2)\, \mathcal{M}_\tau\mathcal{M}^*_{\tau'}\;,
  \label{eqn:dsigma}
\end{equation}
where
\[
\mathrm{d}\Phi_n = \delta^{(4)}\!\left(p_{\rm in}-\sum_{k=1}^n p_k\right)
\prod_{k=1}^n \frac{\mathrm{d}^3 p_k}{(2\pi)^3\,2E_k}
\]
is the differential phase space for the production of $n$ particles and $\hat{s}$ is the partonic centre-of-mass energy squared. Here $C$ is the appropriate factor for summing over the colours and spins of the outgoing particles and averaging over those of the incoming particles, with the exception of the spin of $i$. The matrix element for the process is $\mathcal{M}$, where the helicity indices of all particles except $i$ have been suppressed for clarity. For unpolarized hadrons $H_{i\tau\tau'}(x,Q^2)\to 1/s_i\,\delta_{\tau\tau'}$, where $s_i$ is the number of spin states for parton $i$, and
Eq.~\eqref{eqn:dsigma} reduces to the standard cross section formula.

The spin-density matrix $H_{i\tau\tau'}$ defined here is the object passed to the parton shower evolution discussed in the next subsection. Its generalization to an accepted NLO real-emission state is one of the main ingredients of Section~\ref{sec:polarizeddis}.

\subsection{Parton Shower Evolution}

The differential cross section including a single initial-state emission takes the form
\begin{align}
 \mathrm{d}\sigma_{n+1} ={}& \frac{\mathrm{d}z}z \frac{\mathrm{d}Q^2}{Q^2} \frac{\mathrm{d}\phi}{2\pi}\frac{\alpha_s}{2\pi}\,
 C \frac{(2\pi)^4}{2\hat{s}}\, \mathrm{d}x\, \mathrm{d}\Phi_n\,
 f_i\left(x/z,Q^2\right)\, H_{i\lambda_0\lambda'_0}\left(x/z,Q^2\right)\nonumber\\
 &\times F_{\lambda_0\lambda_1\lambda_2} F^*_{\lambda'_0\lambda'_1\lambda_2}\,
 \mathcal{M}_{\lambda_1}\mathcal{M}^*_{\lambda_1'}\;,
\end{align}
where $F_{\lambda_0\lambda_1\lambda_2}$ are the helicity amplitudes for the branching, defined in Ref.\,\cite{Richardson:2018pvo} such that
\begin{equation}
  P_{i\to jk}(z) = \frac12\sum_{\lambda_0,\lambda_1,\lambda_2}  F_{\lambda_0\lambda_1\lambda_2} F^*_{\lambda_0\lambda_1\lambda_2}\;,
\end{equation}
with $P_{i\to jk}(z)$ the standard spin-averaged splitting function for the branching $i\to jk$. In the backward-evolution notation used here, $j$ is the parton entering the hard subprocess, $i$ is its pre-branching parent, and $k$ is the emitted parton. The variable $z$ is the corresponding momentum fraction for the incoming leg, so that the parton entering the hard subprocess carries momentum fraction $x = z x_{\rm parent}$ and the pre-branching parent distribution is evaluated at $x_{\rm parent}=x/z$.

The differential branching probability is then
\begin{equation}
  \mathrm{d}\mathcal{P} = \frac{\mathrm{d}z}z \frac{\mathrm{d}Q^2}{Q^2} \frac{\mathrm{d}\phi}{2\pi}\frac{\alpha_s}{2\pi}\,
  \frac{f_i\left(x/z,Q^2\right) }{ f_j(x,Q^2) }
  \frac{H_{i\lambda_0\lambda'_0}\left(x/z,Q^2\right) F_{\lambda_0\lambda_1\lambda_2} F^*_{\lambda'_0\lambda'_1\lambda_2} \mathcal{M}_{\lambda_1}\mathcal{M}^*_{\lambda_1'}}{H_{j\tau\tau'}(x,Q^2) \mathcal{M}_\tau\mathcal{M}^*_{\tau'}}\;.
\end{equation}
For unpolarized hadrons this reduces to
\begin{equation}
  \mathrm{d}\mathcal{P} = \frac{\mathrm{d}z}z \frac{\mathrm{d}Q^2}{Q^2} \frac{\mathrm{d}\phi}{2\pi}\frac{\alpha_s}{2\pi}\,
  \frac{f_i\left(x/z,Q^2\right) }{ f_j(x,Q^2) }
  F_{\lambda_0\lambda_1\lambda_2} F^*_{\lambda_0\lambda'_1\lambda_2} D_{\lambda_1\lambda_1'}\;,
\end{equation}
where the decay matrix is given by
\begin{equation}
  D_{\lambda\lambda'} = \frac{\mathcal{M}_{\lambda}\mathcal{M}^*_{\lambda'}}{\mathcal{M}_\tau\mathcal{M}^*_{\tau}}\;.
\end{equation}
If spin correlations between the emission and the hard process are neglected, $D_{\lambda\lambda'}\to \frac12\delta_{\lambda\lambda'}$,
and one recovers the standard result
\begin{equation}
  \mathrm{d}\mathcal{P}_\mathrm{uncorrelated} = \frac{\mathrm{d}z}z \frac{\mathrm{d}Q^2}{Q^2} \frac{\alpha_s}{2\pi}\,
  \frac{f_i\left(x/z,Q^2\right) }{ f_j(x,Q^2) } P_{i\to jk}(z)\;.
\end{equation}
Rewriting the full branching probability in terms of $\mathrm{d}\mathcal{P}_\mathrm{uncorrelated}$,
\begin{equation}
  \mathrm{d}\mathcal{P} = \mathrm{d}\mathcal{P}_\mathrm{uncorrelated} W_\mathrm{correlated}\;,
\end{equation}
where
\begin{equation}
  W_\mathrm{correlated} =
  \frac{\mathrm{d}\phi}{2\pi}\frac1{2H_{j\tau\tau'}(x,Q^2) D_{\tau\tau'}}
  \frac{2H_{i\lambda_0\lambda'_0}\left(x/z,Q^2\right) F_{\lambda_0\lambda_1\lambda_2} F^*_{\lambda'_0\lambda'_1\lambda_2}D_{\lambda_1\lambda_1'}}{P_{i\to jk}(z)}\;,
\end{equation}
where the two ratio factors reduce to unity in the absence of beam polarization and of spin correlations, respectively.
We can write the second ratio in the form
\begin{equation}
   \frac{2H_{i\lambda_0\lambda'_0}\left(x/z,Q^2\right) F_{\lambda_0\lambda_1\lambda_2} F^*_{\lambda'_0\lambda'_1\lambda_2}D_{\lambda_1\lambda_1'}}{P_{i\to jk}(z)} = 1 + A + B + C\,,\label{eqn:corr}
\end{equation}
where $A$ is azimuthally independent and contributes only for polarized hadrons, $B$ is azimuthally dependent and present even for unpolarized beams, and $C$ is azimuthally dependent but contributes only for polarized hadrons.
The values of $A$, $B$, and $C$ are given in Tables~\ref{tab:branch1} and~\ref{tab:branch2}, where the unit-trace condition for both $H_i$ and $D$ has been used to simplify the expressions.

The azimuthal dependence of $B$ and $C$ is proportional to $e^{\pm 2 i\phi}$ and therefore averages to zero upon integration over the full azimuth.
In the present study we restrict attention to longitudinal beam polarization, for which $H_{i+-}=H_{i-+}=0$ for all parton species. In particular, $H_{g\pm\mp}=0$ for spin-$\frac12$ hadrons. The off-diagonal terms therefore do not contribute in the present context. Extending the framework to transverse polarization would require transversity PDFs in the quark channels together with a dedicated rederivation of the corresponding hard and collinear spin structures in a transverse-spin basis. For a spin-$\frac12$ hadron there is no leading-twist gluon transversity, since the corresponding gluonic operator carries two units of helicity flip and therefore requires a target with spin $J\ge 1$~\cite{Jaffe:1989xy,Kumano:2019igu}. We nevertheless retain the off-diagonal terms in the notation, since they would contribute for transversely polarized hadrons or polarized resolved-photon processes.
\begin{table}
  \begin{center}
    \begin{tabular}{|c|c|}
      \hline
      Branching & A \\
      \hline
      $q\to qg$ & $\lambda_i \left(D_{++}-D_{--}\right)+\frac{4 z}{1+z^2} \left(H_{i-+}D_{-+} +H_{i+-}D_{+-} \right)$ \\
      \hline
      $q\to gq$ & $\lambda_i \frac{  (2-z) z}{1+(1-z)^2} \left(D_{++}-D_{--}\right)$ \\
      \hline
      $g\to gg$ & $\lambda_i \frac{z (2-z -2z(1-z))}{(1-(1-z) z)^2} \left(D_{++}-D_{--}\right)$ \\
      & $+\frac{2 z^2}{(1-(1-z)z)^2}\left(D_{-+} H_{i-+}+D_{+-}H_{i+-}\right)$\\
      \hline
      $g\to q\bar{q}$ & $-\lambda_i \frac{1-2 z}{1-2 (1-z) z}\left(D_{++}-D_{--}\right)$ \\
      \hline
  \end{tabular}
    \caption{Azimuthally independent spin weights for the different possible QCD branchings. Here $z$ denotes the longitudinal momentum fraction of the parton entering the hard subprocess, and $\lambda_i$ and $H_{i\pm\mp}$ are evaluated at $(x/z,Q^2)$. In particular, for $q\to gq$ the parton entering the hard subprocess is the gluon and the emitted parton is the quark, so $z$ is the gluon momentum fraction.}
    \label{tab:branch1}
  \end{center}
\end{table}

\begin{table}
  \begin{center}
    \begin{tabular}{|c|c|c|}
      \hline
      Branching & B & C \\
      \hline
      $q\to qg$ & 0 & 0 \\
      \hline
      $q\to gq$ & $-\frac{2 (1-z)}{1+(1-z)^2} \left(e^{2 i \phi } D_{-+}+e^{-2 i \phi } D_{+-}\right)$& 0 \\
      \hline
      $g\to gg$ & $-\frac{(1-z)^2}{(1-(1-z) z)^2} \left(e^{2 i \phi } D_{-+}+e^{-2 i \phi } D_{+-}\right)$ & $-\frac{z^2 (1-z)^2 \left(e^{-2 i \phi } H_{i-+}+e^{2 i \phi }H_{i+-}\right)}{(1-(1-z) z)^2}$\\
      \hline
      $g\to q\bar{q}$ & 0 &  $\frac{2 (1-z) z \left(e^{-2 i \phi } H_{i-+}+e^{2 i \phi } H_{i+-}\right)}{1-2 (1-z) z}$\\
      \hline
  \end{tabular}
    \caption{Azimuthally dependent spin weights for the different possible QCD branchings, with the same conventions for $z$ and for the suppressed arguments of $H_{i\pm\mp}$ as in Table~\ref{tab:branch1}.}
    \label{tab:branch2}
  \end{center}
\end{table}

The branching probability, averaged over $\phi$ as before, can therefore be generated with the additional azimuthally averaged weight
\begin{equation}
  W_{\langle\phi\rangle} = \frac{1+A}{2H_{j\tau\tau'}(x,Q^2) D_{\tau\tau'}}\;,
\label{eq:Wphi}
\end{equation}
which can be included in the normal weight for the PDF term. The azimuthal angle can then be generated using the full distribution in Eq.~\eqref{eqn:corr} rather than
using only the $1+B$ distribution that applies in the unpolarized case. The decay matrix $D$ is then updated to include the branching and used to initialize the next emission. Once the backward evolution terminates, the spin-averaged density matrix $\rho$, in the notation of Ref.\,\cite{Richardson:2018pvo}, is replaced by the full hadronic spin-density matrix $H_{i\lambda\lambda'}$. The chain is then traversed in the forward direction to initialize the final-state showers of the radiated partons.

The formalism developed in this section applies to shower emissions that evolve from a Born-level hard process. In the polarized DIS calculation below this Born-level spin-density initialization is already a new ingredient in the event simulation. In Section~\ref{sec:polarizeddis} we then extend the same idea to the case in which the shower is seeded instead by an accepted NLO real-emission configuration.

\subsection{Correlated Helicity Weights}
\label{sec:correlated-helicity-weights}

The spin-density-matrix formalism can be used either to generate the physical cross section for a specified choice of longitudinal lepton and hadron polarizations, or to evaluate several helicity configurations on the same accepted event kinematics. The latter construction is useful for validation observables in which a polarized cross section is obtained from cancellations among helicity-dependent contributions. For the four longitudinal beam-helicity choices $(++,+-,-+,--)$, where the first sign denotes the lepton helicity and the second the hadron helicity,\footnote{Since the lepton and hadron beams are counter-propagating, equal beam helicities correspond to anti-aligned spins. With the conventions used here, $\Delta\sigma_{LL}$ is therefore positive for a positive $g_1$ in the photon-exchange case.} we define
\begin{equation}
\begin{aligned}
  \sigma_0(O)
  &=
  \frac{
  \sigma^{++}(O)+\sigma^{+-}(O)+\sigma^{-+}(O)+\sigma^{--}(O)}
  {4}\,,
  \\
  \Delta\sigma_{LL}(O)
  &=
  \frac{
  \sigma^{++}(O)+\sigma^{--}(O)-\sigma^{+-}(O)-\sigma^{-+}(O)}
  {4}\,.
\end{aligned}
\label{eq:helicity-combinations}
\end{equation}

These combinations may be formed from independent helicity samples. In the differential validation performed here, we also use a correlated-helicity estimator in which the densities for the four lepton-hadron helicity configurations are evaluated on the same accepted event. If $\rho_{00}(\Phi)$ is the sampled unpolarized density and
$\rho_{\lambda_\ell\lambda_h}(\Phi)$ is the signed density for a target
helicity configuration at the same phase-space point $\Phi$, the auxiliary analysis weight
is
\begin{equation}
  w_{\lambda_\ell\lambda_h}
  =
  w_{\rm evt}\,
  \frac{\rho_{\lambda_\ell\lambda_h}(\Phi)}
       {|\rho_{00}(\Phi)|}\,.
\label{eq:correlated-helicity-weight}
\end{equation}
In the \Herwig\ implementation,
the positive- and negative-weight contributions are generated as separate event
samples, and $w_{\rm evt}$ is the nominal event weight of the sampled contribution. The
densities include both the Born matrix-element ratio and the corresponding
signed NLO weight for the target helicity state. The absolute value in the
sampled denominator keeps the positive- and negative-weight POWHEG pieces in
the same event-weight convention, while the target density itself remains
signed. The differential analysis can therefore be filled with weights for the
four helicity states, together with the derived combinations
\begin{equation}
\begin{aligned}
  w_{\Sigma}
  &=
  \frac{w_{++}+w_{+-}+w_{-+}+w_{--}}{4}\,,
  \\
  w_{\Delta LL}
  &=
  \frac{w_{++}+w_{--}-w_{+-}-w_{-+}}{4}\,,
\end{aligned}
\label{eq:correlated-helicity-combinations}
\end{equation}
used to fill the unpolarized and polarized validation histograms. Because these
weights already contain the signed target densities, the positive- and
negative-weight histograms in the correlated estimator are added after their
separate normalizations have been applied. This differs from the
independent-helicity summaries, where the physical NLO result is formed as the
positive-weight contribution minus the negative-weight contribution for each
helicity sample. The correlated construction is an analysis-level estimator for
the differential validation. It does not reweight the accepted POWHEG Sudakov
factor or the hardest-emission choice.

\section{Polarized Deep-Inelastic Scattering}\label{sec:polarizeddis}

We now present the polarized DIS calculation developed in \Herwig, following the treatment of the unpolarized POWHEG calculation of~\cite{DErrico:2011wfa}. We first
derive the real-emission contributions, and then assemble the virtual terms and
collinear remainders needed for the NLO calculation. We validate these
ingredients in parton-level comparisons before turning to the shower-level
treatment, where the spin information is propagated through the parton shower. For reference, Table~\ref{tab:dis-notation-summary} summarizes the main kinematic variables and channel-dependent labels used in the real-emission derivation and in the assembly of the NLO terms.

\begin{table}[t]
  \centering
  \footnotesize
  \setlength{\tabcolsep}{4pt}
  \begin{tabular}{|p{0.11\textwidth}|p{0.28\textwidth}|p{0.37\textwidth}|}
    \hline
    Symbol & Channel assignment & Meaning \\
    \hline
    $x_B$ & Same in QCDC and BGF & Bjorken scaling variable of the underlying Born DIS configuration. \\
    \hline
    $x_p$ & QCDC channel: incoming quark. BGF channel: incoming gluon. & Real-emission momentum-fraction variable, so that the incoming parton in the kernel carries $x=x_B/x_p$. \\
    \hline
    $z_p$ & In both channels: outgoing-quark fraction & Phase-space variable used to parameterize the real-emission region and its collinear limits. \\
    \hline
    $x_1,x_2,x_3$ & QCDC channel: incoming quark, outgoing quark, gluon. BGF channel: incoming gluon, outgoing quark, outgoing antiquark. & Channel-dependent Breit-frame momentum fractions entering the CALKUL factorization. \\
    \hline
    $r_i,\bar r_i$ & Definitions differ between QCDC and BGF & Auxiliary massless mapped momenta used to factorize the real matrix element onto Born-like DIS kinematics. \\
    \hline
    $\ell$ & Channel independent & Breit-frame lepton energy variable, $\ell=2/y_B-1$, such that $\ell Q/2$ is the incoming-lepton energy in the Breit frame. \\
    \hline
    $D_B,\,N_B$ & Channel independent & Parity-even and parity-odd Born coefficients, including the beam polarizations and crossing signs (Section~\ref{sec:virt-coll-pol}). \\
    \hline
    $\Sigma_B$ & Channel independent & Born spin/angular structure, $\Sigma_B=(1+\ell^2)D_B+\ell N_B$, appearing in the denominators of the NLO kernels. \\
    \hline
  \end{tabular}
  \caption{Summary of the main notation used in Sections~\ref{sec:realconts} and~\ref{sec:virt-coll-pol}. The labels $x_1,x_2,x_3$ and the mapped momenta $r_i,\bar r_i$ refer to different external legs in QCDC and BGF.}
  \label{tab:dis-notation-summary}
\end{table}

\subsection{Real Emission Contributions}
\label{sec:realconts}

\subsubsection{QCD Compton Scattering}

\begin{figure}[htbp]
\includegraphics[width=0.47\linewidth]{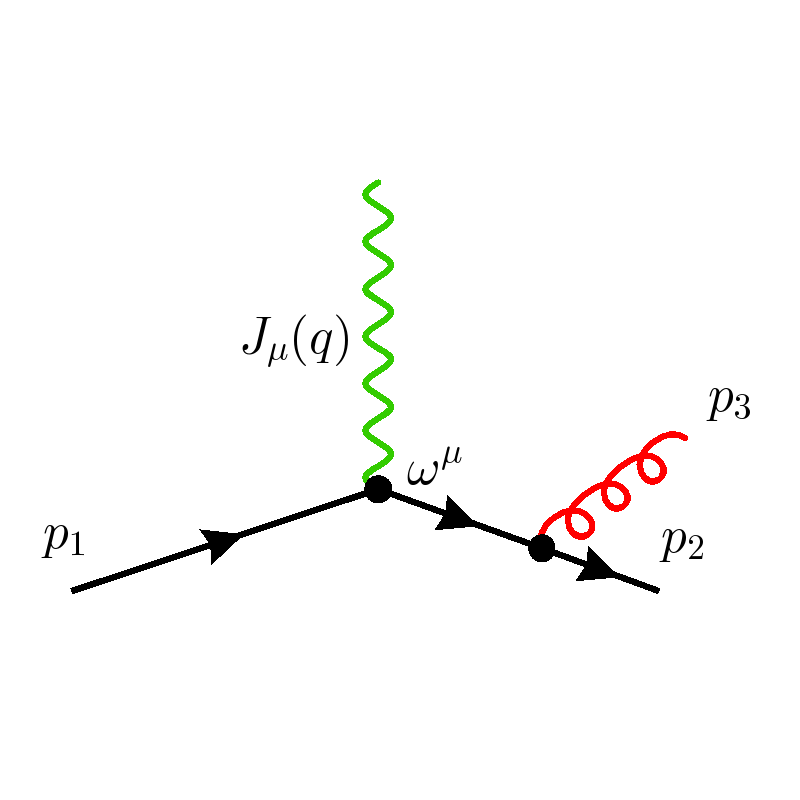}
\includegraphics[width=0.47\linewidth]{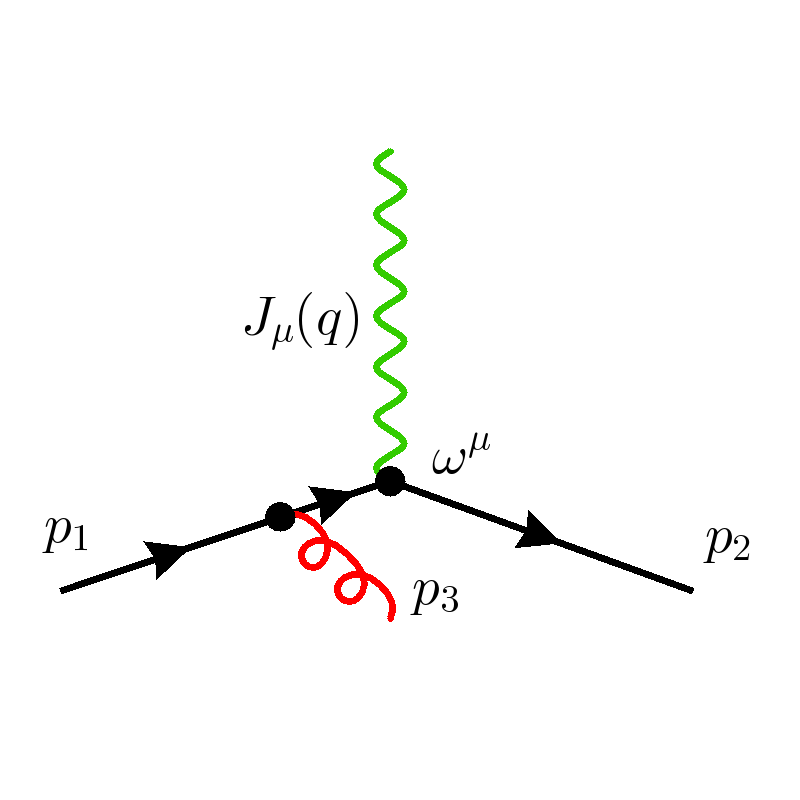}

\par\bigskip
    \caption{The QCD Compton process.}
    \label{fig:realemission}
\end{figure}

For the diagrams shown in Fig.~\ref{fig:realemission}, we write the matrix element for $eq \to eqg$, where the outgoing quark has helicity $\lambda$ and the emitted gluon has helicity $\rho$, as
\begin{equation}
  \mathcal{M} _3(\lambda,\rho) = ig_s\bar{u}_\lambda(p_2)
  \left[\slsh\epsilon_\rho(p_3)\frac{-(\slsh p_2+\slsh p_3)}{2\ldot{p_2}{p_3}}
        \omega^\mu+\omega^\mu\frac{-(\slsh p_1-\slsh p_3)}{-2\ldot{p_1}{p_3}}
        \slsh\epsilon_\rho(p_3)\right]u_\lambda(p_1)J_\mu(q) T^a_{ij}\;,
\end{equation}
where the momentum of the incoming (outgoing) quark is $p_1$ ($p_2$), and that of the gluon is $p_3$. The four-vector $\omega^\mu = (V_q + A_q \gamma^5)\gamma^\mu$ represents a general boson-quark coupling, with $(V_q,A_q)=(Q_q,0)$ for photon exchange and $(V_q,A_q)=(C_{V,q},C_{A,q})$ for $Z$ exchange in the conventions introduced below, and the $T^a_{ij}$ are the SU(3) generator matrix elements. All partons are taken to be massless, so that quark chirality---and hence helicity---is conserved. We adopt the CALKUL gluon-polarization tensor~\cite{DeCausmaecker:1981jtq}, in which the two diagrams contribute to physically distinct helicity configurations, and all interference is absorbed into the definition of $\epsilon_\rho$:
\begin{align}
  \slsh\epsilon_\rho(k, p_1, p_2) &= N\left(
    \half(1+\rho\gamma^5)\slsh k\slsh p_2\slsh p_1 -
    \slsh p_2\slsh p_1\slsh k\half(1+\rho\gamma^5) \right), \\
  N(k,p_1,p_2) &= \left[ 4 \ldot{p_1}{k} \ldot{p_2}{k} \ldot{p_1}{p_2}
    \right]^{-\half}.
\end{align}
Using the massless on-shell conditions and standard Dirac algebra, this gives
\begin{align}
  \mathcal{M}_3^+\equiv\mathcal{M}_3(\lambda,\phantom{-}\lambda) &=
    -ig_sN\bar{u}_\lambda(p_2)\omega^\mu(\slsh p_1-\slsh p_3)\slsh p_2
    u_\lambda(p_1)J_\mu(q)T^a_{ij}\;, \\
  \mathcal{M}_3^-\equiv\mathcal{M}_3(\lambda,-\lambda) &=
    -ig_sN\bar{u}_\lambda(p_2)\slsh p_1(\slsh p_2+\slsh p_3)\omega^\mu
    u_\lambda(p_1)J_\mu(q)T^a_{ij}\;.
\end{align}
We first define the momentum fractions
\begin{align}\label{eq:xi}
x_i \equiv \frac{ 2 \ldot{p_i}{q} } {\ldot{q}{q}}\;,
\end{align}
with $x_1 \leq -1$ and $x_2 \leq 1$, where the equalities are reached only at the Born level. In the real-emission region $x_1<-1$ and $x_2<1$, and $x_2$ may also change sign. Throughout, $x_B = Q^2/(2\ldot{P}{q})$ denotes the Bjorken scaling variable and $y_B = \ldot{P}{q}/\ldot{P}{q_1}$ the DIS inelasticity, where $P$ and $q_1$ are the incoming hadron and lepton momenta, respectively.
In terms of these, we introduce the auxiliary vectors
\begin{align}
  r_1       &\equiv          - p_1/x_1,\\
  \bar{r}_2 &\equiv \phantom{-}r_1+q=p_2+p_3-p_1-p_1/x_1,\\
  r_2       &\equiv \phantom{-}p_2/x_2,\\
  \bar{r}_1 &\equiv \phantom{-}r_2-q=p_1-p_3-p_2+p_2/x_2.
\end{align}
The vectors $r_1$ and $\bar{r}_2$ coincide with the incoming and outgoing quark momenta at leading order (where $x_1=-1$ and $p_3=0$) for the same values of $y_B$ and $Q^2$. By contrast, $r_2$ remains collinear with the physical outgoing quark momentum $p_2$, while $\bar{r}_1$ carries the compensating transverse recoil relative to the current direction.

These vectors allow the matrix elements to be cast in a factorized form. Momentum conservation at the boson-quark vertex gives $p_1+q=p_2+p_3$. These relations give the identities
\begin{align}
      u(\alpha p) &\cong \sqrt\alpha\,u(p)\qquad (\alpha>0), \\
      (\slsh p_1 - \slsh p_3) \slsh p_2 &= \left(\slsh p_1 - \slsh p_3 - \slsh p_2 + \frac{\slsh p_2 }{x_2} \right) \slsh p_2\\
      &= \left(-\slsh q + \frac{\slsh p_2}{x_2}\right)\slsh p_2
      = \slashed{\bar{r}_1} \slsh p_2 = u_\lambda(\bar{r}_1)\, \bar{u}_\lambda(\bar{r}_1)\, \slsh p_2\;,\\
      \slsh p_1 (\slsh p_2 + \slsh p_3) &= \slsh p_1 \left( \slsh p_2 + \slsh p_3 - \slsh p_1 - \frac{\slsh p_1 }{ x_1 } \right) \\
      &= \slsh p_1 \left(\slsh q - \frac{\slsh p_1}{x_1}\right) =\slsh p_1 \slashed{\bar{r}_2} \\
      &= \slsh p_1\, u_\lambda(\bar{r}_2)\, \bar{u}_\lambda(\bar{r}_2)\;,
\end{align}
where $\cong$ denotes equality up to an overall complex phase that cancels in $|\mathcal{M}|^2$. For the crossed initial-state factor with $x_1<0$, the spinor rescaling is understood through its crossed analytic continuation. We therefore write the modulus explicitly as $\sqrt{|x_1|}$ and absorb the associated phase into the coefficient $C^-$. The same convention is used for the $x_2$ rescaling in the real-emission region where $x_2$ changes sign: the factor $\sqrt{x_2}$ below is literal in the positive-$x_2$ sector and otherwise denotes the corresponding analytically continued spinor rescaling. The associated phases drop out of the squared coefficients, so the kernels depend only on $x_1^2$ and $x_2^2$. Inserting these into the matrix elements we obtain
\begin{align}
  \mathcal{M}_3^+ &\cong g_sN\sqrt{x_2}\;\bar{u}_\lambda(r_2)
    \omega^\mu u_\lambda(\bar{r}_1)\bar{u}_\lambda(\bar{r}_1)\slsh p_2
    u_\lambda(p_1)J_\mu(q)\,T^a_{ij} \\
  &\equiv C^+\widetilde{\mathcal{M}}_2(\bar{r}_1,r_2)\,T^a_{ij} ,\\
  C^+ &= g_sN\sqrt{x_2}\;\bar{u}_\lambda(\bar{r}_1)
    \slsh p_2u_\lambda(p_1)\;, \\
  \mathcal{M}_3^- &\cong C^-\widetilde{\mathcal{M}}_2(r_1,\bar{r}_2)\,T^a_{ij} ,\\
  C^- &= g_sN\sqrt{|x_1|}\,e^{i\varphi_1}\;\bar{u}_\lambda(p_2)
    \slsh p_1u_\lambda(\bar{r}_2)\;,
\end{align}
where $\widetilde{\mathcal{M}}_2(q_1,q_2)=\bar{u}_\lambda(q_2) \omega^\mu u_\lambda(q_1)
 J_\mu(q)$ is the colour-stripped Born matrix element for the current to scatter
 an incoming quark $q_1$ to an outgoing quark $q_2$, and $e^{i\varphi_1}$ denotes the crossing phase associated with $x_1<0$. The real-emission colour factor $T^a_{ij}$ is kept outside $\widetilde{\mathcal{M}}_2$ and yields the overall $C_F$ only after the colour trace in the squared matrix element. Finally, using the relation\footnote{See, e.g., Ref.\,\cite{DeCausmaecker:1981jtq}.}
\begin{align}
u_\pm(p) \bar{u}_\pm(p) = \half (1 \pm \gamma^5)\,\slsh p\;,
\end{align}
we obtain the explicit expressions
\begin{align}
  |C^+|^2 &= \frac{8\pi\alpha_s}{(-1-x_1)(1-x_2)Q^2} x_2^2\;,\\
  |C^-|^2 &= \frac{8\pi\alpha_s}{(-1-x_1)(1-x_2)Q^2} x_1^2\;.
\end{align}

In the Breit frame, defined by $q=(0;0,0,-Q)$ or, equivalently, $2x_B \vec{P}+\vec{q}=0$, the four-momenta of the leading-order partons and exchanged boson are:
\begin{align}
\tilde{p}_1 &= \frac{Q}2(1;0,0,1)\;;\\
\tilde{p}_2 &= \frac{Q}2(1;0,0,-1)\;;\\
q &= Q(0;0,0,-1)\;,
\end{align}
where the tilde denotes leading-order quantities. In this frame, the four-momenta of the real-emission process are:
\begin{align}
p_1 &= \frac{Q}2(-x_1;0,0,-x_1)\;;\\
p_2 &= \frac{Q}2(\sqrt{x^2_2+x_\perp^2};\phantom{-}x_\perp\cos\phi,
                                        \phantom{-}x_\perp\sin\phi,-x_2)\;;\\
p_3 &= \frac{Q}2(\sqrt{x^2_3+x_\perp^2};          -x_\perp\cos\phi,
                                                  -x_\perp\sin\phi,-x_3)\;,
\end{align}
where the $x_i$ have been defined in Eq.~\eqref{eq:xi}. Imposing momentum conservation, $x_{3}=2+x_{1}-x_{2}$, the transverse component is
\begin{equation}
x_\perp^2 = \frac{(x_3^2-x_1^2-x_2^2)^2}{4x_1^2}-x^2_2\;.
\end{equation}

With these kinematics, the leading-order cross section takes the form\footnote{At Born level the exchanged electroweak boson is colour singlet, so the colour structure is just $\delta_{ij}$ and its sum/average reduces to unity. The colour dependence is therefore implicit in $\mathrm{d}\Gamma_2$ and does not appear as a separate factor multiplying $|\widetilde{\mathcal{M}}_2|^2$.}
\begin{align}
  \mathrm{d} \sigma_2 &= \frac{1}{64\pi}\;\frac{1}{s^2 x_B^2}f_q(x_B,Q^2)
    \mathrm{d} Q^2\, \mathrm{d} x_B\, \frac{\mathrm{d}\varphi}{2\pi}\; |\widetilde{\mathcal{M}}_2(\tilde{p}_1,\tilde{p}_2)|^2 \\
  &\equiv \mathrm{d} \Gamma_2 |\widetilde{\mathcal{M}}_2(\tilde{p}_1,\tilde{p}_2)|^2\;.
\end{align}
Here $\mathrm{d}\Gamma_2$ is a compact shorthand for the Born-level flux, PDF, and phase-space measure appearing in the first line, rather than a pure invariant phase-space element alone. We use $\varphi$ for the azimuthal orientation of the Born or underlying-Born DIS scattering plane in the Breit frame and $\phi$ for the real-emission radiation azimuth, reserving $\Phi$ for phase-space measures and phase-space points.
The real-emission cross section is
\begin{align}
  \mathrm{d}\sigma_3 &= \frac{C_F}{128(2\pi)^3}\;\frac{1}{s^2x_B^2}
    f_q(-x_B x_1,Q^2)
    \mathrm{d} Q^2\, \mathrm{d} x_B\,\frac{\mathrm{d}\varphi}{2\pi}\frac{\mathrm{d} x_1}{x_1^2}\mathrm{d}x_2\frac{\mathrm{d} \phi}{2\pi}\;
    Q^2 |\mathcal{M}_3|^2 \\
  &= \frac{C_F}{4(2\pi)^2}\mathrm{d}\Gamma_2
    \frac{-x_B x_1f_q(-x_B x_1,Q^2)}
         { x_B    f_q( x_B    ,Q^2)}
    \frac{\mathrm{d} x_1}{-x_1^3}\mathrm{d}x_2
    \frac{\mathrm{d}\phi}{2\pi}\;
    Q^2 |\mathcal{M}_3|^2 \\
  &= \frac{C_F\alpha_s}{2\pi}\mathrm{d}\Gamma_2
    \frac{-x_B x_1f_q(-x_B x_1,Q^2)}
         { x_B    f_q( x_B    ,Q^2)}
    \frac{\mathrm{d} x_1 \mathrm{d}x_2}{-x_1^3(-1-x_1)(1-x_2)}\;\frac{\mathrm{d}\phi}{2\pi}
\nonumber\\&
   \times \left[ x_1^2|\widetilde{\mathcal{M}}_2(r_1,\bar{r}_2)|^2 +
           x_2^2|\widetilde{\mathcal{M}}_2(\bar{r}_1,r_2)|^2 \right]\;,
\label{eq:QCDC1}
\end{align}
where we have performed the colour trace. We now introduce the phase-space variables $x_p \equiv -1/x_1$ and
\begin{equation}
z_p \equiv \frac{\ldot{p_2}{P}}{\ldot{q}{P}} = 1 + \frac{1-x_2}{x_1}\;,
\end{equation}
where $P$ is the incoming hadron momentum. In terms of these variables, the phase-space limits factorize:
\begin{equation}
  x_B < x_p < 1\;, \qquad 0 < z_p < 1\;.
\end{equation}
Inverting these relations gives
\begin{align}
  x_1 &= -\frac{1}{x_p}\;, \\
  x_2 &= 1-\frac{1-z_p}{x_p}\;.
\end{align}
The third momentum fraction follows from momentum conservation as
\begin{equation}
  x_3 = 2+x_1-x_2 = 1-\frac{z_p}{x_p}\;.
\end{equation}
The Jacobian for the transformation is
\begin{equation}
  \left|\frac{\partial(x_1,x_2)}{\partial(x_p,z_p)}\right| = \frac{1}{x_p^3}\;,
\end{equation}
so that
\begin{equation}
  \frac{\mathrm{d} x_1 \mathrm{d}x_2}{-x_1^3(-1-x_1)(1-x_2)}
  = x_p^2\,\frac{\mathrm{d} x_p \mathrm{d}z_p}{(1-x_p)(1-z_p)}\;,
\end{equation}
because $-x_1^3=1/x_p^3$, $-1-x_1=(1-x_p)/x_p$, and $1-x_2=(1-z_p)/x_p$. Moreover,
\begin{equation}
  x_3^2-x_1^2-x_2^2 = -\,\frac{2(2x_pz_p-x_p-z_p+1)}{x_p^2}\;.
\end{equation}
The transverse momentum in units of $Q/2$ is
\begin{equation}
  x_\perp^2 = \frac{4(1-x_p)(1-z_p)z_p}{x_p}\;,
\end{equation}
and therefore
\begin{equation}
  x_p^2(x_2^2+x_\perp^2) = (2x_pz_p-x_p-z_p+1)^2\;.
\end{equation}
Substituting these relations into Eq.~\eqref{eq:QCDC1} yields the real-emission cross section:
\begin{align}
  \mathrm{d}\sigma_3 &= \frac{C_F\alpha_s}{2\pi}\frac{\mathrm{d}\phi}{2\pi}
    \frac{\frac{x_B}{x_p} f_q(x_B/x_p,Q^2)}
         { x_B    f_q( x_B    ,Q^2)}
    \frac{\mathrm{d} x_p \mathrm{d}z_p}{(1-x_p)(1-z_p)}\;
\nonumber\\&
   \times \mathrm{d} \Gamma_2 \left\{ |\widetilde{\mathcal{M}}_2(r_1,\bar{r}_2)|^2 +
        x_p^2 x_2^2 |\widetilde{\mathcal{M}}_2(\bar{r}_1,r_2)|^2 \right\}\;.
\label{eq:QCDC2}
\end{align}
Since $r_1 = -p_1/x_1$ and $\bar{r}_2 = r_1 + q$, the pair $(r_1,\bar{r}_2)$ defines a Born-like DIS configuration with the same $Q^2$, $y_B$, and azimuthal angle as the mapped leading-order point. The factor $\mathrm{d} \Gamma_2 |\widetilde{\mathcal{M}}_2(r_1,\bar{r}_2)|^2$ is therefore precisely the corresponding leading-order differential cross section, which we denote again by $\mathrm{d}\sigma_2$. Defining
\begin{align}\label{eq:R2}
R_2 \equiv \frac{x_2^2}{x^2_2+x^2_\perp} \frac{|\widetilde{\mathcal{M}}_2(\bar{r}_1,r_2)|^2}{|\widetilde{\mathcal{M}}_2(r_1,\bar{r}_2)|^2}\;,
\end{align}
so that
\begin{equation}
  x_2^2 |\widetilde{\mathcal{M}}_2(\bar{r}_1,r_2)|^2
  = (x_2^2+x_\perp^2) R_2 |\widetilde{\mathcal{M}}_2(r_1,\bar{r}_2)|^2\;,
\end{equation}
the cross section can be written as
\begin{align}
  \mathrm{d}\sigma_3 &= \frac{C_F\alpha_s}{2\pi}\frac{\mathrm{d}\phi}{2\pi}
    \frac{\frac{x_B}{x_p} f_q(x_B/x_p,Q^2)}
         { x_B    f_q( x_B    ,Q^2)}
    \frac{\mathrm{d} x_p \mathrm{d}z_p}{(1-x_p)(1-z_p)} \left\{ 1 +
         x_p^2 (x_2^2 + x_\perp^2) R_2 \right\} \mathrm{d} \sigma_2\;.
\label{eq:QCDC3}
\end{align}
Specializing to $ep$ scattering, the four-momenta of the incoming and outgoing leptons in the Breit frame are, respectively,
\begin{align}
q_1 &= \frac{Q}2(\ell;\sqrt{\ell^2-1},0,-1)\;;\\
q_2 &= \frac{Q}2(\ell;\sqrt{\ell^2-1},0,1)\;.
\end{align}
Equivalently, at the parton level, $y_B=\ldot{p_1}{q}/\ldot{p_1}{q_1}$ gives $\ell = 2/y_B -1$, and $\ell Q/2$ is the energy of the incoming lepton in the Breit frame. Defining $\cos\theta_2 = x_2 / \sqrt{x_2^2 + x_\perp^2}$, we then obtain
\begin{align}\label{eq:R2expression}
R_2 = \frac{\cos^2\theta_2+\mathcal{A}\cos\theta_2\left(\ell-\sqrt{\ell^2-1}\sin\theta_2\cos\phi\right)+\left(\ell-\sqrt{\ell^2-1}\sin\theta_2\cos\phi\right)^2}{1+\mathcal{A}\ell+\ell^2}\;.
\end{align}
The denominator is the Born DIS angular factor: the lowest-order
matrix element squared expressed in the same Breit-frame variables,
with the overall coupling normalization factored out. Here
$\mathcal{A}=0$ for pure photon exchange, while for charged-current
DIS one has $\mathcal{A}=2$ in the default $e^-q$ orientation, with
the sign flipping upon crossing either the lepton or quark
line. Unless stated otherwise, the neutral-current formulas displayed
in this subsection are written for the default $e^-q$ orientation.
Crossed channels are obtained later through the explicit
$\eta_\ell,\eta_q$ substitutions. We define the couplings
\begin{align}
C_{V,i} &= \frac{1}{\sin\theta_W \cos\theta_W} \left(\frac{T_{3,i}}{2} - Q_i \sin^2 \theta_W\right)\;,\\
C_{A,i} &= \frac{1}{\sin\theta_W \cos\theta_W} \frac{T_{3,i}}{2}\;,
\end{align}
where $\theta_W$ is the Weinberg angle, $T_{3,i}$ and $Q_i$ are the
weak isospin and charge of the fermion, respectively. For the full
neutral-current case, including $\gamma/Z$ interference, we then have
\begin{align}\label{eq:Aunpol}
\mathcal{A} = \frac{4 \kappa_Z C_{A,\ell} C_{A,q} \left(Q_\ell Q_q+2\kappa_Z C_{V,\ell}C_{V,q} \right)}
{\left(Q_\ell^2Q_q^2+2Q_\ell Q_q \kappa_Z C_{V,\ell}C_{V,q}
+\kappa_Z^2\left(C^2_{V,\ell}+C^2_{A,\ell}\right)\left(C^2_{V,q   }+C^2_{A,q   }\right)\right)}\;,
\end{align}
with the propagator ratio $\kappa_Z=\frac{Q^2}{(Q^2+m^2_Z)}$, where $m_Z$ is the $Z$ boson mass. In the numerical calculation the exact finite-width propagator factors
\begin{equation}
  \chi_{\gamma Z}(Q^2)=\frac{Q^2(Q^2+m_Z^2)}{(Q^2+m_Z^2)^2+m_Z^2\Gamma_Z^2}\,,
  \qquad
  \chi_{ZZ}(Q^2)=\frac{Q^4}{(Q^2+m_Z^2)^2+m_Z^2\Gamma_Z^2}
\end{equation}
are used instead.\footnote{For spacelike momentum transfer the propagator cannot resonate, so the width term is formally a higher-order effect. It is retained here for definiteness, and shifts the $Z$-exchange contributions only at the sub-permille level.} With the present conventions for $C_{V,i}$ and $C_{A,i}$, the exact-width result is obtained algebraically by replacing every interference-term factor $\kappa_Z$ by $\chi_{\gamma Z}(Q^2)$ and every pure-$Z$ factor $\kappa_Z^2$ by $\chi_{ZZ}(Q^2)$, without introducing any additional powers of $\sin\theta_W\cos\theta_W$. A similar treatment of the finite-width effect follows for the charged-current case.

\paragraph{Polarized case.}
So far we have worked with unpolarized incoming partons. The same procedure applies to definite-helicity incoming partons with polarizations $P_\ell, P_q=\pm 1$, yielding $\mathrm{d} \sigma_3^{P_\ell,P_q}$ with $\mathrm{d}\sigma_2^{P_\ell,P_q}$ appearing in Eq.~\eqref{eq:QCDC3}. The same functional form for $R_2$ arises as in Eq.~\eqref{eq:R2expression}. We evaluated the relevant amplitudes using the \texttt{FeynCalc} package~\cite{Mertig:1990an,Shtabovenko:2016sxi,Shtabovenko:2020gxv,Shtabovenko:2023idz}, which gives, for the full neutral-current case,
\begin{align}\label{eq:APol}
\mathcal{A}^{NC}_{P_\ell,P_q} = \frac{ N_0 + P_\ell N_\ell + P_q N_q + P_\ell P_q N_{\ell q}} {D_0 + P_\ell D_\ell + P_q D_q + P_\ell P_q D_{\ell q} }\;,
\end{align}
where
\begin{align}\label{eq:NDsA}
N_0 &= 4 \kappa_Z C_{A,\ell} C_{A,q} \left(Q_\ell Q_q+2\kappa_Z C_{V,\ell}C_{V,q} \right)\;;\\
N_\ell &= -4\kappa_Z C_{A,q}  (C_{V,q} \kappa_Z (C_{A,\ell}^2+C_{V,\ell}^2)+C_{V,\ell} Q_\ell Q_q)\;;\\
N_q &= -4\kappa_Z C_{A,\ell} (C_{V,\ell} \kappa_Z (C_{A,q}^2+C_{V,q}^2)+C_{V,q} Q_\ell Q_q)\;; \\
N_{\ell q} &= 2 \left(\kappa_Z^2 (C_{A,\ell}^2+C_{V,\ell}^2) (C_{A,q}^2+C_{V,q}^2)+2\kappa_Z C_{V,\ell} C_{V,q} Q_\ell Q_q  + Q_\ell^2 Q_q^2\right)\;;\\
D_0 &= \left(Q_\ell^2Q_q^2+2Q_\ell Q_q \kappa_Z C_{V,\ell}C_{V,q}
+\kappa_Z^2\left(C^2_{V,\ell}+C^2_{A,\ell}\right)\left(C^2_{V,q   }+C^2_{A,q   }\right)\right)\;; \\
D_\ell &= -2\kappa_Z C_{A,\ell} \left(\kappa_Z C_{V,\ell} (C_{A,q}^2+C_{V,q}^2)+C_{V,q} Q_\ell Q_q\right)\;; \\
D_q &= -2\kappa_Z C_{A,q} \left( \kappa_Z C_{V,q} (C_{A,\ell}^2+C_{V,\ell}^2)+C_{V,\ell} Q_\ell Q_q\right)\;;\\
D_{\ell q} &= 2\kappa_Z C_{A,\ell} C_{A,q} (2\kappa_Z C_{V,\ell} C_{V,q} + Q_\ell Q_q)\;.\label{eq:NDsAFinal}
\end{align}
By linearity of the spin-density-matrix decomposition, the same coefficient structure can be used for mixed longitudinal polarization states by promoting $P_\ell$ and $P_q$ from helicity eigenvalues to continuous polarization parameters in the interval $[-1,1]$. This is the form used later when the incoming parton polarization is reconstructed from $P_q(x)=P_z\,\Delta f_q(x)/f_q(x)$. As a consistency check, the unpolarized case is recovered by setting $P_\ell = P_q = 0$, giving $\mathcal{A}^{NC}_{0,0}=\mathcal{A}$ as in Eq.~\eqref{eq:Aunpol}. The pure photon contribution follows from setting $C_{A,\ell}, C_{A,q}, C_{V,\ell}, C_{V,q}=0$, which yields $\mathcal{A}^{\gamma}_{P_\ell, P_q} = 2 P_\ell P_q$. For the charged-current process in the default $e^-q$ orientation, the angular analysing power of the non-vanishing left-handed helicity channel is $\mathcal{A}^{CC}=2$. The explicit beam and parton polarizations enter instead through the overall Born normalization and chiral projectors, which vanish for wrong-chirality configurations. Crossing either the lepton or quark line flips the sign.

Having established the QCD Compton corrections, we now turn to the boson-gluon fusion channel. This channel introduces a qualitatively different flavour structure and supplies the gluon-initiated contribution to the NLO DIS calculation.

\subsubsection{Boson-Gluon Fusion}
\label{sec:BGfusion}

\begin{figure}[htbp]
\includegraphics[width=0.47\linewidth]{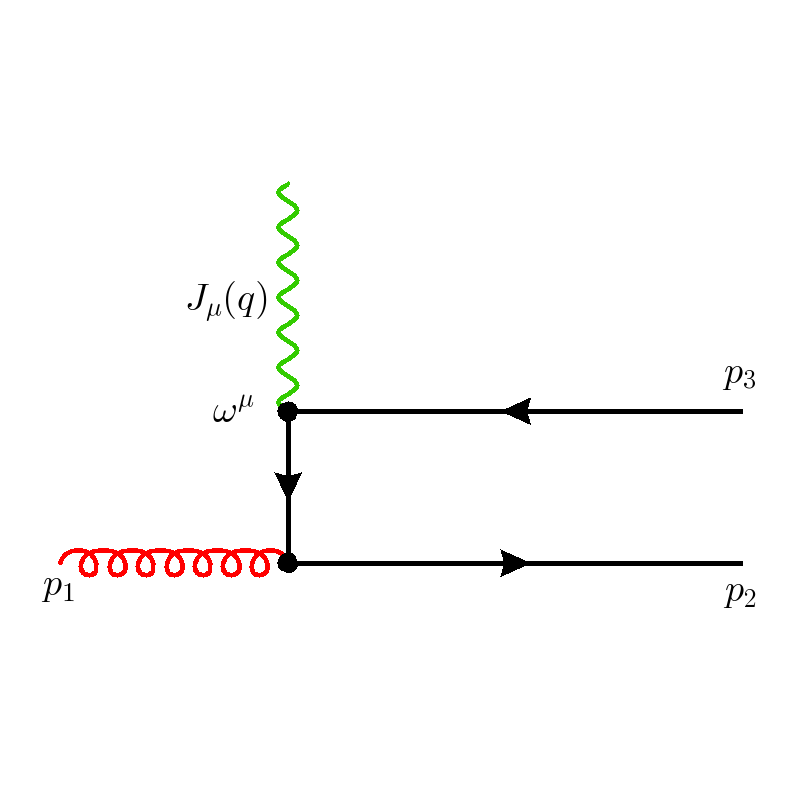}
\includegraphics[width=0.47\linewidth]{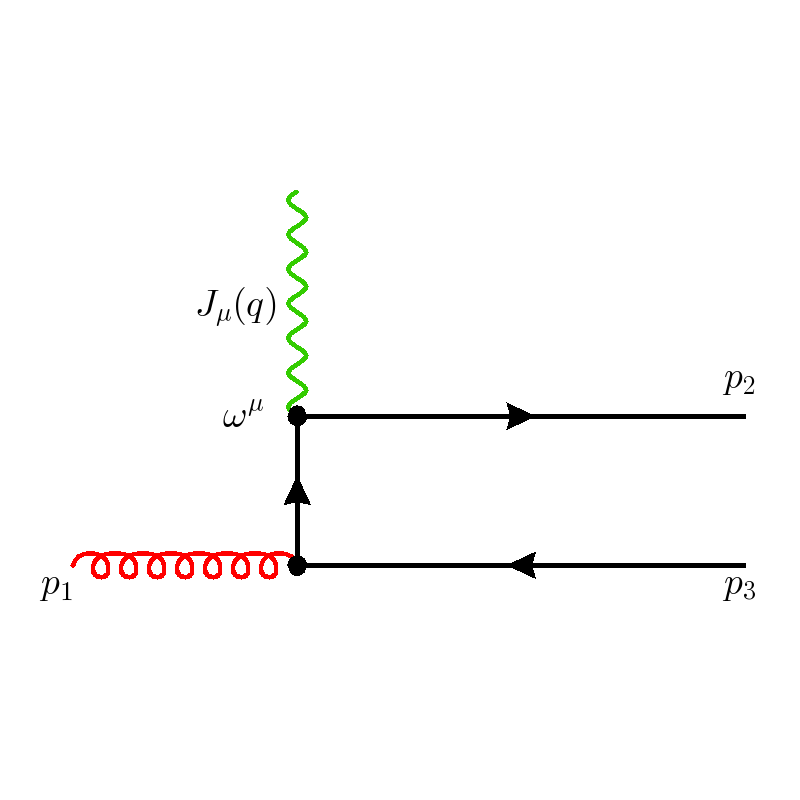}
    \caption{The boson-gluon fusion process.}
    \label{fig:realemissionBGF}
\end{figure}

The boson-gluon fusion matrix element (Fig.~\ref{fig:realemissionBGF}) can be related to the QCD Compton amplitude by crossing. We nevertheless present the derivation for the physical $V^\ast g\to q\bar q$ process explicitly, since this form fixes the channel assignments used throughout the calculation. Written in standard quark/antiquark spinor notation, the matrix element is
\begin{equation}
  \mathcal{M}_3(\lambda,\rho) = ig_s\bar{u}_\lambda(p_2)
  \left[\slsh\epsilon_\rho(p_1)\frac{-(\slsh p_2-\slsh p_1)}{-2\ldot{p_1}{p_2}}
        \omega^\mu+\omega^\mu\frac{ (\slsh p_3-\slsh p_1)}{-2\ldot{p_1}{p_3}}
        \slsh\epsilon_\rho(p_1)\right]v_\lambda(p_3)J_\mu(q)\,T^a_{ij}\;,
\end{equation}
where $p_1$ denotes the incoming gluon momentum and $p_2$, $p_3$ are the momenta of the outgoing quark and antiquark, respectively. For the crossed-spinor manipulations used below, the outgoing antiquark is represented as an incoming quark line of momentum $-p_3$. Equivalently, the algebra is written in terms of the crossed spinor chain with $u_\lambda(p_3)$, and the corresponding crossing phase is absorbed into the scalar coefficients $C^\pm$. The auxiliary vectors are
\begin{align}
\label{minus}
  r_3       &\equiv - p_3/x_3\;,\\
  \bar{r}_2 &\equiv \phantom{-}r_3+q=p_2-p_1+p_3-p_3/x_3\;,\\
  r_2       &\equiv \phantom{-}p_2/x_2\;,\\
  \bar{r}_3 &\equiv \phantom{-}r_2-q=-p_3+p_1-p_2+p_2/x_2\;.
\end{align}

These vectors again allow the matrix elements to be cast in a factorized form. Momentum conservation for $V^\ast g\to q\bar q$ gives
\begin{equation}
  q+p_1=p_2+p_3\;,
\end{equation}
or equivalently $q=p_2+p_3-p_1$, so the combinations $-p_3+p_1-p_2$ and $p_2-p_1+p_3$ are simply $-q$ and $q$, respectively. Using this together with the definitions of $\bar r_2$ and $\bar r_3$, $\slsh p_3^2=0$, and $u(\alpha p)\cong \sqrt{\alpha}\,u(p)$ for lightlike momenta, the Dirac chains may be rewritten as
\begin{align}
      (\slsh p_1 - \slsh p_3) \slsh p_2 &= \left(-\slsh p_3 + \slsh p_1 - \slsh p_2 + \frac{\slsh p_2 }{x_2} \right) \slsh p_2\\
      &= \left(-\slsh q + \frac{\slsh p_2}{x_2}\right)\slsh p_2
      = \slashed{\bar{r}_3} \slsh p_2 = u_\lambda(\bar{r}_3)\, \bar{u}_\lambda(\bar{r}_3)\, \slsh p_2\;,\\
      \slsh p_3 (\slsh p_2 - \slsh p_1) &= \slsh p_3 \left( \slsh p_2 - \slsh p_1 - \slsh p_3 - \frac{\slsh p_3 }{ x_3 } \right) \\
      &= \slsh p_3 \left(\slsh q - 2\slsh p_3 - \frac{\slsh p_3}{x_3}\right)
      = \slsh p_3 \left(\slsh q - \frac{\slsh p_3}{x_3}\right) =\slsh p_3 \slashed{\bar{r}_2} \\
      &= \slsh p_3\, u_\lambda(\bar{r}_2)\, \bar{u}_\lambda(\bar{r}_2)\;,
\end{align}
where, in the second line, the term proportional to $\slsh p_3^2$ vanishes because $p_3^2=0$. As in the QCDC case, the crossed spinor rescaling for the antiquark channel is understood via analytic continuation when $x_3<0$. We therefore write the modulus explicitly and absorb the associated phase into $C^-$. This leads to
\begin{align}
  \mathcal{M}_3^+ &\cong g_sN\sqrt{x_2}\;\bar{u}_\lambda(r_2)
    \omega^\mu u_\lambda(\bar{r}_3)\bar{u}_\lambda(\bar{r}_3)\slsh p_2
    u_\lambda(p_3)J_\mu(q)\,T^a_{ij} \\
  &\equiv C^+\widetilde{\mathcal{M}}_2(\bar{r}_3,r_2)\,T^a_{ij} ,\\
  C^+ &= g_sN\sqrt{x_2}\;\bar{u}_\lambda(\bar{r}_3)
    \slsh p_2 u_\lambda(p_3)\;, \\
  \mathcal{M}_3^- &\cong C^-\widetilde{\mathcal{M}}_2(r_3,\bar{r}_2)\,T^a_{ij} ,\\
  C^- &= g_sN\sqrt{|x_3|}\,e^{i\varphi_3}\;\bar{u}_\lambda(p_2)
    \slsh p_1u_\lambda(\bar{r}_2)\;.
\end{align}
Here $x_3$ refers to the outgoing antiquark, in contrast to the QCDC case where it labels the gluon, and the phase $e^{i\varphi_3}$ drops out of $|C^-|^2$. We obtain
\begin{align}
  |C^+|^2 &= \frac{8\pi\alpha_s}{(1-x_3)(1-x_2)Q^2} x_2^2 ,\\
  |C^-|^2 &= \frac{8\pi\alpha_s}{(1-x_3)(1-x_2)Q^2} x_3^2\;.
\end{align}

We adopt the same parametrization in terms of $x_p$ and $z_p$, where $z_p$ continues to denote the outgoing-quark momentum fraction and $x_p$ now refers to the incoming gluon. We then obtain
\begin{align}
  x_3 &= 1-\frac{z_p}{x_p}, \\
  x_2 &= 1-\frac{1-z_p}{x_p}.
\end{align}
Hence
\begin{equation}
  1-x_3=\frac{z_p}{x_p},\qquad
  1-x_2=\frac{1-z_p}{x_p},\qquad
  \frac{1}{(1-x_3)(1-x_2)}=\frac{x_p^2}{z_p(1-z_p)}\;.
\end{equation}
This gives
\begin{align}
  |C^+|^2 &= \frac{8\pi\alpha_s}{z_p(1-z_p)Q^2}
    \left\{x_p^2(x_2^2+x_\perp^2)\right\}
    \frac{x_2^2}{x_2^2+x_\perp^2}, \\
  |C^-|^2 &= \frac{8\pi\alpha_s}{z_p(1-z_p)Q^2}
    \left\{x_p^2(x_3^2+x_\perp^2)\right\}
    \frac{x_3^2}{x_3^2+x_\perp^2}.
\end{align}
After the colour trace, which now yields $T_R$, the real-emission cross section is then
\begin{align}
  \mathrm{d}\sigma_3
  &= \frac{T_R\alpha_s}{2\pi} \mathrm{d}\Gamma_2
    \frac{\myfrac{x_B}{x_p}f_g(\myfrac{x_B}{x_p},Q^2)}
         {        x_B      f_q(        x_B      ,Q^2)}
    \frac{\mathrm{d} x_p \mathrm{d}z_p}{z_p(1-z_p)}\;\frac{\mathrm{d}\phi}{2\pi}
\nonumber\\&\times
  \left\{
    \left(x_p^2(x_3^2+x_\perp^2)\right)R_3+
    \left(x_p^2(x_2^2+x_\perp^2)\right)R_2\right\}
    |\widetilde{\mathcal{M}}_2(\tilde{p}_1,\tilde{p}_2)|^2,
\end{align}
where $\tilde{p}_{1,2}$ are the Born-level momenta, and
\begin{align}
  R_3 &= \frac{x_3^2}{x_3^2+x_\perp^2}
    \;\frac{|\widetilde{\mathcal{M}}_2(r_3,\bar{r}_2)|^2}{|\widetilde{\mathcal{M}}_2(\tilde{p}_1,\tilde{p}_2)|^2}, \\
  R_2 &= \frac{x_2^2}{x_2^2+x_\perp^2}
    \;\frac{|\widetilde{\mathcal{M}}_2(\bar{r}_3,r_2)|^2}{|\widetilde{\mathcal{M}}_2(\tilde{p}_1,\tilde{p}_2)|^2},
\end{align}
where $R_2$ and $R_3$ now denote the corresponding BGF ratios normalized to the unmapped Born configuration. For the explicit DIS case, these expressions may be rewritten in the same form
as before. One obtains the same expression for $R_2$, while $R_3$ is obtained
by the corresponding angular replacement $\cos\theta_2 \to -\cos\theta_3$,
$\sin\theta_2 \to \sin\theta_3$, and $\phi \to \pi-\phi$.
\begin{align}\label{eq:BGF}
  \mathrm{d}\sigma_3 &= \frac{T_R\alpha_s}{2\pi}\frac{\mathrm{d}\phi}{2\pi}
    \frac{\frac{x_B}{x_p} f_g\!\left(\frac{x_B}{x_p},Q^2\right)}{ x_B f_q\!\left( x_B ,Q^2\right)}
    \frac{\mathrm{d} x_p \mathrm{d}z_p}{z_p(1-z_p)} \\\nonumber
    & \times \left\{ x_p^2 ( x_3^2 + x_\perp^2 ) R_3 +
        x_p^2 (x_2^2 + x_\perp^2) R_2 \right\} \mathrm{d} \sigma_2\;.
\end{align}
At this stage the expression does not yet distinguish quark from antiquark scattering. The two endpoint poles arise from the two Born limits identified above: $z_p\to1$ implies $1-x_2\to0$ and therefore the quark-collinear limit, while $z_p\to0$ implies $1-x_3\to0$ and therefore the antiquark-collinear limit. If we view Eq.~\eqref{eq:BGF} as a correction to a given lowest-order process, partons and antipartons should be treated equivalently. The $z_p = 1$ singularity corresponds to configurations collinear to the quark-initiated Born process, while $z_p = 0$ corresponds to the antiquark-initiated one. We therefore separate
\begin{equation}
\frac{1}{z_{p}(1-z_{p})}=\frac{1}{z_{p}}+\frac{1}{1-z_{p}}\;,
\label{eq:zp-partfrac}
\end{equation}
and rewrite the cross section as
\begin{align}
\mathrm{d}\sigma_3^{(q)} = \frac{T_R\alpha_s}{2\pi}\frac{\mathrm{d}\phi}{2\pi}
\frac{\mathrm{d}x_p\mathrm{d}z_p}{(1-z_p)} \frac{\frac{x_B}{x_p}f_g(\frac{x_B}{x_p}, Q^2)}{x_Bf_q(x_B, Q^2)}
\left\{x_p^2(x_2^2+x_\perp^2)R_2+x_p^2(x_3^2+x_\perp^2)R_3\right\}\mathrm{d} \sigma_2\;,
\end{align}
\begin{align}
\mathrm{d}\sigma_3^{(\bar q)} = \frac{T_R\alpha_s}{2\pi}\frac{\mathrm{d}\phi}{2\pi}
\frac{\mathrm{d}x_p\mathrm{d}z_p}{z_p} \frac{\frac{x_B}{x_p}f_g(\frac{x_B}{x_p}, Q^2)}{x_Bf_q(x_B, Q^2)}
\left\{x_p^2(x_2^2+x_\perp^2)R_2+x_p^2(x_3^2+x_\perp^2)R_3\right\}\mathrm{d} \sigma_2\;,
\end{align}
where $\mathrm{d}\sigma_3^{(q)}$ is attached to the quark-initiated underlying Born configuration and $\mathrm{d}\sigma_3^{(\bar q)}$ to the charge-conjugate antiquark one~\cite{Seymour:1994we,Seymour:1994ti}. The same kernel braces therefore appear in both channels. Only the endpoint factor and the associated underlying Born flavour assignment differ.

The final expression for $R_2$ is as for the QCD Compton case, but with the substitution $P_q \rightarrow P_g$ in the polarized analysing power, where $P_g(x)=P_z\,\Delta f_g(x)/f_g(x)$ denotes the longitudinal gluon polarization induced by the polarized hadron. This follows from the CALKUL decomposition~\cite{DeCausmaecker:1981jtq}: for each gluon helicity $\lambda_g$, chirality conservation along the massless quark propagator selects a unique quark helicity at the electroweak vertex, and QCD parity invariance ensures that the two gluon helicities contribute with equal weight. For massless quarks, the net effect in the CALKUL factorized analysing-power numerator is that the effective quark polarization at the DIS vertex equals the gluon polarization, $P_q^{\mathrm{eff}}=P_g$. This statement refers to the analysing power entering the factorized kernels, in which the $z_p$ dependence of the polarization transfer is carried by the prefactors; it should not be confused with the collinear polarization-transfer ratio $\Delta P_{qg}(z)/P_{qg}(z)$, which emerges only after azimuthal averaging in the collinear limit.

$R_3$ is given by
\begin{align}\label{eq:R3expression}
R_3 = \frac{\cos^2\theta_3-\mathcal{A}\cos\theta_3\left(\ell+\sqrt{\ell^2-1}\sin\theta_3\cos\phi\right)+\left(\ell+\sqrt{\ell^2-1}\sin\theta_3\cos\phi\right)^2}{1+\mathcal{A}\ell+\ell^2}\;,
\end{align}
where $\cos\theta_3 = x_3 / \sqrt{x_3^2 + x_\perp^2}$ and $\mathcal{A}$ is the unpolarized analysing power of Eq.~\eqref{eq:Aunpol}. For polarized beams, the $R_3$ analysing power is evaluated with the charge-conjugate incoming and outgoing Born-parton labels and with the substitution $P_q \rightarrow - P_g$. The two BGF limits are therefore treated as
\begin{equation}
  R_2:\ \mathcal{A}_{R_2}=\mathcal{A}^{NC}(q_{\rm Born},+P_g)\,,
  \qquad
  R_3:\ \mathcal{A}_{R_3}=\mathcal{A}^{NC}(q_{\rm Born}^{\rm conj},-P_g)\,,
\end{equation}
where $q_{\rm Born}^{\rm conj}$ denotes the charge-conjugate of the Born quark flavour. In the photon-only limit this reduces to $\mathcal{A}_{R_2}=2P_\ell P_g$ and $\mathcal{A}_{R_3}=-2P_\ell P_g$. For $\gamma+Z$ or $Z$ exchange, the charge-conjugated flavour labels are kept when evaluating Eqs.~\eqref{eq:APol} and~\eqref{eq:NDsA}-\eqref{eq:NDsAFinal}. In the full neutral-current case the same mapped polarizations also enter the parity-even denominators inside the Born-matrix-element ratios defining $R_2$ and $R_3$, so the shorthand in Eq.~\eqref{eq:R3expression} changes the full ratio, not only the numerator analysing power.

The real-emission contributions derived here contain soft and collinear singularities. We combine them with the virtual corrections and collinear counterterms in the next subsection, obtaining the finite ingredients used in the POWHEG calculation.

\subsection{Virtual Contribution and Collinear Remainders}
\label{sec:virt-coll-pol}

We use dimensional regularization in $d=4-2\epsilon$ and the $\overline{\text{MS}}$ scheme for UV renormalization and collinear factorization. For $\gamma_5$ we employ the HVBM/Larin prescription~\cite{tHooft:1972tcz,Breitenlohner:1977hr,Larin:1993tq}. In the polarized case this requires the standard finite Larin-to-polarized-$\overline{\text{MS}}$ scheme transformation, which at this order is given by $\Delta K^{\text{F.S.}}_{qq}(x)=4C_F(1-x)$ and $\Delta K^{\text{F.S.}}_{ab}(x)=0$ otherwise~\cite{Mertig:1995ny,Vogelsang:1996im}.\footnote{This is the scheme assumed in Eqs.~\eqref{eq:Dfm-pol} and \eqref{eq:DA-kernel} below: after this finite transformation the polarized quark kernels coincide with the unpolarized ones, so the difference between the polarized and unpolarized collinear luminosities enters only through $\Delta f_i$ and the replacement $A\to\Delta A$.}

We work with $Q^2\equiv -q^2>0$ in the Breit frame. In massless QCD the one-loop quark form factor is helicity conserving, so the renormalized virtual correction multiplies the Born spin projector and remains diagonal in helicity.

\paragraph{Flux convention and Born decomposition.}
We primarily generate the \emph{physical cross section for polarized beams} rather than the spin-difference observable $\Delta\sigma$.\footnote{For selected validation samples the four longitudinal helicity combinations and the derived $\sigma$ and $\Delta\sigma$ combinations are evaluated on the same accepted event, as described in Section~\ref{sec:correlated-helicity-weights}.}
The hadron spin-density matrix of Sec.~\ref{sec:formalism}, together with the lepton-helicity dependence of the Born amplitude, carries the spin dependence of the Born cross section. We collect this dependence into the exact Born structure $\Sigma_B(\Phi_B)$.

The virtual correction is evaluated at the Born momentum fraction $x_B$, is helicity-diagonal, and therefore multiplies the polarized Born uniformly.
However, the finite collinear counterterms and the real-emission PDF flux ratios involve convolutions with PDFs at the mapped fraction $x_B/x_p$.  Because the polarized and unpolarized gluon splitting kernels differ ($\Delta P_{gq}(z) = 2z-1$ vs.\ $P_{gq}(z) = z^2+(1{-}z)^2$)~\cite{Altarelli:1977zs,Mertig:1995ny},\footnote{Our notation $\Delta P_{gq}$ follows the branching assignment for $g\to q\bar q$ used here. In the standard DGLAP convention the same kernels are usually denoted as $\Delta P_{qg}$ and $P_{qg}$.} these convolutions must be decomposed into unpolarized and polarized parts, each with the appropriate splitting functions and PDFs.
We follow the arrangement of Ref.~\cite{DErrico:2011wfa} when organizing the finite virtual and collinear pieces.

Specializing the general polarized cross-section formula of Eq.~\eqref{eqn:dsigma} to the Born-level DIS configuration, we write
\begin{equation}
  \mathrm{d}\sigma_B^{\text{pol}}
  =
  \mathrm{d}x_B\, f_i(x_B,\mu_F)\,
  \frac{\mathrm{d}\hat\sigma_B^{\text{pol}}}{\mathrm{d}\Phi_2}\,
  \mathrm{d}\Phi_2 ,
\end{equation}
where $i=q,\bar q$ denotes the incoming Born parton flavour, and $\mathrm{d}\hat\sigma_B^{\text{pol}}/\mathrm{d}\Phi_2$ is the Born-level partonic cross section, with the hadron polarization carried by the partonic spin-density matrix $H_{i\tau\tau'}(x_B,\mu_F)$ and the lepton polarization entering through the Born helicity amplitude. In the notation used below, this polarization dependence is summarized by
\begin{equation}
  \Sigma_B(\Phi_B)\equiv (1+\ell^2)\,D_B+\ell\,N_B\,,
\label{eq:born-decomp}
\end{equation}
where $D_B$ and $N_B$ denote the parity-even and parity-odd Born coefficients defined explicitly below. Equivalently, the Born partonic cross section may be written as
\begin{equation}
  \frac{\mathrm{d}\hat\sigma_B^{\text{pol}}}{\mathrm{d}\Phi_2}
  =
  \mathcal{N}_B(\Phi_B)\,\Sigma_B(\Phi_B)\,,
\end{equation}
so that
\begin{equation}
  \mathrm{d}\sigma_B^{\text{pol}}
  =
  \mathrm{d}x_B\, f_i(x_B,\mu_F)\,\mathrm{d}\Phi_2\,
  \mathcal{N}_B(\Phi_B)\,\Sigma_B(\Phi_B)\,.
\end{equation}
Here $\mathcal{N}_B$ collects the common Born normalization inherited from Eq.~\eqref{eqn:dsigma}, namely the flux factor together with the overall electroweak prefactor common to all helicity configurations. The reduced coefficients $D_B$ and $N_B$ therefore contain the non-trivial $\gamma/Z$ coupling combinations and polarization dependence, while $\Sigma_B$ is the spin/angular Born factor that survives in the ratios entering the NLO kernels.

For quark-initiated DIS the renormalized finite virtual correction reads
\begin{equation}
\frac{\mathrm{d}\hat\sigma_V^{\text{pol}}}{\mathrm{d}\Phi_2}
=
\frac{\alpha_s(\mu_R)}{2\pi}\,C_F\;
V(x_B;\mu_F)\;
\frac{\mathrm{d}\hat\sigma_B^{\text{pol}}}{\mathrm{d}\Phi_2}\,,
\label{eq:virt-pol}
\end{equation}
with
\begin{equation}
V(x_B;\mu_F)=
-\frac{\pi^2}{3}-\frac{9}{2}
+\frac{3}{2}\ln\!\frac{Q^2}{\mu_F^2(1-x_B)}
+2\ln(1-x_B)\,\ln\!\frac{Q^2}{\mu_F^2}
+\ln^2(1-x_B)\,.
\label{eq:VxB}
\end{equation}
The factor $V(x_B;\mu_F)$ therefore multiplies the full helicity structure of the Born term.

The initial-state collinear remnants are organized through modified luminosities that collect the quark- and gluon-initiated collinear limits in correspondence with the QCDC and BGF real-emission channels.
We first define the unpolarized quark and gluon modified luminosities. For an incoming quark flavour $q$,
\begin{align}
f^{\,q}_m(x_B,\mu_F)
&=\int_{x_B}^{1}\frac{\mathrm{d}x_p}{x_p}\Bigg\{
\Big[f_q\!\left(\frac{x_B}{x_p},\mu_F\right)-x_p\,f_q(x_B,\mu_F)\Big]\,B(x_p)
+\ f_q\!\left(\frac{x_B}{x_p},\mu_F\right)\,C(x_p)
\Bigg\}.
\label{eq:fm-pol}
\end{align}
The corresponding unpolarized gluon modified luminosity is
\begin{align}
f^{\,g}_m(x_B,\mu_F)
&=\int_{x_B}^{1}\frac{\mathrm{d}x_p}{x_p}\,
f_g\!\left(\frac{x_B}{x_p},\mu_F\right)\,A(x_p)\,.
\label{eq:fmg-pol}
\end{align}
The kernels entering the collinear remainder coincide with those used in the standard unpolarized DIS treatment and are given explicitly by
\begin{align}
A(x_p)&=\frac{T_R}{C_F}\Big[x_p^2+(1-x_p)^2\Big]\,
\ln\!\frac{Q^2(1-x_p)}{\mu_F^2\,x_p}
+\frac{2T_R}{C_F}\,x_p(1-x_p),
\label{eq:A-kernel}\\[1mm]
B(x_p)&=\left[\frac{2}{1-x_p}\,\ln\!\frac{Q^2(1-x_p)}{\mu_F^2}-\frac{3}{2}\,\frac{1}{1-x_p}\right],
\label{eq:B-kernel}\\[1mm]
C(x_p)&=\left[(1-x_p)-\frac{2}{1-x_p}\ln x_p-(1+x_p)\,\ln\!\frac{Q^2(1-x_p)}{\mu_F^2\,x_p}\right].
\label{eq:C-kernel}
\end{align}
The kernel $A$ therefore represents the $g\!\to\! q\bar q$ collinear configuration that is the initial-state limit of the BGF channel in Sec.~\ref{sec:BGfusion}. The kernels $B$ and $C$ organize the quark-initiated remnants by separating the end-point subtraction from the non-local logarithms, so that the convolution in Eq.~\eqref{eq:fm-pol} is finite in the plus-distribution sense and directly comparable to the integrated subtraction used on the real side.
Throughout, endpoint terms are understood in the standard plus-distribution sense,
\begin{equation}
  \int_{x_B}^{1}\!\mathrm{d}x_p\,\frac{\phi(x_p)}{(1-x_p)_+}
  \equiv
  \int_{x_B}^{1}\!\mathrm{d}x_p\,\frac{\phi(x_p)-\phi(1)}{1-x_p}
  +\phi(1)\ln(1-x_B)\,,
\end{equation}
for any smooth test function $\phi(x_p)$.
In Eq.~\eqref{eq:fm-pol} this boundary term vanishes in the quark-subtracted piece because the corresponding test function,
\begin{equation}
  \phi_q(x_p)=\frac{1}{x_p}\Big[f_q\!\left(\frac{x_B}{x_p},\mu_F\right)-x_p\,f_q(x_B,\mu_F)\Big],
\end{equation}
satisfies $\phi_q(1)=0$. The same statement holds for the polarized replacement $f_q\to\Delta f_q$ in Eq.~\eqref{eq:Dfm-pol}.

\paragraph{Polarized collinear counterterms.}
For the polarized component of the cross section, the corresponding quark and gluon modified luminosities are
\begin{align}
\Delta f^{\,q}_m(x_B,\mu_F)
&=\int_{x_B}^{1}\frac{\mathrm{d}x_p}{x_p}\Bigg\{
\Big[\Delta f_q\!\left(\frac{x_B}{x_p},\mu_F\right)-x_p\,\Delta f_q(x_B,\mu_F)\Big]\,B(x_p)
\nonumber\\&\hspace{4.2cm}
+\ \Delta f_q\!\left(\frac{x_B}{x_p},\mu_F\right)\,C(x_p)
\Bigg\},
\label{eq:Dfm-pol}
\end{align}
and
\begin{align}
\Delta f^{\,g}_m(x_B,\mu_F)
&=\int_{x_B}^{1}\frac{\mathrm{d}x_p}{x_p}\,
\Delta f_g\!\left(\frac{x_B}{x_p},\mu_F\right)\,\Delta A(x_p)\,,
\label{eq:Dfmg-pol}
\end{align}
where the polarized gluon kernel is
\begin{equation}
\Delta A(x_p)=\frac{T_R}{C_F}\big(2x_p-1\big)\,
\ln\!\frac{Q^2(1-x_p)}{\mu_F^2\,x_p}
+\frac{2T_R}{C_F}\,(1-x_p)\,.
\label{eq:DA-kernel}
\end{equation}
This is the polarized \emph{gluon collinear} kernel in the modified-luminosity normalization used here. Together with the finite BGF remainder below it reproduces the standard polarized DIS gluon coefficient function. It contains the polarized splitting function $\Delta P_{gq}(z) = 2z - 1$ (with associated finite part $2(1{-}z)$) in place of the unpolarized $P_{gq}(z)=z^2+(1{-}z)^2$ (with finite part $2z(1{-}z)$). The quark kernels $B$ and $C$ are identical for the polarized and unpolarized luminosities, since $\Delta P_{qq}(z) = P_{qq}(z)$ after the finite Larin-to-polarized-$\overline{\text{MS}}$ scheme transformation discussed at the start of this subsection.

\paragraph{Decomposition of the collinear contribution.}
In the event-generator calculation, the Born partonic cross section for polarized beams at fixed lepton kinematics $(\ell \equiv 2/y_B - 1)$ is therefore written in terms of the Born structure introduced above as
\begin{equation}
  \frac{\mathrm{d}\hat\sigma_B^{\text{pol}}}{\mathrm{d}\Phi_2} \propto \Sigma_B(\Phi_B)\,,
\end{equation}
where
\begin{equation}
  D_B=D_0+\eta_\ell P_\ell D_\ell+\eta_q P_q D_q+\eta_\ell\eta_q P_\ell P_q D_{\ell q}\,,
  \qquad
  N_B=\eta_\ell\eta_q N_0+\eta_q P_\ell N_\ell+\eta_\ell P_q N_q+P_\ell P_q N_{\ell q}\,,
\end{equation}
and the lepton and quark crossing signs are
\begin{equation}
  \eta_\ell =
  \begin{cases}
    +1 & \text{for } e^- \\
    -1 & \text{for } e^+
  \end{cases},
  \qquad
  \eta_q =
  \begin{cases}
    +1 & \text{for } q \\
    -1 & \text{for } \bar q
  \end{cases}.
\end{equation}
The exact helicity decomposition of the Born matrix element fixes these signs term by term: a single overall crossing sign is not sufficient for the neutral-current spin-dependent coefficients.
This decomposition has been verified numerically against the exact helicity matrix element, and the resulting coefficient-level closure holds to numerical precision once the term-by-term crossing pattern above is used.
More generally,
\begin{equation}
  \Sigma_B = D_B\bigl(1+\mathcal{A}_{\text{born}}\,\ell+\ell^2\bigr)\,,
\end{equation}
where
\begin{equation}
  \mathcal{A}_{\text{born}}\equiv \frac{N_B}{D_B}\,.
\end{equation}
For pure photon exchange this reduces to the usual single-projector form, with $\mathcal{A}_{\text{born}}=2P_\ell P_q$ and $\Sigma_B=D_0(1+\mathcal{A}_{\text{born}}\,\ell+\ell^2)$. For full neutral-current DIS this rewriting remains valid for the Born denominator, but the separate parity-even and parity-odd structures must still be retained in the exact collinear projector decomposition below.

We therefore project the quark and gluon collinear terms onto the exact Born structures.  Defining
\begin{equation}
  D_{\rm even}=D_0+\eta_\ell P_\ell D_\ell,\qquad
  D_{\rm spin}=\eta_q P_q(D_q+\eta_\ell P_\ell D_{\ell q}),
\end{equation}
\begin{equation}
  N_{\rm even}=\eta_\ell\eta_q N_0+\eta_q P_\ell N_\ell,\qquad
  N_{\rm spin}=P_q(\eta_\ell N_q+P_\ell N_{\ell q}),
\end{equation}
we introduce the four projector weights
\begin{equation}
  w_q^{u}=\frac{(1+\ell^2)D_{\rm even}+\ell N_{\rm even}}{\Sigma_B},\qquad
  w_q^{p}=\frac{(1+\ell^2)D_{\rm spin}+\ell N_{\rm spin}}{\Sigma_B},
\end{equation}
\begin{equation}
  w_g^{u}=\frac{(1+\ell^2)D_{\rm even}}{\Sigma_B},\qquad
  w_g^{p}=\frac{\ell N_{\rm spin}}{\Sigma_B}.
\end{equation}
Using $P_q(x_B)=P_z\,\Delta f_q(x_B,\mu_F)/f_q(x_B,\mu_F)$, it is convenient to introduce the finite reduced polarized weights
\begin{equation}
  \widetilde w_q^{p}=
  \frac{(1+\ell^2)\eta_q\!\left(D_q+\eta_\ell P_\ell D_{\ell q}\right)+\ell\!\left(\eta_\ell N_q+P_\ell N_{\ell q}\right)}{\Sigma_B},
  \qquad
  \widetilde w_g^{p}=
  \frac{\ell\!\left(\eta_\ell N_q+P_\ell N_{\ell q}\right)}{\Sigma_B},
\end{equation}
so that $w_q^{p}=P_q(x_B)\,\widetilde w_q^{p}$ and $w_g^{p}=P_q(x_B)\,\widetilde w_g^{p}$.
Then the quark-collinear contribution is
\begin{equation}
  \frac{\mathrm{d}\hat\sigma_{\text{coll},q}^{\text{pol}}}{\mathrm{d}\Phi_2}
  =
  \frac{\alpha_s(\mu_R)}{2\pi}\,C_F\,
  \left[
  w_q^{u}\,\frac{f_m^{\,q}(x_B,\mu_F)}{f_q(x_B,\mu_F)}
  +
  \widetilde w_q^{p}\,P_z\,\frac{\Delta f_m^{\,q}(x_B,\mu_F)}{f_q(x_B,\mu_F)}
  \right]
  \frac{\mathrm{d}\hat\sigma_B^{\text{pol}}}{\mathrm{d}\Phi_2}\,,
  \label{eq:coll-pol-q}
\end{equation}
while the gluon-collinear contribution is
\begin{equation}
  \frac{\mathrm{d}\hat\sigma_{\text{coll},g}^{\text{pol}}}{\mathrm{d}\Phi_2}
  =
  \frac{\alpha_s(\mu_R)}{2\pi}\,C_F\,
  \left[
  w_g^{u}\,\frac{f_m^{\,g}(x_B,\mu_F)}{f_q(x_B,\mu_F)}
  +
  \widetilde w_g^{p}\,P_z\,\frac{\Delta f_m^{\,g}(x_B,\mu_F)}{f_q(x_B,\mu_F)}
  \right]
  \frac{\mathrm{d}\hat\sigma_B^{\text{pol}}}{\mathrm{d}\Phi_2}\,.
  \label{eq:coll-pol-g}
\end{equation}
Written in this way, the polarized terms are manifestly finite: the factors multiplying the modified luminosities are proportional to $P_z\,\Delta f_m/f_q$, and there is no separate division by $\Delta f_q(x_B)$ when the polarized PDF crosses zero.
The quark contribution is therefore fully contained in the $B,C$ luminosities $f_m^{\,q}$ and $\Delta f_m^{\,q}$, while the gluon contribution is carried by the corresponding gluon luminosities $f_m^{\,g}$ and $\Delta f_m^{\,g}$. Since $A$ and $\Delta A$ contain the ratio $T_R/C_F$, the gluon channel reproduces the expected overall factor $T_R\,\alpha_s/(2\pi)$ despite the common prefactor written in Eq.~\eqref{eq:coll-pol-g}.
The absence of $N_{\rm even}$ and $D_{\rm spin}$ from the gluon weights is exact for the massless light-parton channel considered here.  Projecting the charge-conjugate $R_2$ and $R_3$ contributions onto the Born basis $(1+\ell^2)D+\ell N$ shows that the parity-odd spin-independent gluon piece is antisymmetric under charge conjugation and cancels identically, while the standard light-parton coefficient functions contain no gluon contribution to the parity-even spin-dependent channel.  The gluon channel therefore contributes only to the spin-independent parity-even piece and to the spin-dependent parity-odd piece.

\paragraph{Real-emission contributions.}
The exact kernels below are obtained by integrating the subtracted real-emission matrix elements over the unresolved variables at fixed Born phase space. For the integrated real emission the treatment differs from the collinear counterterms.  The NLO weight computes $\sigma_{\text{real}}/\sigma_B$, so the denominator is the unmapped Born structure $\Sigma_B=(1+\ell^2)D_B+\ell N_B$, equivalently $(1+\mathcal{A}_{\text{born}}\,\ell+\ell^2)$ after factoring out $D_B$, and must use $\mathcal{A}_{\text{born}}$ to cancel the Born exactly.  The \emph{kernel} in the numerator, however, describes the real-emission matrix element with an incoming parton at $x=x_B/x_p$, and therefore uses the \emph{mapped} coefficients at that momentum fraction.  For QCDC ($q\to qg$) there is a single Born subprocess, so a single mapped analysing power suffices:
\begin{equation}
  \mathcal{A}_{\text{map}}^{(q)} = \mathcal{A}^{NC}_{P_\ell,\,P_q(x_B/x_p)}\,,
\label{eq:a-mapped-q}
\end{equation}
where $P_q(x)=P_z\,\Delta f_q(x)/f_q(x)$.
For full neutral-current exchange the parity-even Born coefficient also changes under the mapping $P_q(x_B)\to P_q(x_B/x_p)$. Defining
\begin{equation}
  P_{q,m}\equiv P_q(x_B/x_p)\,,
\end{equation}
and denoting
\begin{align}
  D_m=D_0+\eta_\ell P_\ell D_\ell+\eta_q P_{q,m}D_q+\eta_\ell\eta_q P_\ell P_{q,m}D_{\ell q}\,,\nonumber \\
  N_m=\eta_\ell\eta_q N_0+\eta_q P_\ell N_\ell+\eta_\ell P_{q,m}N_q+P_\ell P_{q,m}N_{\ell q}\,,
\end{align}
and
\begin{equation}
  K_q(x_p,\ell)=2+2\ell^2-x_p+3x_p\ell^2,\qquad
  r_D=\frac{D_m}{D_B},
\end{equation}
the exact QCDC real kernel is
\begin{equation}
  \mathcal{R}_{\text{QCDC}}
  =
  \frac{C_F}{x_p}\frac{f_q(x_B/x_p)}{f_q(x_B)}
  \frac{D_m\,K_q(x_p,\ell)+\ell N_m\,(2x_p+1)}
       {(1+\ell^2)D_B+\ell N_B}\,.
  \label{eq:realq-exact}
\end{equation}
Equivalently,
\begin{equation}
  \mathcal{R}_{\text{QCDC}}
  =
  r_D\,
  \frac{C_F}{x_p}\frac{f_q(x_B/x_p)}{f_q(x_B)}
  \frac{K_q(x_p,\ell)+\mathcal{A}_{\text{map}}^{(q)}\,\ell\,(2x_p+1)}
       {1+\mathcal{A}_{\text{born}}\ell+\ell^2}\,.
  \label{eq:realq-ratio}
\end{equation}
For photon exchange, $D_m=D_B$ and Eq.~\eqref{eq:realq-ratio} reduces to the familiar form used in the unpolarized treatment of Ref.~\cite{DErrico:2011wfa}.

For BGF ($g\to q\bar q$), the exact real-emission matrix element still factorizes onto two collinear limits, $R_2$ and $R_3$, with charge-conjugate Born labels for the antiquark channel.  In the integrated NLO weight, we obtain the finite gluon remainder by integrating the subtracted BGF real-emission kernel over the unresolved variables. The result can then be organized in the same exact basis as the collinear counterterm: the charge-conjugate $R_2/R_3$ contributions cancel in the parity-odd spin-independent channel, while the parity-even spin-dependent gluon channel is absent for massless light quarks.  Only the spin-independent parity-even and spin-dependent parity-odd structures therefore survive.

Accordingly, the finite BGF remainder is written with the same exact gluon-projector split as the collinear term,
\begin{equation}
  \mathcal{R}_{\text{BGF}}
  =
  w_g^{u}\,\mathcal{R}_{\text{BGF}}^{u}
  +
  \widetilde w_g^{p}\,\mathcal{R}_{\text{BGF}}^{p}\,,
  \label{eq:realg-proj}
\end{equation}
where
\begin{equation}
  \mathcal{R}_{\text{BGF}}^{u}
  =
  -\frac{T_R}{x_p}\frac{f_g(x_B/x_p)}{f_q(x_B)}
  \frac{1+\ell^2+2(1-3\ell^2)x_p(1-x_p)}{1+\ell^2}\,,
\end{equation}
\begin{equation}
  \mathcal{R}_{\text{BGF}}^{p}
  =
  -2\,\frac{T_R}{x_p}\,P_z\,\frac{\Delta f_g(x_B/x_p)}{f_q(x_B)}
  \bigl(2x_p-1\bigr)\,.
  \label{eq:realg-pol}
\end{equation}
\noindent
The factor of $2$ in Eq.~\eqref{eq:realg-pol} refers to the combined odd remainder of the charge-conjugate $R_2+R_3$ pair before the underlying-Born split of Eq.~\eqref{eq:zp-partfrac} is applied. For a fixed underlying Born flavour, the $1/(1-z_p)$ and $1/z_p$ branches are assigned to the quark and antiquark channels separately, so each channel receives one half of this odd remainder. The single-channel gluon kernels are therefore
\begin{equation}
  K_{g,\text{coll}}^{(q)}(x_p)
  =
  \bigl(2x_p-1\bigr)\ln\!\frac{Q^2(1-x_p)}{\mu_F^2\,x_p}
  +2(1-x_p)\,,
  \qquad
  K_{g,\text{real}}^{(q)}(x_p)
  =
  -\bigl(2x_p-1\bigr)\,,
\end{equation}
and hence
\begin{align}
  K_{g}^{(q)}(x_p)
  &=
  K_{g,\text{coll}}^{(q)}(x_p)+K_{g,\text{real}}^{(q)}(x_p)
  \nonumber\\
  &=
  \bigl(2x_p-1\bigr)\ln\!\frac{Q^2(1-x_p)}{\mu_F^2\,x_p}
  +3-4x_p\,.
\end{align}
At $\mu_F^2=Q^2$ this reproduces the standard $\overline{\text{MS}}$ polarized gluon coefficient-function shape familiar from polarized DIS~\cite{Kodaira:1980qp,Zijlstra:1993sh,deFlorian:1994wp},
\begin{equation}
  C^{\rm spin}_{g}(z) = T_R\Bigl[(2z-1)\ln\!\frac{1-z}{z} + 3 - 4z\Bigr]\,.
  \label{eq:Cgspin}
\end{equation}
The charge-conjugate antiquark Born channel contributes the same kernel, so the conventional inclusive normalization is recovered after summing both underlying Born channels.
In the photon limit, $w_g^{u}=(1+\ell^2)/\Sigma_B$ and $\widetilde w_g^{p}=2P_\ell\,\ell/\Sigma_B$, so Eq.~\eqref{eq:realg-proj} reduces exactly to the familiar photon-only expression.

For later reference, the real-emission kernels depend on the mapped PDFs through the ratios
\(
f_q(x_B/x_p)/f_q(x_B)
\),
\(
f_g(x_B/x_p)/f_q(x_B)
\),
and
\(
P_z\,\Delta f_g(x_B/x_p)/f_q(x_B)
\).
The mapped BGF analysing powers $\mathcal{A}_{R_2}$ and $\mathcal{A}_{R_3}$ remain useful as diagnostic quantities for the exact $R_2/R_3$ crossing structure, but they no longer appear explicitly in the integrated finite remainder.
The same quark/antiquark assignment is used in the gluon collinear term: the $1/z_p$ and $1/(1-z_p)$ pieces are attached to the antiquark and quark Born limits, respectively, so the collinear partition matches the BGF real-emission channel assignment.

Combining the virtual correction~\eqref{eq:virt-pol} with the collinear remainders of Eqs.~\eqref{eq:coll-pol-q} and \eqref{eq:coll-pol-g} yields a single finite factor multiplying the Born helicity projector,
\begin{equation}
\frac{\mathrm{d}\hat\sigma_{V+\text{coll}}^{\text{pol}}}{\mathrm{d}\Phi_2}
=
\frac{\alpha_s(\mu_R)}{2\pi}\,C_F\;
\mathcal{V}(\Phi_B)\;
\frac{\mathrm{d}\hat\sigma_B^{\text{pol}}}{\mathrm{d}\Phi_2}\,,
\label{eq:Vpluscoll-pol}
\end{equation}
where
\begin{equation}
\mathcal{V}(\Phi_B)\equiv V(x_B;\mu_F)
+\mathcal{C}_q(\Phi_B)+\mathcal{C}_g(\Phi_B)\,,
\label{eq:Vcal}
\end{equation}
with $V(x_B;\mu_F)$ from Eq.~\eqref{eq:VxB} and $\mathcal{C}_{q,g}$ the bracketed projector-weighted ratios multiplying $\mathrm{d}\hat\sigma_B^{\text{pol}}/\mathrm{d}\Phi_2$ in Eqs.~\eqref{eq:coll-pol-q} and~\eqref{eq:coll-pol-g}. In the photon limit this reduces to the corresponding single-projector form.
Writing $B^{\text{pol}}(\Phi_B)$ for the polarized Born function in the POWHEG notation, with spin dependence appearing in $\Sigma_B$, and denoting by $\mathcal{R}_I(\Phi_B,\Phi_R)$ the channel-dependent real-emission kernels and by $\mathcal{D}_I(\Phi_B,\Phi_R)$ the corresponding local POWHEG subtraction terms for singular region $I$, we take both $\mathcal{R}_I$ and $\mathcal{D}_I$ to include the channel colour factor but not the common $\alpha_s/(2\pi)$ prefactor. With this convention the corresponding $\bar B$ function used in POWHEG/MC@NLO reads
\begin{align}
\bar B^{\text{pol}}(\Phi_B)
&=B^{\text{pol}}(\Phi_B)\Bigg[
1+\frac{\alpha_s(\mu_R)}{2\pi}\,C_F\,\mathcal{V}(\Phi_B)
\nonumber \\[1mm]
&+\sum_{I\in\{\mathrm{QCDC},\,\mathrm{BGF}\}}
\frac{\alpha_s(\mu_R)}{2\pi}
\int\!\big(\mathcal{R}_I(\Phi_B,\Phi_R)-\mathcal{D}_I(\Phi_B,\Phi_R)\big)\,\mathrm{d}\Phi_R
\Bigg].
\label{eq:Bbar-pol}
\end{align}
The real kernels $\mathcal{R}_I$ use the exact QCDC ratio of Eq.~\eqref{eq:realq-ratio} together with the exact gluon-projector decomposition of Eq.~\eqref{eq:realg-proj}: the QCDC kernel carries its single factor of $C_F$ and the mapped quark PDF ratio $f_q(x_B/x_p)/f_q(x_B)$, while the BGF remainder carries its single factor of $T_R$ and is split into an even part proportional to $f_g(x_B/x_p)/f_q(x_B)$ and an odd part proportional to $P_z\,\Delta f_g(x_B/x_p)/f_q(x_B)$.  In both channels the denominator is the Born factor built from $\mathcal{A}_{\text{born}}$. If instead one defines colour-stripped real and subtraction kernels, the corresponding channel factors $C_{\mathrm{QCDC}}=C_F$ and $C_{\mathrm{BGF}}=T_R$ must be restored outside the integral.

Specifically, the QCDC numerator kernel uses the single mapped analysing power $\mathcal{A}_{\text{map}}^{(q)}$ together with the exact mapped-even ratio $r_D$ of Eq.~\eqref{eq:realq-ratio}. The BGF kernel uses the exact gluon-projector split of Eq.~\eqref{eq:realg-proj}. For the BGF odd term, Eq.~\eqref{eq:realg-pol} denotes the combined charge-conjugate $R_2+R_3$ remainder, while the kernel attached to a fixed underlying Born flavour uses the single-channel form displayed immediately above Eq.~\eqref{eq:Cgspin}, i.e.\ one half of the pair-summed odd remainder after the endpoint split of Eq.~\eqref{eq:zp-partfrac}.  In practice the subtracted real-emission integrals in Eq.~\eqref{eq:Bbar-pol} are not evaluated through a numerical local subtraction. The $z_p$ and azimuthal integrals are performed analytically channel by channel, and every singular structure of the polarized real-emission cross section is removed by a counterterm of matching spin structure: in the QCDC channel the soft and collinear subtractions are identical for the spin-independent and spin-dependent projector structures, since $\Delta P_{qq}=P_{qq}$ after the finite scheme transformation discussed at the start of this subsection, while in the BGF channel the spin-independent and spin-dependent collinear singularities are removed by the $A$ and $\Delta A$ kernels of Eqs.~\eqref{eq:A-kernel} and~\eqref{eq:DA-kernel}, respectively. Only the finite remainders displayed above therefore enter $\bar B^{\text{pol}}$, and the formal difference $\mathcal{R}_I-\mathcal{D}_I$ is implemented in closed form.

In the generation of the hardest emission, the radiation variables are distributed according to the exact polarized ratio of the real-emission and Born cross sections, including the mapped analysing powers and the mapped denominator ratio $r_D$. The spin-averaged envelope and veto machinery of the unpolarized algorithm remains valid for this ratio: the singular endpoint factors $1/[(1-x_p)(1-z_p)]$ and $1/[z_p(1-z_p)]$ are polarization independent, and the spin dependence enters only through bounded factors, since the analysing powers obey $|\mathcal{A}|\le 2$ for physical longitudinal spin states\footnote{For $|P_\ell|,|P_q|\le1$ the Born angular structure is a convex combination of definite-helicity contributions, each of the form $c_+(1+\ell)^2+c_-(1-\ell)^2$ with $c_\pm\ge0$, so that $|\mathcal{A}|=2\,|c_+-c_-|/(c_++c_-)\le2$.} and the mapped Born-projector ratios are finite. In particular, the factorized QCDC structure $|\widetilde{\mathcal{M}}_2(r_1,\bar r_2)|^2$ maps onto the polarized Born configuration exactly, so the Born spin projectors cancel in the corresponding part of the ratio, while in the soft limit $x_p\to1$ the mapped quantities reduce to their Born values. Near zeros of the polarized PDFs, where the reconstruction of the mapped parton polarization becomes numerically delicate, the implementation clamps the polarization to the physical interval $[-1,1]$ and falls back to the Born-level analysing power for the affected veto step. No associated instability is observed in the validation samples below.
In the unpolarized limit ($P_q\to0$), the spin-dependent projector pieces vanish and Eq.~\eqref{eq:Bbar-pol} reduces exactly to the standard \Herwig\ NLO DIS result.

With the real-emission, virtual, and collinear NLO ingredients in place, we proceed to a comparison with the fixed-order program \POLDIS\ in the next subsection.

\subsection{Fixed-Order Benchmarks and Parton-Level Validation}

\subsubsection{Validation Setup}
We perform the fixed-order comparison using the \textsc{Rivet} analysis framework~\cite{Bierlich:2024vqo} (version 4.1.1), with the \Herwig/\textsc{Rivet} event selection matched to the \POLDIS\ setup. We use $E_e=18~\mathrm{GeV}$ and $E_p=275~\mathrm{GeV}$, with the DIS window
$100~\mathrm{GeV}^2 < Q^2 \le 2500~\mathrm{GeV}^2$ and $0.2 \le y \le 0.6$, and with
incoming-leg momentum fractions allowed down to $10^{-5}$. No additional
restriction is imposed on the hadronic invariant mass $W^2 \equiv (P + q)^2$
beyond $W^2 \ge 0$, and the minimum partonic invariant mass is set to
$\hat{M}_{\mathrm{min}}=0$. The same $Q^2$ window is used for the
neutral-current and charged-current validation samples.

To isolate the perturbative content of the comparison, we disable hadronization
and decays, neglect constituent masses, keep only the hardest POWHEG emission in
the analysed final state, and reconstruct jets from the two hardest final-state
partons. The observables shown after the dijet selection should therefore be
read as hardest-emission comparison observables in this setup, rather than as a
statement that every jet observable is formally NLO accurate.

On the analysis side, we follow the \POLDIS\ dijet definition. The jet inputs,
which exclude the scattered lepton, are boosted to the Breit frame and
clustered there with the inclusive anti-$k_T$ algorithm~\cite{Cacciari:2008gp}
with radius parameter $R=1.0$ and $E$-scheme recombination, as implemented in
\textsc{FastJet}~\cite{Cacciari:2011ma} (version 3.5.1). The resulting jets are ordered in
Breit-frame transverse momentum, and the laboratory-frame momentum of each jet
is obtained by summing the laboratory momenta of the same constituents. The two
leading Breit-frame jets are identified first, and the acceptance cuts are then
applied to those same jets after boosting them back to the laboratory frame. The
dijet-selected observables therefore require
$p_{T,1}^{B}>5~\mathrm{GeV}$ and
$p_{T,2}^{B}>4~\mathrm{GeV}$, where the superscript $B$ denotes the Breit frame,
while both selected jets must lie in
the laboratory-frame rapidity range $-3.5<y_{\mathrm{jet}}<3.5$, where
$y_{\mathrm{jet}}$ denotes the jet rapidity and not the DIS inelasticity. In addition,
we also consider a set of inclusive observables before the dijet selection is applied, i.e.\ after the DIS
kinematic selection and the identification of the two leading Breit-frame jets,
but before the Breit-frame $p_T$ thresholds and laboratory-frame rapidity cuts
are imposed.

We choose the PDFs to bring the \Herwig\ \textsc{POWHEG} setup as
close as possible to the \POLDIS\ fixed-order reference. For the main
validation presented here we use the unpolarized proton set
\texttt{NNPDF40\_nlo\_pch\_as\_01180}, based on the NNPDF4.0 determination
of Ref.~\cite{NNPDF:2021njg}, together with the polarized-difference set
\texttt{NNPDFpol20\_nlo\_as\_01180}, based on the NNPDFpol2.0 determination
of Ref.~\cite{Cruz-Martinez:2025ahf} through the \textsc{LHAPDF}~\cite{Buckley:2014ana} interface (version 6.5.5). This choice gives a consistent
NNPDF-family input combination for the LO and NLO comparisons against
\POLDIS, while remaining close to the default \Herwig\ inputs. As in the rest of the validation, we intentionally
use the same PDF inputs at LO and NLO, since our aim is to isolate
calculation-level differences rather than perform an order-matched PDF
phenomenology study.
This setup also gives a useful stress test of the polarized projector
construction, since the chosen polarized PDFs can change sign across the
relevant $x$ range. In the calculation the polarized terms enter through the
finite combinations proportional to $P_z\,\Delta f_m/f_q$, with no separate
division by $\Delta f_q(x_B)$, and we observe no numerical instability
associated with polarized-PDF nodes in the validation samples.
Wherever a reconstructed parton polarization $P_z\,\Delta f_i(x)/f_i(x)$
enters an analysing power or a spin-density matrix, it is clamped to the
physical interval $[-1,1]$. The clamp can act only where the polarized and
unpolarized PDF inputs are locally inconsistent.
As an integrated-rate-only cross-check, we also repeated the validation with the
alternative hybrid PDF combination
\texttt{PDF4LHC15\_nnlo\_100\_pdfas}~\cite{Butterworth:2015oua} and
\texttt{BDSSV24-NNLO}~\cite{Borsa:2024mss}. The resulting integrated cross
sections show comparable agreement, so we interpret the remaining residual
offsets as compatible with residual input-level convention differences rather
than with a substantive physics discrepancy. The polarized and unpolarized PDFs
are still not obtained from a single simultaneous fit, so the ratio
$\Delta f/f$ should not be interpreted as a strictly bounded event-level
polarization variable throughout the full sampled phase space.

We match the electroweak and coupling inputs in the same spirit. For this
validation setup, we use a fixed electromagnetic coupling
$\alpha_{\mathrm{EM}}=1/137 \simeq 0.00729927$ to match the \POLDIS\
conventions, and evaluate the photon and $Z$ fermion vertices with fixed
Standard Model couplings. For the channels involving $Z$ exchange, we also set
the sine-squared of the Weinberg angle to $\sin^2\theta_W=0.22164$ and the $W$
boson mass to $m_W=80.45~\mathrm{GeV}$, matching these in \POLDIS. For QCD, the LO samples use one-loop running of the strong coupling and
the NLO samples two-loop running, in both cases with $\alpha_s(M_Z)=0.118$
and $M_Z = 91.1876$~GeV.

We do not expect exact agreement with a pure fixed-order calculation, since \textsc{POWHEG}
remains an NLO+Sudakov framework rather than a strict fixed-order one.
The Sudakov suppression and resummation built into the hardest-emission
generation induce residual differences, especially for
emission-sensitive differential observables.
For the scale bands in the no-shower POWHEG differential comparisons, the
\Herwig\ variation is performed around the generator-native POWHEG emission
scales.  The Born and $\bar B$ contributions use the DIS scale
$\mu_B^2=Q^2$, up to the common scale factor $\kappa^2$.  For an accepted real
emission, the coupling in the POWHEG emission probability is evaluated at the
Breit-frame emission scale
$\mu_{R,\mathrm{em}}^2=(p_{T}^{B})^2=Q^2 x_\perp^2/4$, while the numerator PDF in the
real-emission ratio is evaluated at
$\mu_{F,\mathrm{em}}^2=Q^2+(p_{T}^{B})^2
=Q^2[1+(1-x_p)(1-z_p)z_p/x_p]$. The denominator PDF remains at the
underlying-Born scale $Q^2$.  The mapped incoming-parton polarization that
multiplies this unpolarized carrier is constructed from
$P_z\,\Delta f/f$ at the hard scale $Q^2$ and is kept fixed at this scale when
$\kappa$ is varied.  Changing the scale of this ratio would first modify the
$\mathcal{O}(\alpha_s)$ emission kernel beyond NLO.  The displayed \Herwig\
scale envelope is therefore obtained by applying the same factors
$\kappa=1/2,1,2$ to the generator-native coupling and ordinary-PDF scales,
while retaining the hard-$Q^2$ polarized fraction.  This is different from the
\POLDIS\ and independent fixed-order reference convention, where the central
scale is $\mu_R^2=\mu_F^2=Q^2$ and the whole calculation, including the
polarized PDFs, is varied uniformly.
The integrated cross sections below are formed from the positive- and
negative-weight samples for each helicity configuration. The correlated
helicity weights used for the differential parton-level validation plots are
defined in Section~\ref{sec:correlated-helicity-weights}. The shower-level
comparison instead uses independently generated physical-helicity samples, as
described below.

\subsubsection{Total DIS Cross Section}
We validate the total DIS cross section at fixed cuts for the three
neutral-current setups $\gamma$, $Z$, and $\gamma Z$. At NLO, each helicity
configuration is assembled from the positive- and negative-weight POWHEG
samples, reflecting the real-emission and subtraction contributions, as
\begin{equation}
  \sigma_h^{\mathrm{NLO}}
  =
  \sigma_h^{(+)}
  -
  \sigma_h^{(-)}\,.
\end{equation}
Here $h$ labels the beam-helicity configuration.
For the unpolarized comparison we use
\begin{align}
  \sigma_{0,\gamma} &\equiv \sigma_{00,\gamma}\,,
  \\
  \sigma_{0,X} &\equiv
  \frac{\sigma_{++,X}+\sigma_{+-,X}+\sigma_{-+,X}+\sigma_{--,X}}{4}\,,
  \qquad X\in\{Z,\gamma Z\}\,,
\end{align}
while the polarized observables are
\begin{align}
  \Delta\sigma_{\gamma}
  &\equiv
  \frac{\sigma_{++,\gamma}-\sigma_{+-,\gamma}}{2}\,,
  \\
  \Delta\sigma_{LL,X}
  &\equiv
  \frac{\sigma_{++,X}+\sigma_{--,X}-\sigma_{+-,X}-\sigma_{-+,X}}{4}\,,
  \qquad X\in\{Z,\gamma Z\}\,.
\end{align}
For convenience we also define the interference contribution by subtracting the
pure-photon and pure-$Z$ results from the full neutral-current result,
\begin{align}
  \sigma_{0,\mathrm{int}}
  &\equiv
  \sigma_{0,\gamma Z}-\sigma_{0,\gamma}-\sigma_{0,Z}\,,
  \\
  \Delta\sigma_{LL,\mathrm{int}}
  &\equiv
  \Delta\sigma_{LL,\gamma Z}-\Delta\sigma_{\gamma}-\Delta\sigma_{LL,Z}\,.
\end{align}
\paragraph{Neutral-current cross section.}
In the pure-photon case, parity implies that the two lepton-helicity choices
are equivalent, so that $\Delta\sigma_\gamma$ may be evaluated from either
$(++, +-)$ or $(--,-+)$. Tables~\ref{tab:validation-total-lo}
and~\ref{tab:validation-total-nlo} summarize the integrated fixed-order
comparison to \POLDIS. The quoted \Herwig\ values are taken
directly from the extracted sample summaries, with all entries expressed in pb.

\begin{table}[t]
  \centering
  \footnotesize
  \setlength{\tabcolsep}{4.5pt}
  \begin{tabular*}{\textwidth}{@{\extracolsep{\fill}}|l|l|ccc|@{}}
    \hline
    Setup & Observable & \Herwig\ [pb] & \POLDIS\ [pb] & HW/POLDIS \\
    \hline
    $\gamma$   & $\sigma_{0,\gamma}$              & $1182.89 \pm 0.01$        & $1182.92 \pm 0.05$        & $0.99997 \pm 0.00005$ \\
    $\gamma$   & $\Delta\sigma_{\gamma}$          & $37.333 \pm 0.005$        & $37.3364 \pm 0.0016$      & $0.99992 \pm 0.00014$ \\
    $Z$        & $\sigma_{0,Z}$                   & $1.148486 \pm 0.000005$   & $1.148406 \pm 0.000072$   & $1.00007 \pm 0.00006$ \\
    $Z$        & $\Delta\sigma_{LL,Z}$            & $0.048401 \pm 0.000005$   & $0.048398 \pm 0.000005$   & $1.00006 \pm 0.00016$ \\
    $\gamma Z$ & $\sigma_{0,\gamma Z}$            & $1193.732 \pm 0.004$      & $1193.771 \pm 0.053$      & $0.99997 \pm 0.00005$ \\
    $\gamma Z$ & $\Delta\sigma_{LL,\gamma Z}$     & $42.019 \pm 0.004$        & $42.0107 \pm 0.0017$      & $1.00019 \pm 0.00009$ \\
    int.       & $\sigma_{0,\mathrm{int}}$        & $9.698 \pm 0.008$         & $9.706 \pm 0.076$         & $0.9992 \pm 0.0078$ \\
    int.       & $\Delta\sigma_{LL,\mathrm{int}}$ & $4.637 \pm 0.006$         & $4.6258 \pm 0.0023$       & $1.0024 \pm 0.0014$ \\
    \hline
  \end{tabular*}
  \caption{Integrated LO cross sections and ratios.  The interference entries
  are defined by subtracting the pure-photon and pure-$Z$ contributions from
  the full neutral-current result, and are computed from the unrounded cross
  sections.}
  \label{tab:validation-total-lo}
\end{table}

\begin{table}[t]
  \centering
  \footnotesize
  \setlength{\tabcolsep}{4.5pt}
  \begin{tabular*}{\textwidth}{@{\extracolsep{\fill}}|l|l|ccc|@{}}
    \hline
    Setup & Observable & \Herwig\ [pb] & \POLDIS\ [pb] & HW/POLDIS \\
    \hline
    $\gamma$   & $\sigma_{0,\gamma}$              & $1104.80 \pm 0.01$        & $1104.80 \pm 0.06$        & $0.99999 \pm 0.00005$ \\
    $\gamma$   & $\Delta\sigma_{\gamma}$          & $34.454 \pm 0.005$        & $34.4513 \pm 0.0016$      & $1.00008 \pm 0.00015$ \\
    $Z$        & $\sigma_{0,Z}$                   & $1.098113 \pm 0.000005$   & $1.098050 \pm 0.000072$   & $1.00006 \pm 0.00007$ \\
    $Z$        & $\Delta\sigma_{LL,Z}$            & $0.044271 \pm 0.000005$   & $0.044267 \pm 0.000005$   & $1.00009 \pm 0.00017$ \\
    $\gamma Z$ & $\sigma_{0,\gamma Z}$            & $1114.949 \pm 0.004$      & $1114.972 \pm 0.056$      & $0.99998 \pm 0.00005$ \\
    $\gamma Z$ & $\Delta\sigma_{LL,\gamma Z}$     & $38.999 \pm 0.004$        & $38.9955 \pm 0.0017$      & $1.00010 \pm 0.00010$ \\
    int.       & $\sigma_{0,\mathrm{int}}$        & $9.055 \pm 0.008$         & $9.071 \pm 0.079$         & $0.9982 \pm 0.0087$ \\
    int.       & $\Delta\sigma_{LL,\mathrm{int}}$ & $4.501 \pm 0.006$         & $4.5000 \pm 0.0023$       & $1.0002 \pm 0.0015$ \\
    \hline
  \end{tabular*}
  \caption{Integrated NLO cross sections and ratios.  The NLO values are formed
  from the positive- and negative-weight samples according to
  $\sigma_h^{\mathrm{NLO}}=\sigma_h^{(+)}-\sigma_h^{(-)}$. The interference
  entries are computed from the unrounded cross sections.}
  \label{tab:validation-total-nlo}
\end{table}

The integrated
fixed-order cross sections agree with \POLDIS\ within the quoted uncertainties
across all neutral-current channels. For the dominant pure-photon, pure-$Z$, and full
neutral-current cross sections, the ratios HW/POLDIS are consistent with unity
at the permille level or better at both LO and NLO. The polarized
combinations are likewise statistically compatible in the pure-photon and
pure-$Z$ setups. The interference contributions remain the most
cancellation-sensitive observables, especially in the polarized channel, but
they also show no statistically significant tension in the present sample. We
therefore interpret the remaining differences as compatible with the quoted
statistical uncertainties, with small convention-level input mismatches, and with small differences in the evolution of $\alpha_s$, which is performed internally in \Herwig\ instead of using the PDF $\alpha_s$.

\paragraph{Charged-current cross section.}
For the charged-current validation we use the same fixed cuts and compare the
helicity-averaged unpolarized cross section
\begin{equation}
  \sigma_{0,W}
  \equiv
  \frac{\sigma_{++,W}+\sigma_{+-,W}+\sigma_{-+,W}+\sigma_{--,W}}{4}
\end{equation}
and the double-spin combination
\begin{equation}
  \Delta\sigma_{LL,W}
  \equiv
  \frac{\sigma_{++,W}+\sigma_{--,W}-\sigma_{+-,W}-\sigma_{-+,W}}{4}\,.
\end{equation}
The charged-current comparison is summarized in
Table~\ref{tab:validation-total-cc}. The helicity-averaged and double-spin
combinations agree with \POLDIS\ to better than permille accuracy at both LO
and NLO, with small residual pulls.

\begin{table}[!hpt]
  \centering
  \footnotesize
  \setlength{\tabcolsep}{4.5pt}
  \begin{tabular*}{\textwidth}{@{\extracolsep{\fill}}|l|l|ccc|@{}}
    \hline
    Order & Observable & \Herwig\ [pb] & \POLDIS\ [pb] & HW/POLDIS \\
    \hline
    LO  & $\sigma_{0,W}$              & $9.00689 \pm 0.00009$ & $9.00749 \pm 0.00063$ & $0.99993 \pm 0.00007$ \\
    LO  & $\Delta\sigma_{LL,W}$       & $2.97415 \pm 0.00009$ & $2.97434 \pm 0.00026$ & $0.99994 \pm 0.00009$ \\
    NLO & $\sigma_{0,W}$              & $8.64411 \pm 0.00007$ & $8.64435 \pm 0.00063$ & $0.99997 \pm 0.00007$ \\
    NLO & $\Delta\sigma_{LL,W}$       & $2.86504 \pm 0.00007$ & $2.86541 \pm 0.00026$ & $0.99987 \pm 0.00009$ \\
    \hline
  \end{tabular*}
  \caption{Integrated charged-current LO and NLO cross sections and ratios.
  The NLO values are formed from the positive- and negative-weight samples
  according to $\sigma_h^{\mathrm{NLO}}=\sigma_h^{(+)}-\sigma_h^{(-)}$.
  The quoted uncertainties are statistical; those of $\sigma_{0,W}$ and
  $\Delta\sigma_{LL,W}$ coincide because both observables are formed from the
  same four helicity samples.}
  \label{tab:validation-total-cc}
\end{table}

\subsubsection{Differential Distributions}
For the differential validation we use a parton-level comparison without
subsequent shower evolution. This is neither a physical showered prediction nor
the strict fixed-order integration path used internally by fixed-order programs.
Instead, the accepted POWHEG real-emission configuration is analysed directly at
matrix-element parton level, with correlated helicity weights filling the
unpolarized and polarized histograms for the same accepted event, as described
in Section~\ref{sec:correlated-helicity-weights}. The comparison therefore
isolates the realized real-emission kinematics and helicity weights before
shower-interface reconstruction, subsequent shower evolution, hadronization, and
decays are applied.

We show the standard DIS preselection variables and the dijet-selected
kinematic distributions. The preselection variables are
\begin{equation}
  Q^2=-q^2,\qquad
  x_{\mathrm{Bj}}=\frac{Q^2}{2P\cdot q},\qquad
  y=\frac{P\cdot q}{P\cdot q_1}\,,
\end{equation}
where $P$ is the incoming hadron momentum, $q_1$ is the incoming lepton momentum,
and $q$ is the exchanged electroweak-boson momentum. The variables
$x_{\mathrm{Bj}}$ and $y$ coincide with the $x_B$ and $y_B$ of the preceding
sections, and we retain the $x_{\mathrm{Bj}}$ label in the analysis figures. The quantities
$p_{T,1}^{B}$ and $p_{T,2}^{B}$ denote the transverse momenta of the two
leading Breit-frame partons before the final dijet selection is applied.
For the two-parton matrix-element configurations used in this comparison,
transverse-momentum conservation in the Breit frame gives
$p_{T,1}^{B}=p_{T,2}^{B}$. We therefore show only $p_{T,1}^{B}$. The same
distribution also represents the second parton before subsequent shower
evolution. Events in which the POWHEG generation yields no resolvable
emission consist of a single parton along the Breit-frame axis. Following the
\POLDIS\ pre-cut convention, such events still enter the preselection
distributions with their full weight, with the missing jet transverse momenta
recorded as zero, and subsequently fail the dijet selection.

For the dijet-selected distributions, the two leading Breit-frame jets are
required to satisfy the same transverse-momentum thresholds and
laboratory-frame rapidity acceptance used by \POLDIS. We use
\begin{equation}
\begin{array}{rcl}
M_{jj}^2 &\equiv& (p_{j_1}+p_{j_2})^2\,, \\[2mm]
\eta^\ast &\equiv& \dfrac{1}{2}\left|\eta_1^B-\eta_2^B\right|\,, \\[2mm]
\xi &\equiv& x_{\mathrm{Bj}}\left(1+\dfrac{M_{jj}^2}{Q^2}\right)\,,
\end{array}
\end{equation}
where $p_{j_1}$ and $p_{j_2}$ are the selected jet four-momenta, and
$\eta_i^B$ are their Breit-frame pseudorapidities. The displayed $\xi$
distribution is binned in $\log_{10}(\xi)$.

\begin{figure}[!htbp]
  \centering
  \includegraphics[width=0.32\linewidth]{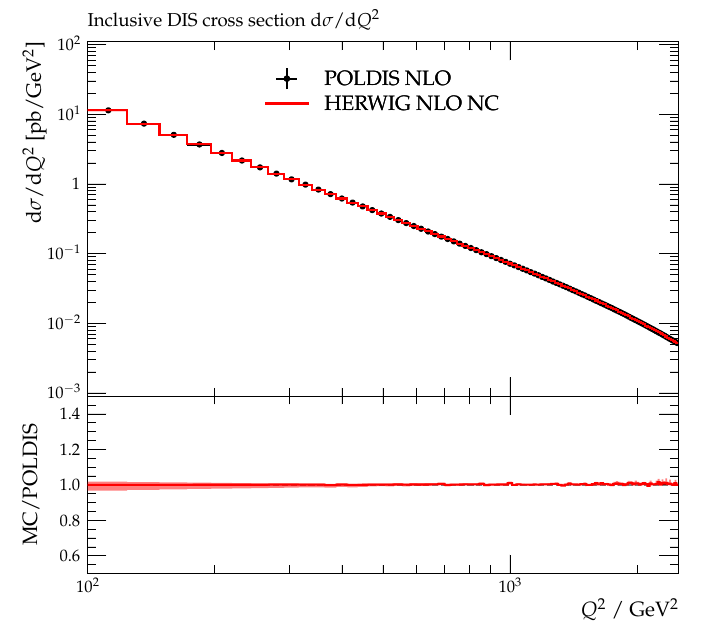}
  \includegraphics[width=0.32\linewidth]{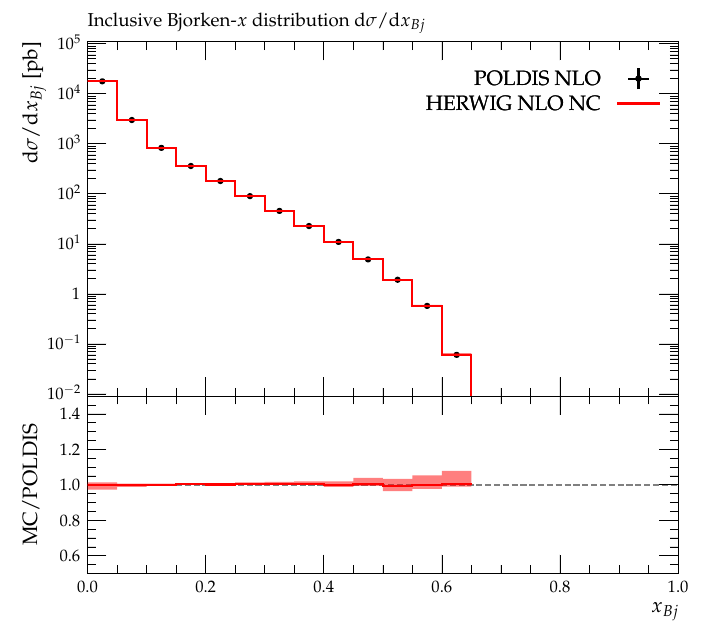}
  \includegraphics[width=0.32\linewidth]{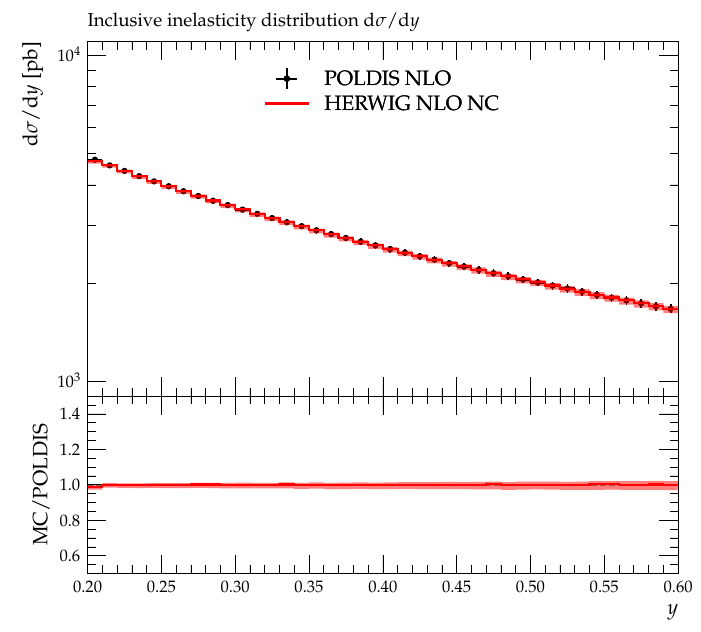}

  \includegraphics[width=0.32\linewidth]{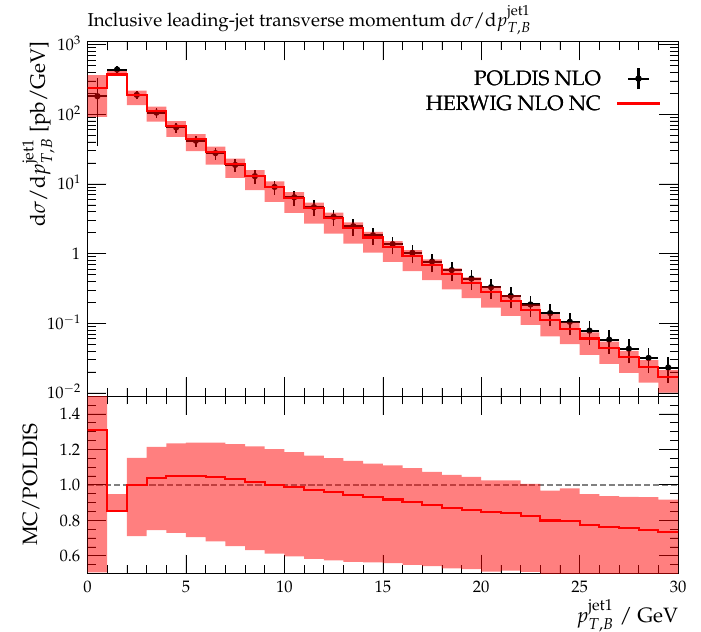}
  \caption{Parton-level validation without shower evolution of inclusive unpolarized differential
  distributions before dijet cuts:
  $\mathrm{d}\sigma/\mathrm{d}Q^2$,
  $\mathrm{d}\sigma/\mathrm{d}x_{\mathrm{Bj}}$,
  $\mathrm{d}\sigma/\mathrm{d}y$,
  and $\mathrm{d}\sigma/\mathrm{d}p_{T,1}^{B}$.}
  \label{fig:validation-partonlevel-inclusive-unpol}
\end{figure}

\begin{figure}[!htbp]
  \centering
  \includegraphics[width=0.32\linewidth]{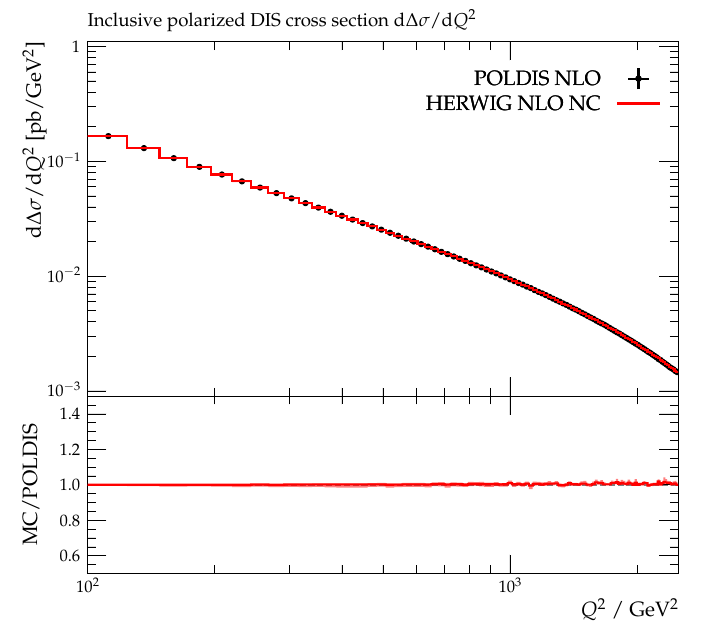}
  \includegraphics[width=0.32\linewidth]{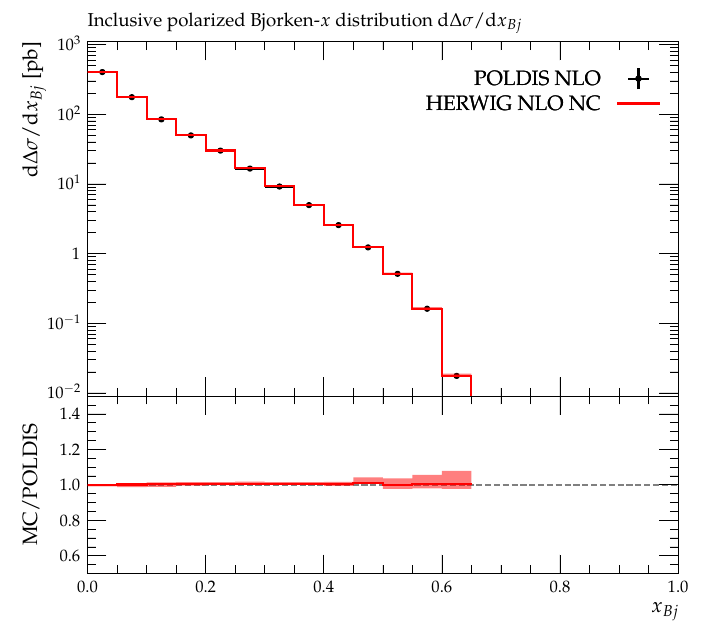}
  \includegraphics[width=0.32\linewidth]{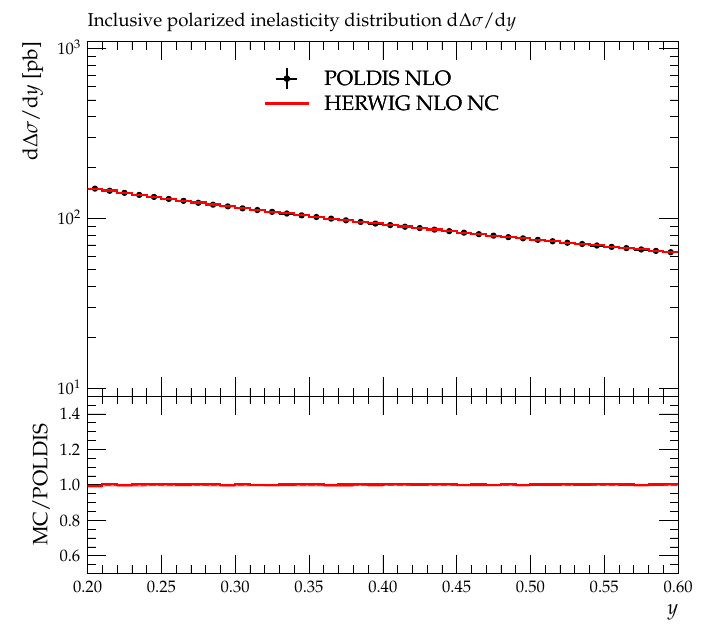}

  \includegraphics[width=0.32\linewidth]{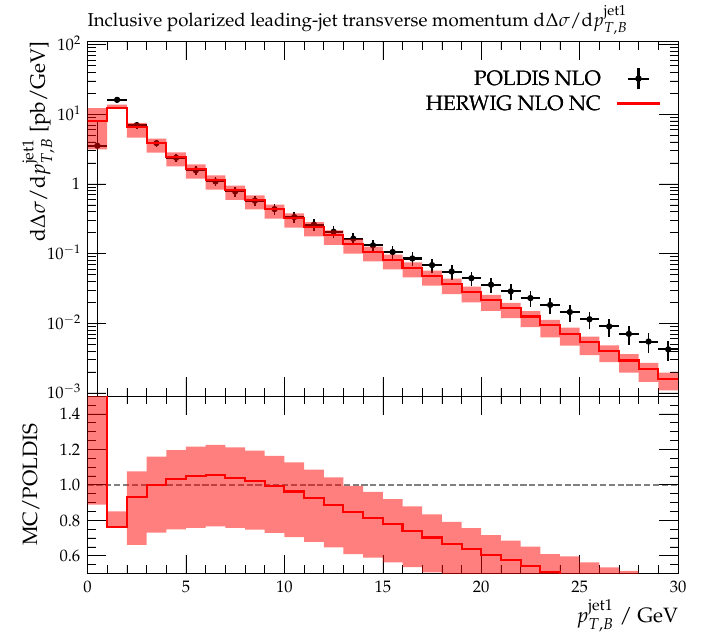}
  \caption{Parton-level validation without shower evolution of inclusive polarized differential
  distributions before dijet cuts:
  $\mathrm{d}\Delta\sigma/\mathrm{d}Q^2$,
  $\mathrm{d}\Delta\sigma/\mathrm{d}x_{\mathrm{Bj}}$,
  $\mathrm{d}\Delta\sigma/\mathrm{d}y$,
  and $\mathrm{d}\Delta\sigma/\mathrm{d}p_{T,1}^{B}$.}
  \label{fig:validation-partonlevel-inclusive-pol}
\end{figure}

\begin{figure}[!htbp]
  \centering
  \includegraphics[width=0.32\linewidth]{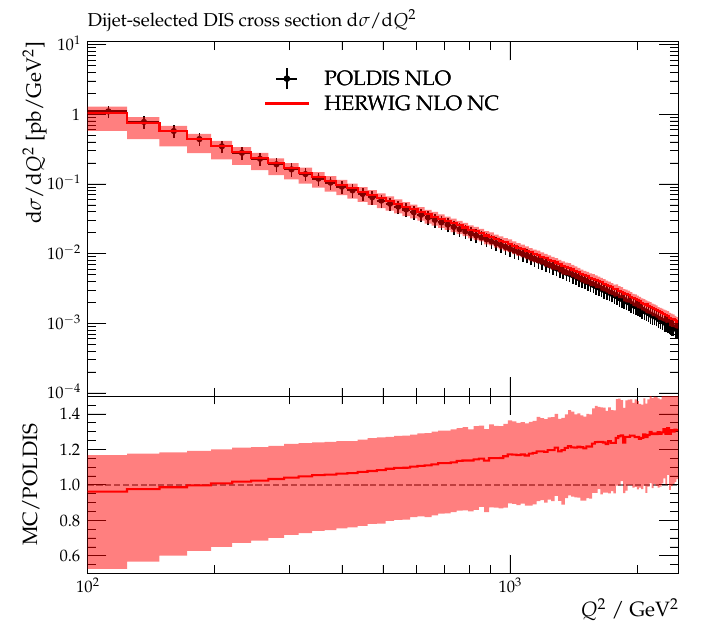}
  \includegraphics[width=0.32\linewidth]{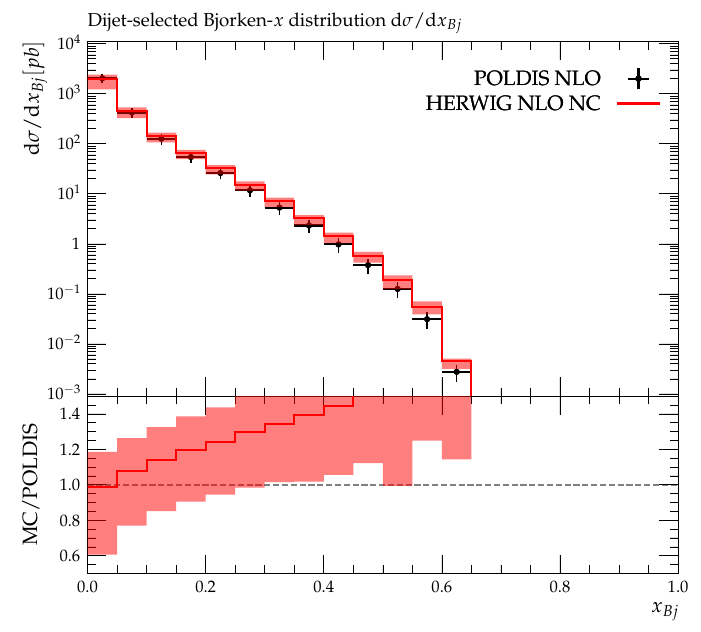}
  \includegraphics[width=0.32\linewidth]{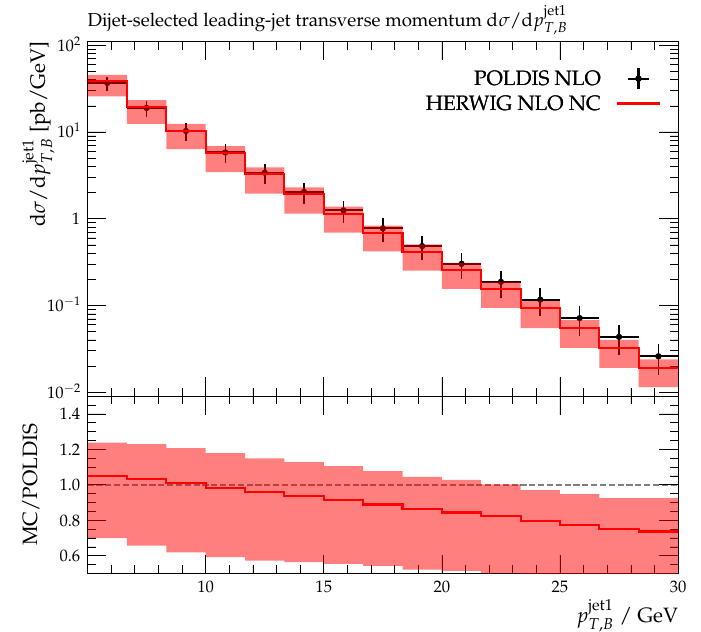}

  \includegraphics[width=0.32\linewidth]{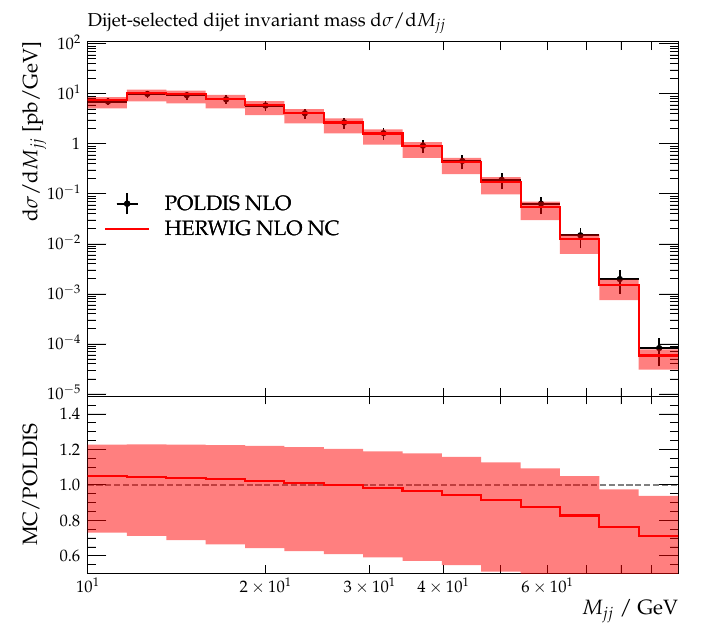}
  \includegraphics[width=0.32\linewidth]{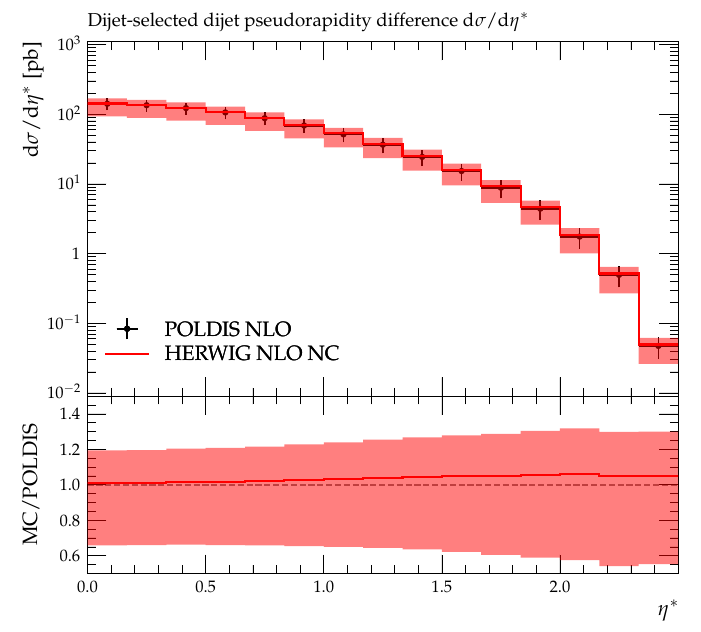}
  \includegraphics[width=0.32\linewidth]{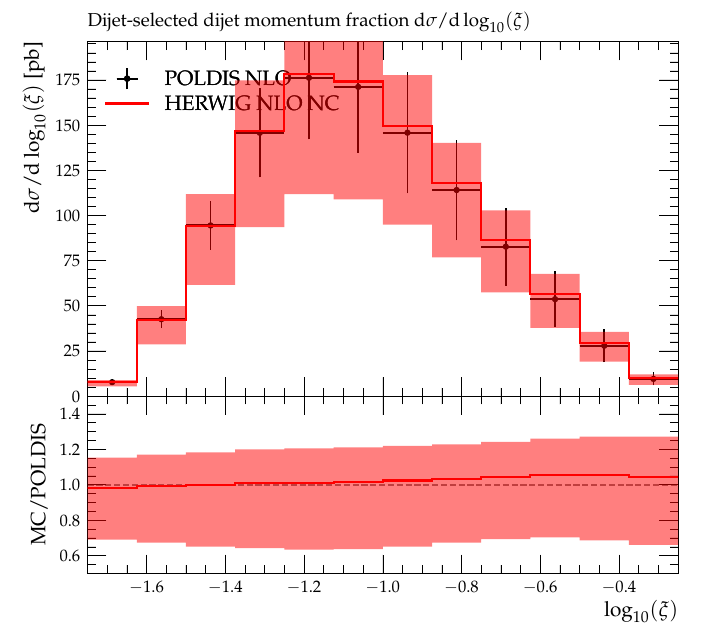}
  \caption{Parton-level validation without shower evolution of dijet-selected unpolarized kinematic
  distributions after dijet cuts:
  $\mathrm{d}\sigma/\mathrm{d}Q^2$,
  $\mathrm{d}\sigma/\mathrm{d}x_{\mathrm{Bj}}$,
  $\mathrm{d}\sigma/\mathrm{d}p_{T,1}^{B}$,
  $\mathrm{d}\sigma/\mathrm{d}M_{jj}$,
  $\mathrm{d}\sigma/\mathrm{d}\eta^\ast$, and
  $\mathrm{d}\sigma/\mathrm{d}\log_{10}(\xi)$.}
  \label{fig:validation-partonlevel-dijet-unpol-kin}
\end{figure}

\begin{figure}[!htbp]
  \centering
  \includegraphics[width=0.32\linewidth]{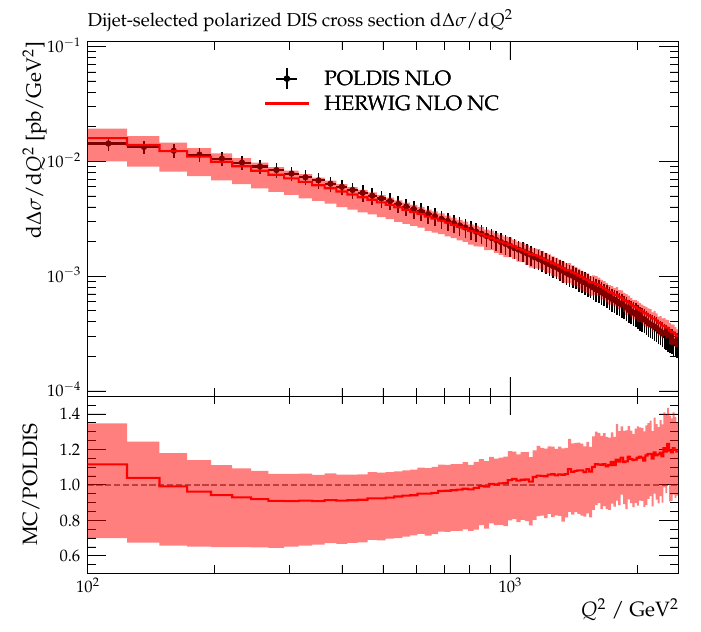}
  \includegraphics[width=0.32\linewidth]{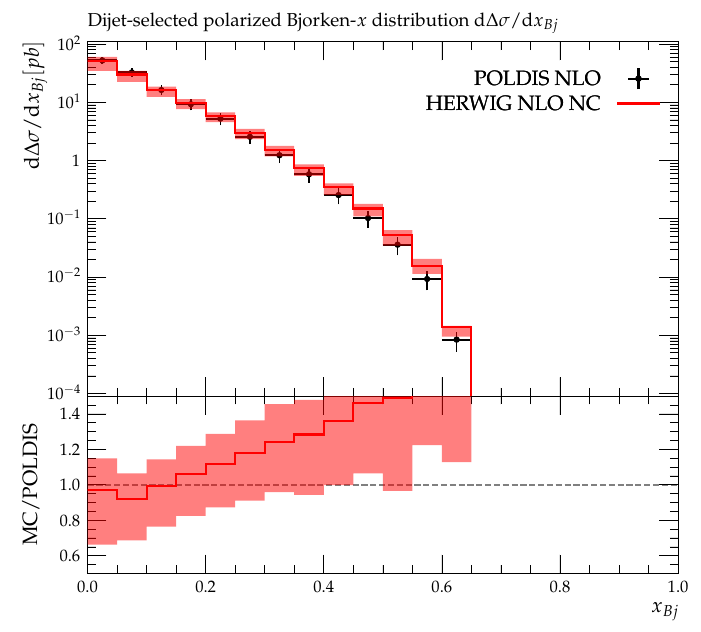}
  \includegraphics[width=0.32\linewidth]{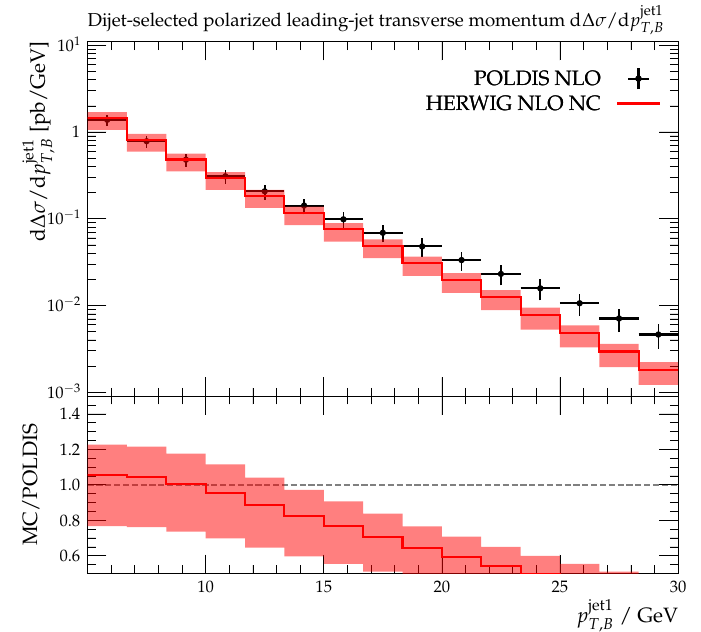}

  \includegraphics[width=0.32\linewidth]{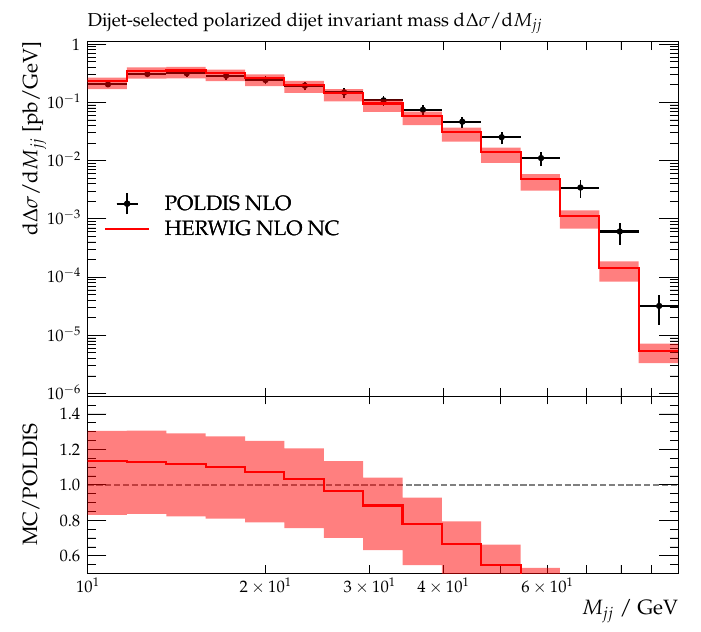}
  \includegraphics[width=0.32\linewidth]{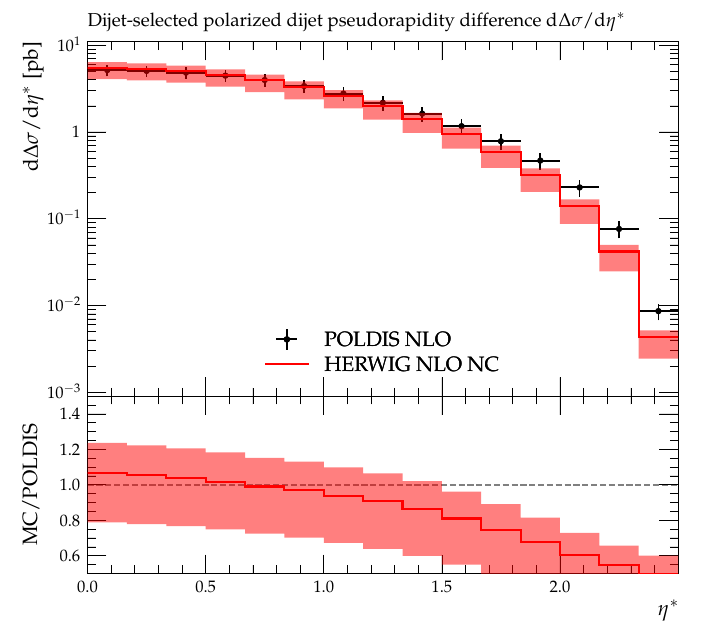}
  \includegraphics[width=0.32\linewidth]{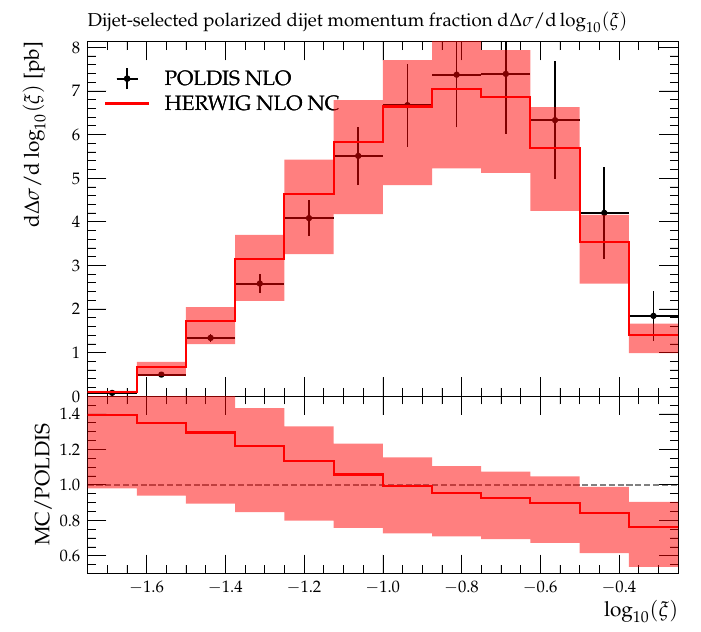}
  \caption{Parton-level validation without shower evolution of dijet-selected polarized kinematic
  distributions after dijet cuts:
  $\mathrm{d}\Delta\sigma/\mathrm{d}Q^2$,
  $\mathrm{d}\Delta\sigma/\mathrm{d}x_{\mathrm{Bj}}$,
  $\mathrm{d}\Delta\sigma/\mathrm{d}p_{T,1}^{B}$,
  $\mathrm{d}\Delta\sigma/\mathrm{d}M_{jj}$,
  $\mathrm{d}\Delta\sigma/\mathrm{d}\eta^\ast$, and
  $\mathrm{d}\Delta\sigma/\mathrm{d}\log_{10}(\xi)$.}
  \label{fig:validation-partonlevel-dijet-pol-kin}
\end{figure}

The unpolarized and polarized preselection distributions are shown in
Figs.~\ref{fig:validation-partonlevel-inclusive-unpol}
and~\ref{fig:validation-partonlevel-inclusive-pol}. The dijet-selected
kinematic distributions are shown in
Figs.~\ref{fig:validation-partonlevel-dijet-unpol-kin}
and~\ref{fig:validation-partonlevel-dijet-pol-kin}. The corresponding
charged-current differential distributions are collected in
Appendix~\ref{app:charged-current-differential-validation}.
These figures use the \Herwig\ no-shower POWHEG construction and therefore
test the parton-level hardest-emission kinematics that are passed to the
analysis.  Their red scale bands follow the \Herwig-native POWHEG emission-scale
choice described above, rather than the fixed-order $Q^2$-central scale choice.
In Appendix~\ref{app:standalone-fixed-order-validation} we present a
complementary, independent fixed-order calculation based on the same Born,
virtual, collinear, QCDC, and BGF ingredients.  That comparison removes the
POWHEG Sudakov factor and isolates the fixed-order terms implemented in
\Herwig.

The inclusive preselection distributions show close shape agreement across
$Q^2$, $x_{\mathrm{Bj}}$, and $y$, as expected for observables controlled
primarily by the DIS leptonic kinematics and the common phase-space cuts. One
should also not expect a large fixed-order scale variation for the inclusive
preselection $Q^2$ distributions: before the dijet selection they are
structure-function-like DIS observables whose leading contribution is
electroweak and carries no explicit factor of $\alpha_s$, so the residual scale
dependence enters through PDF evolution and NLO coefficient-function terms. After
the dijet selection the selected $Q^2$ spectrum is instead tied to resolved QCD
radiation, and its scale dependence is correspondingly larger.  The comparison
remains stable for the invariant-mass,
pseudorapidity-separation, and $\xi$ distributions, with residual shape differences
that are largest in observables directly tied to the accepted hardest emission.
This is most visible in the Breit-frame transverse-momentum spectrum. The
$p_{T,1}^{B}$ distribution is a Sudakov observable: in the matched calculation
it is generated as the transverse momentum of the hardest POWHEG emission, with
the unresolved region suppressed by the Sudakov form factor, rather than as a
pure fixed-order real-emission histogram. Its agreement with \POLDIS\ should
therefore be read as a validation of the hardest-emission kernel and of its
parton-level analysis interpretation. The fixed-order validation of the
underlying NLO components is given separately in
Appendix~\ref{app:standalone-fixed-order-validation}. The polarized
counterparts follow the same pattern, with somewhat larger bin-to-bin
fluctuations in the regions where the helicity-difference cross section is
smaller.

\FloatBarrier
\subsection{Real-Emission Spin Information in the Shower}
\label{sec:real-emission-spin-shower}
\subsubsection{Spin-Density Construction for the Accepted Real Emission}
In the present \Herwig\ polarized-shower treatment, the spin information of
the incoming lepton and the Born-level hard process is propagated through the
subsequent branching history, following the approach of Section~\ref{sec:formalism}. At NLO, however, the
hardest POWHEG emission selects a definite $2\to 3$ partonic configuration
(i.e.\ the QCD Compton or boson-gluon fusion state) before the shower begins,
and this configuration carries spin information beyond a purely Born-level
description. To retain this information during showering, we supplement the
accepted real-emission event with an auxiliary production vertex constructed
from the exact helicity amplitudes for the corresponding neutral-current DIS
channels,
\begin{equation}
  e\,q \to e\,q\,g
  \qquad\text{and}\qquad
  e\,g \to e\,q\,\bar q\,.
\end{equation}
We evaluate these amplitudes with the same electroweak couplings and channel
assignments as in the validated polarized DIS matrix elements, so the spin
information attached to the event is consistent with the underlying NLO calculation. The charged-current calculation uses the same hard-process framework, but the shower-spin study shown
in this subsection is restricted to the neutral-current case. In the construction
used for the present study, the Born-level polarized DIS spin-density matrix
provides the reference shower initialization, while the accepted real-emission
kernels reconstruct the incoming parton polarization at the mapped momentum
fraction $x_B/x_p$. For
neutral-current DIS this includes the exact mapped-even QCDC denominator factor
and separate BGF $R_2/R_3$ analysing powers for the two charge-conjugate limits.

This additional vertex supplies only the spin-density information needed by the
shower. The hardest-emission kinematics, flavour channel, and colour flow remain
those already selected by the POWHEG event generation, and the accepted
hard-event weights, the $\bar B$ function, and the POWHEG hardest-emission
selection are unchanged. We use the helicity amplitudes solely to construct the
spin-density matrix of the outgoing particles. The spin-dependent shower evolution therefore
can affect only subleading post-hardest-emission correlations and does not alter
the NLO accuracy of the matched hard cross section. This separation allows us
to test whether the additional real-emission spin information produces a
resolvable effect in shower-sensitive observables without obscuring the
fixed-order validation of the hard process.

In this construction, the incoming lepton density matrix is inherited from the
DIS state, while the incoming parton polarization is reconstructed from the beam
polarization and the polarized PDFs at the momentum fraction associated with
the accepted emission. The resulting spin vertex then defines the density
matrices carried by the outgoing lepton and by the two outgoing coloured legs.
This extends the shower initialization procedure described in
Section~\ref{sec:formalism}. In the standard algorithm, spin correlations are
initialized from the Born-level density matrix. Here that object is replaced by
the density matrix of the full real-emission state.

\subsubsection{Validation of the Spin Correlations}
The fixed-order comparisons discussed above validate the polarized DIS matrix
elements and their combination into inclusive and dijet-selected cross
sections. A separate NLO+PS validation first quantifies the effect of the
Born-level polarized DIS spin-density initialization relative to a spin-disabled
shower, and then tests whether propagating the spin information from the exact
real-emission configuration produces a distinguishable additional effect. For
this purpose we use the same post-dijet event sample as in the Breit-frame
dijet analysis: the event must satisfy the DIS window
$100~\mathrm{GeV}^2 < Q^2 \le 2500~\mathrm{GeV}^2$ and $0.2 \le y \le 0.6$, and the two
leading Breit-frame jets must pass
$p_{T,1}^{B}>5~\mathrm{GeV}$,
$p_{T,2}^{B}>4~\mathrm{GeV}$, and
have laboratory-frame rapidity $-3.5<y_{\mathrm{jet}}<3.5$. In the showered
events the jets are clustered from all final-state particles excluding the
scattered lepton, using the same Breit-frame anti-$k_T$ definition with
$R=1.0$ as in the parton-level validation.

The spin-correlation validation then compares three shower configurations:
\begin{itemize}
\item \textbf{Full}: both the POWHEG real-spin vertex and the shower spin
  correlations are enabled.
\item \textbf{Born-Only}: the shower is initialized with the polarized DIS
  Born-level spin-density matrix, while the additional POWHEG real-spin vertex
  is disabled.
\item \textbf{None}: both the POWHEG real-spin vertex and the shower spin
  correlations are disabled.
\end{itemize}
This separation allows us to disentangle two contributions. The comparison of
\textbf{Born-Only} with \textbf{None} tests the effect of the Born-level
polarized DIS spin-density initialization itself, while the comparison of
\textbf{Full} with \textbf{Born-Only} tests the additional spin information
carried by the accepted real-emission state.
For the shower-level analysis, the four physical beam-helicity configurations
are generated as independent hard-process samples for each of the three shower
settings. This is required so that the hard process maintains the polarization
information needed for the correct initialization of the polarized parton
shower. An analysis-level correlated helicity weight on a shared hard-event
history changes the weight used to fill a histogram, but it does not replace
the spin-density matrix with which the shower was initialized and is therefore
not a substitute for this test.

For each physical helicity $h$, the NLO prediction is first formed as
$\sigma_h=\sigma_h^{(+)}-\sigma_h^{(-)}$, after which the combinations in
Eq.~\eqref{eq:helicity-combinations} are constructed. Statistical uncertainties
are evaluated by subdividing each simulation into equivalent subsamples. In
ratios and cumulative moments, the numerator and denominator are formed within
each subsample before the ratio-of-means uncertainty is evaluated, retaining
their Monte Carlo covariance. The three shower settings use the same generator
definitions and run conditions but separate event samples; their realized
event kinematics are not identified event by event.

The shower-level observables are chosen to test the part of the event in which
additional spin information from the accepted POWHEG emission would have the
best chance of becoming visible. We therefore keep the leading dijet system as
the reference object and probe the radiation that resolves a third Breit-frame
jet.

For ordinary differential observables we construct the double-spin asymmetry
\begin{equation}
  A_{LL}(O)=\frac{\Delta\sigma(O)}{\sigma(O)}\,,
\end{equation}
where, in the analysis presented here,
\begin{equation}
  \Delta\sigma(O)=\sigma^{++}(O)-\sigma^{+-}(O)-\sigma^{-+}(O)+\sigma^{--}(O)\;,
\end{equation}
with the first superscript referring to the lepton helicity and the second to
the hadron helicity, while $\sigma(O)$ is the sum of the same four helicity
configurations. Relative to the integrated observables defined earlier, the
common factor of $1/4$ is omitted here because it cancels identically in the
ratio $A_{LL}$.

The first observable is the resolved-emission hardness ratio
\begin{equation}
  r_3 \equiv \frac{p_{T,3}^{B}}{p_{T,1}^{B}}\,,
\end{equation}
where the jets are ordered by Breit-frame transverse momentum after the dijet
selection described above. Jets beyond the leading two are accepted if they
have Breit-frame transverse momentum $p_T^{B}>1~\mathrm{GeV}$ and
laboratory-frame rapidity $-3.5<y_{\mathrm{jet}}<3.5$, and the third jet is
the hardest such additional jet. Events without a third accepted jet do not
contribute to the third-jet observables. The corresponding asymmetry $A_{LL}(r_3)$ tests
whether the additional spin information changes the rate at which the shower
populates hard third-jet configurations.  Since the third jet is tied directly
to radiation beyond the Born-like dijet system, this observable provides a
simple, deliberately targeted handle on the part of the final state most
closely connected to the accepted real emission.

To probe the azimuthal structure more directly, we define the hard-emission
reference plane as the plane spanned by the exchanged boson direction and the
leading Breit-frame jet of the dijet selection.  The third-jet angle relative to this plane is
\begin{equation}
  \Delta\phi_{\mathrm{hard},3}
  =\left|\phi_B(j_3)-\phi_B(j_1)\right|_{\pi}\,,
\end{equation}
where $|\cdots|_{\pi}$ denotes folding into the interval $[0,\pi]$.  The
distributions
$\mathrm{d}\sigma/\mathrm{d}\Delta\phi_{\mathrm{hard},3}$ and
$\mathrm{d}\Delta\sigma/\mathrm{d}\Delta\phi_{\mathrm{hard},3}$ test whether
the real-emission spin-density matrix modifies the orientation of the third jet
with respect to the leading hard plane.  This is designed to be a more direct
spin-correlation handle than a single-particle azimuth measured only with
respect to the lepton plane.

We also consider a $\cos 2\phi$ moment using the hard-plane third-jet angle,
integrated over the tail in $r_3$,
\begin{equation}
  C_{2}^{\mathrm{hard}}(r_{\mathrm{cut}})
  =
  \frac{\int_{r_3>r_{\mathrm{cut}}}\mathrm{d}\sigma\,
        \cos(2\Delta\phi_{\mathrm{hard},3})}
       {\int_{r_3>r_{\mathrm{cut}}}\mathrm{d}\sigma}\,.
\end{equation}
For the spin-dependent moment we use the same unpolarized denominator,
\begin{equation}
  \Delta C_{2}^{\mathrm{hard}}(r_{\mathrm{cut}})
  =
  \frac{\int_{r_3>r_{\mathrm{cut}}}\mathrm{d}\Delta\sigma\,
        \cos(2\Delta\phi_{\mathrm{hard},3})}
       {\int_{r_3>r_{\mathrm{cut}}}\mathrm{d}\sigma}\,.
\end{equation}
The cumulative form reduces the sensitivity to individual low-statistics tail
bins while preserving the dependence on increasingly hard third-jet radiation.
Each cumulative point contains all events above its threshold. Neighbouring
points therefore share many of the same events and are statistically
correlated; the uncertainty calculation accounts for this overlap.

The comparison shown below is organized so that each observable is displayed
both before and after hadronization, with the three configurations overlaid in
every panel.

\begin{figure}[!htbp]
  \centering
  \includegraphics[width=0.47\linewidth]{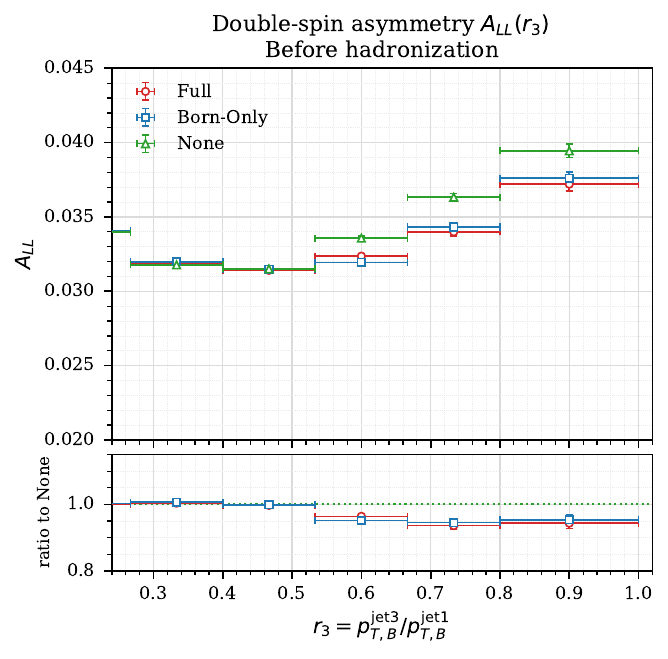}
  \hfill
  \includegraphics[width=0.47\linewidth]{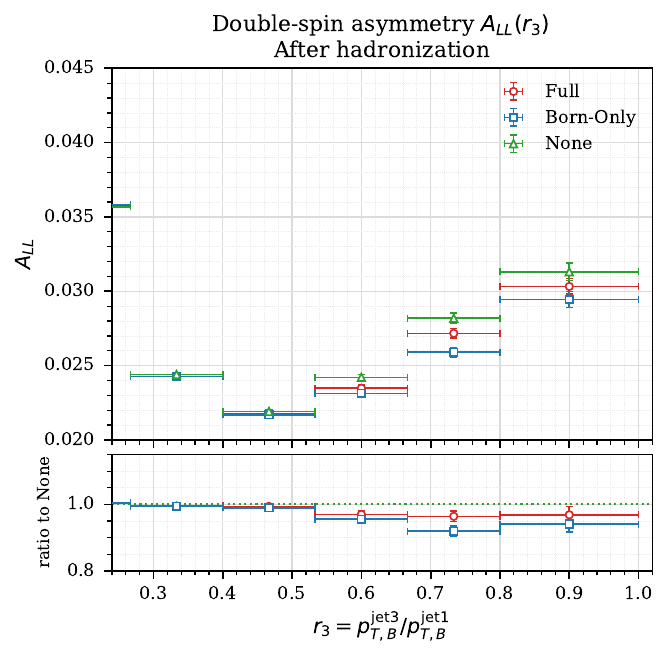}
  \caption{Double-spin asymmetry in the resolved-emission hardness ratio
  $r_3=p_{T,3}^{B}/p_{T,1}^{B}$. Left: shower-level comparison before
  hadronization. Right: the same comparison after hadronization. The lower
  panels give the ratio to \textbf{None}.}
  \label{fig:spincomp-pt3ratio}
\end{figure}

\begin{figure}[!htbp]
  \centering
  \includegraphics[width=0.47\linewidth]{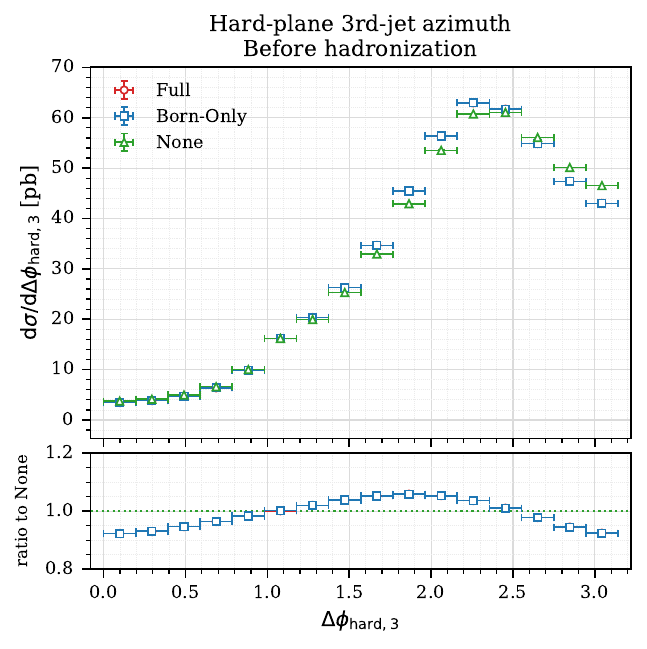}
  \hfill
  \includegraphics[width=0.47\linewidth]{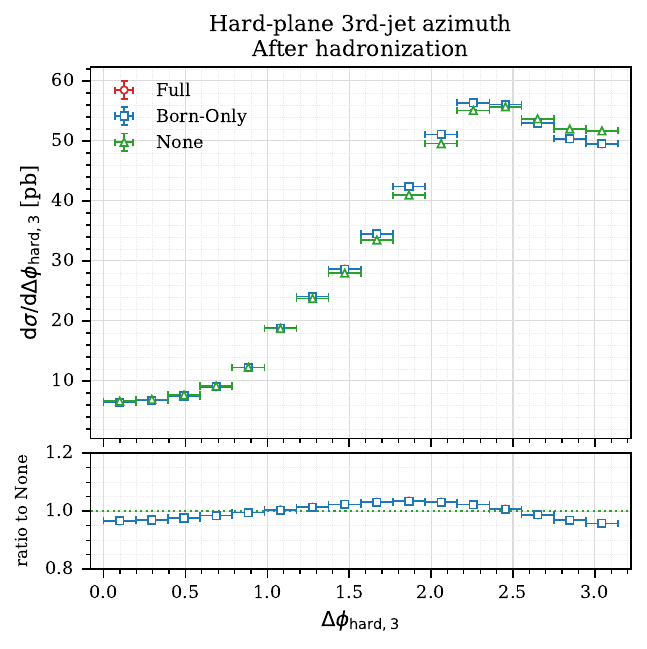}
  \vspace{0.5em}

  \includegraphics[width=0.47\linewidth]{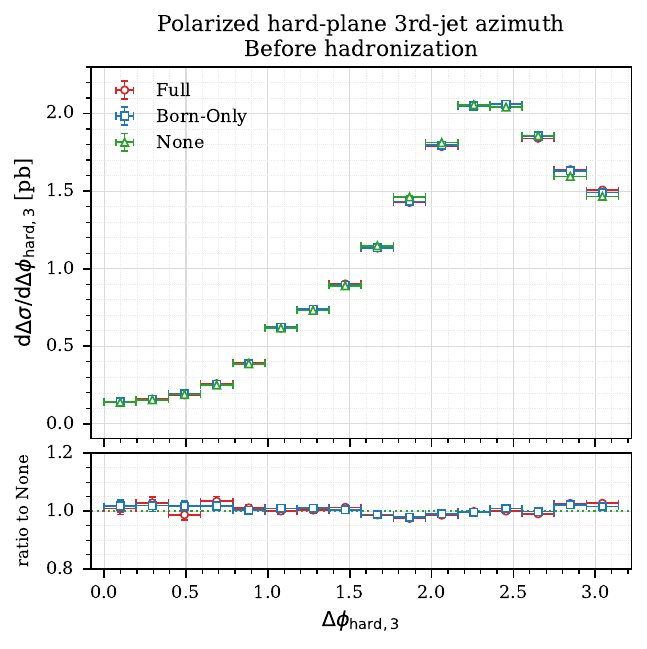}
  \hfill
  \includegraphics[width=0.47\linewidth]{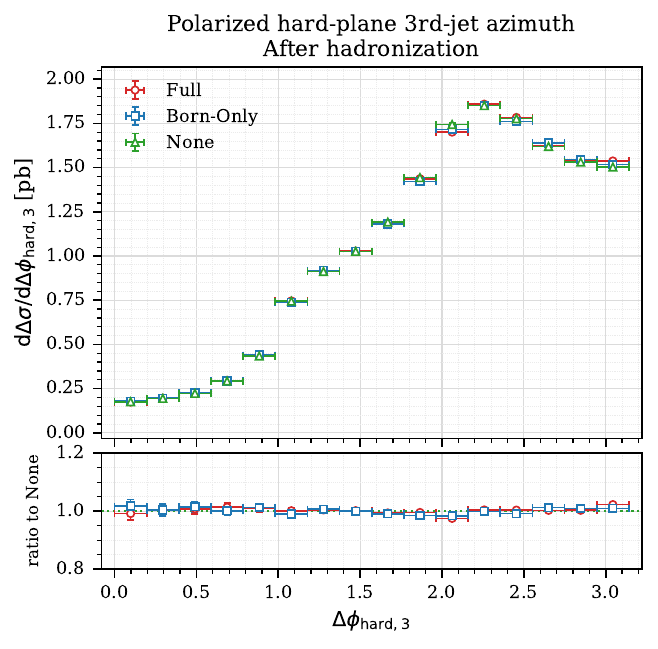}
  \caption{Hard-plane third-jet azimuthal distributions. The upper row shows
  $\mathrm{d}\sigma/\mathrm{d}\Delta\phi_{\mathrm{hard},3}$ and the lower row
  shows
  $\mathrm{d}\Delta\sigma/\mathrm{d}\Delta\phi_{\mathrm{hard},3}$. Left panels
  are before hadronization. Right panels are after hadronization. The lower
  panel within each plot gives the ratio to \textbf{None}. In the upper row,
  the \textbf{Full} markers lie almost exactly underneath the
  \textbf{Born-Only} markers and are therefore largely obscured.}
  \label{fig:spincomp-hardplane-angle}
\end{figure}

\begin{figure}[!htbp]
  \centering
  \includegraphics[width=0.47\linewidth]{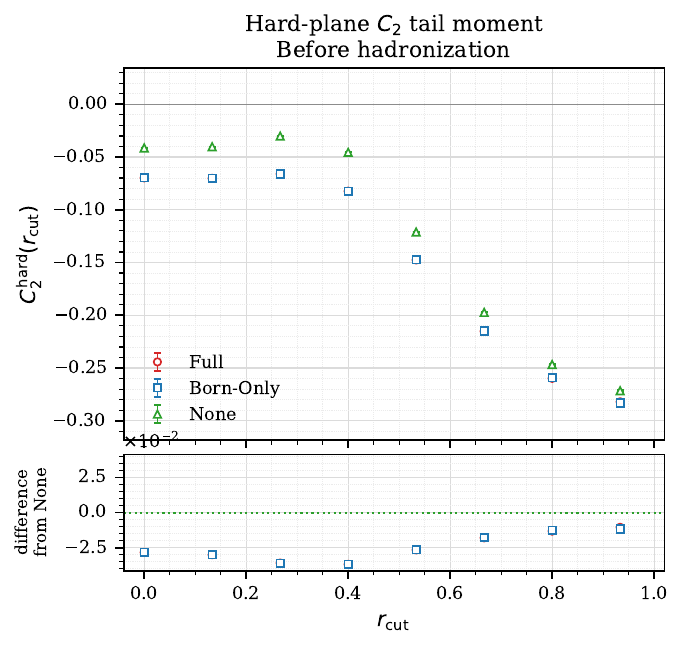}
  \hfill
  \includegraphics[width=0.47\linewidth]{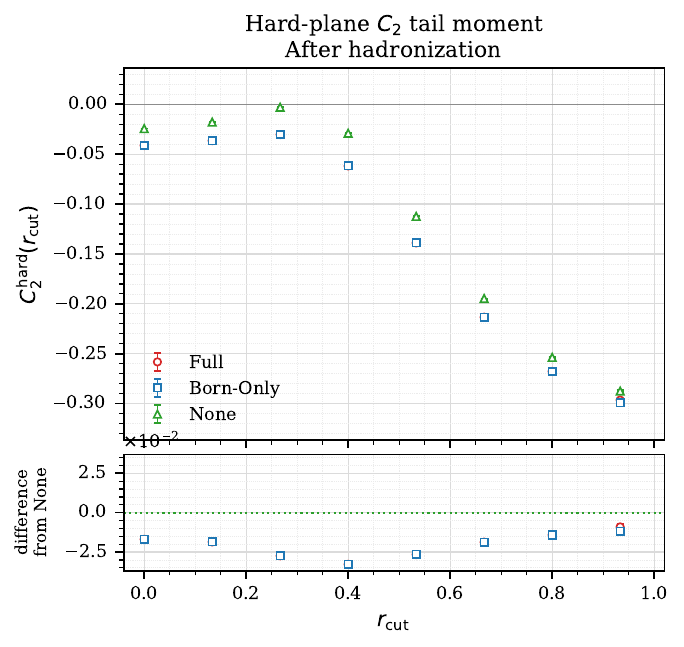}
  \vspace{0.5em}

  \includegraphics[width=0.47\linewidth]{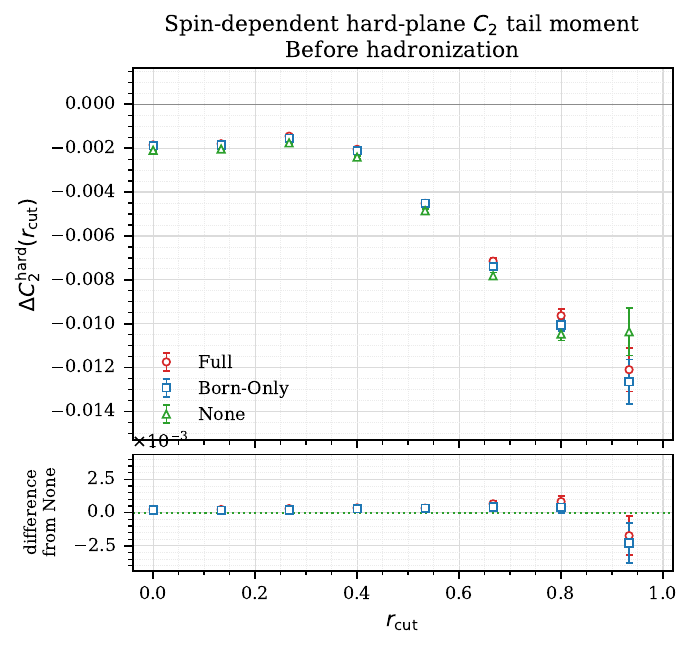}
  \hfill
  \includegraphics[width=0.47\linewidth]{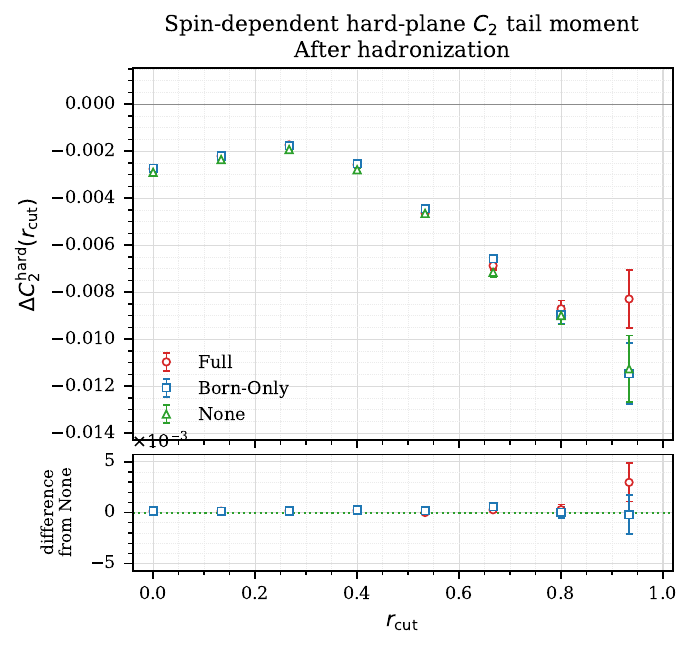}
  \caption{Cumulative hard-plane $\cos(2\Delta\phi_{\mathrm{hard},3})$ moments
  as a function of the third-jet hardness threshold. The upper row shows the
  unpolarized moment $C_{2}^{\mathrm{hard}}(r_{\mathrm{cut}})$ and the lower row
  shows the spin-dependent moment
  $\Delta C_{2}^{\mathrm{hard}}(r_{\mathrm{cut}})$. Left panels are before
  hadronization. Right panels are after hadronization. Since the reference
  moments can pass through zero, the lower panel within each plot shows the
  absolute difference from \textbf{None}, rather than a ratio. These are
  absolute shifts of the dimensionless moments: for example, $10^{-2}$ denotes
  a shift of $0.01$, not a one-per-cent relative change. Neighbouring threshold
  points are statistically correlated because their cumulative event samples
  overlap.}
  \label{fig:spincomp-hardplane-moments}
\end{figure}

The three configurations are compared in
Figs.~\ref{fig:spincomp-pt3ratio}--\ref{fig:spincomp-hardplane-moments}. In
several of these third-jet-sensitive observables both the \textbf{Born-Only}
and the \textbf{Full} predictions depart visibly from \textbf{None}, showing
that the polarized DIS Born-level spin-density initialization changes the
showered prediction relative to a spin-disabled simulation. In the three
displayed bins above $r_3=8/15$, \textbf{Born-Only} reduces $A_{LL}(r_3)$ by
$4.6$--$5.5\%$ relative to \textbf{None} before hadronization and by
$4.5$--$8.1\%$ after hadronization. The covariance-aware comparison of the full
$r_3$ distributions does not establish a statistically significant separation
between \textbf{Full} and \textbf{Born-Only}. The largest local excursion is in
the post-hadronization bin $2/3<r_3<4/5$, where \textbf{Full} is $4.9\%$ above
\textbf{Born-Only}, corresponding to $2.9$ standard deviations before any
account of the several correlated bins and observables examined. It is not
reproduced before hadronization and is therefore not interpreted as a robust
separate real-emission-spin effect. The additional fourth-jet angular and
asymmetry observables that we examined (not shown) likewise reveal no coherent
separation, although their lower rates limit the strength of that statement.
For longitudinally polarized DIS in this kinematic region, the Born-level
shower-spin initialization therefore captures the numerically relevant spin
correlations.

The polarized hard-plane azimuthal distribution in the lower row of
Fig.~\ref{fig:spincomp-hardplane-angle} is particularly stable: within the
Monte Carlo precision, it shows no appreciable shape separation among the
three shower settings. This weak sensitivity is consistent with longitudinal
helicity states providing no transverse spin direction at a single branching,
while the third-jet sample is dominated by soft radiation whose leading
contribution is spin independent. A helicity-dependent azimuthal reshaping must
therefore be transmitted through at least one subsequent non-soft branching
and is further diluted by shower evolution, jet clustering, and hadronization.

The visible differences between the \textbf{Born-Only} and \textbf{None}
configurations are of the expected type. Enabling the Born-seeded
spin chain activates the azimuthally averaged branching weight of
Eq.~\eqref{eq:Wphi}, which corrects each initial-state branching for the
helicity dependence of the parent parton distribution through the spin
weights of Table~\ref{tab:branch1}. This changes the amount of radiation
rather than its azimuthal orientation: the weight survives azimuthal
integration, its
coefficients are of order unity across the sampled $z$ range, and its
effect accumulates along the backward evolution, which probes the
polarized distributions at successively shifted momentum fractions where
$\Delta f/f$ varies appreciably. Enabling the chain also switches on the
azimuthal correlations between successive shower emissions, which are
present even for unpolarized beams. Both mechanisms act together in the
comparisons shown here. Since the dijet system is fixed by the hard
process definition, which is common to all three configurations, these effects
first become visible in the radiation that resolves a third jet, consistent
with the pattern observed above.

By contrast, the absence of a resolvable difference between the \textbf{Full} and
\textbf{Born-Only} configurations is consistent with the structure of the
construction. At the level of the generator probability law, the two settings
use the same hard-event construction, flavour channels, event-weight
prescription, and POWHEG hardest-emission distribution; only the spin-density
information passed to the subsequent shower differs. They are generated as
separate samples and therefore do not share identical realized event
kinematics. A distribution-level difference must arise from correlations
between the radiation that resolves a third jet and the hardest-emission
system, and these are suppressed for several independent reasons. The
small-$r_3$ region that dominates the
third-jet rate is governed by soft-gluon emission, which at leading power
is independent of the spin state of the emitters. The additional
information in the accepted real-emission state resides mainly in the
polarization of the emitted parton, whose azimuthal imprint on a subsequent
branching carries the kernel coefficients of Table~\ref{tab:branch2},
which vanish in the soft region and are sizable only for rare symmetric
splittings. In the polarized observables the residual effect is reduced further. The
component of the gluon polarization that can reorient subsequent radiation
is, by parity, independent of the beam helicities and cancels in
$\Delta\sigma$, while the component that does depend on the beam helicities
does not affect the angular distribution of a single splitting. A
beam-helicity-dependent azimuthal effect therefore requires at least two
successive branchings, with longitudinal polarization offering no
transverse spin direction ($H_{i+-}=H_{i-+}=0$) that could provide it in
one step. Finally, only third jets
descending from the hardest-emission parton carry the new correlation,
which angular-ordered evolution, jet clustering, and hadronization further
dilute.

\FloatBarrier
\section{Summary and Outlook}\label{sec:conclusions}

We have presented a treatment of polarized deep-inelastic scattering at
next-to-leading order in \Herwig\ using the POWHEG matching scheme. The
calculation covers the full neutral-current process, including $\gamma/Z$
interference, as well as the charged-current channel. We derived the
real-emission contributions for both the QCD Compton and boson-gluon fusion
channels, together with the virtual corrections and collinear counterterms
needed for NLO event generation. Integrated cross sections agree with the
fixed-order program \POLDIS\ at the permille or better level for the dominant
channels and, within the quoted uncertainties, at the sub-percent to percent
level for the smaller polarized interference contributions.
The parton-level comparison without subsequent shower evolution separates the
accepted POWHEG real-emission kinematics from the shower-interface
reconstruction, and provides the differential validation used for the inclusive
DIS, dijet-selected kinematic, and Breit-frame transverse-momentum observables
shown here.

We have also shown how to initialize the parton shower with the spin-density
information supplied by the polarized DIS hard process. Already at Born level
this gives a new spin-aware shower baseline for polarized DIS, and the explicit
comparisons show that this baseline can differ from a fully spin-disabled
simulation in third-jet-sensitive observables. We then extended the construction
to the NLO real-emission state. By initializing the shower with the exact
spin-density matrix of the accepted real-emission configuration, the spin
information of the POWHEG event can be retained without modifying the accepted
hard-event normalization or POWHEG emission choice. We introduced third-jet
hardness observables, hard-plane third-jet azimuths, and azimuthal
$\cos 2\phi$ moments as exploratory probes of this additional spin information
in showered events. For the longitudinally polarized DIS setup considered here,
the Born-level spin-density initialization produces a visible effect in several
third-jet-sensitive observables, while the additional spin information from the
accepted real-emission configuration is not resolved as a robust separate
effect within the independent-sample Monte Carlo precision of the present
observable set. A local post-hadronization fluctuation in one hard-$r_3$ bin is
not accompanied by a significant distribution-level separation or a coherent
pre-hadronization counterpart.
As discussed in Section~\ref{sec:real-emission-spin-shower}, this outcome
is expected from the structure of the construction, in which the hardest
emission follows the same probability law in the two settings and the residual
correlations are soft-suppressed, kernel-suppressed for longitudinal
polarization, and further diluted by showering, clustering, and hadronization.

A broader set of exclusive shower and jet observables would pin down where
measurements are most sensitive to the Born-level spin-density initialization,
while varying the shower and hadronization settings would test the robustness
of the spin-on versus spin-disabled differences. More broadly, the formalism
developed here extends naturally beyond DIS, opening the way to polarized
processes at hadron-hadron and lepton-hadron colliders. The polarized DIS hard
process and its spin-correlation treatment will be released in a future
\Herwig\ version, enabling fully exclusive, spin-aware simulations for
polarized DIS phenomenology at the EIC.

\section*{Acknowledgments}
We thank Daniel de Florian for useful discussions and for providing the \POLDIS\ reference calculation used for validation. AP acknowledges support from the US Department of Energy, Office of Science, Office of Nuclear Physics under Award Number DE-SC0025728 and from the National Science Foundation under Grant No.\ PHY 2210161.
MRM is supported by the UK Science and Technology Facilities Council (grant numbers ST/T001011/1 and ST/X000745/1).

\appendix

\section{Independent Fixed-Order Validation}
\label{app:standalone-fixed-order-validation}

The parton-level comparisons in Section~\ref{sec:polarizeddis} use the
\Herwig\ POWHEG construction with subsequent shower evolution disabled.  This
is the appropriate comparison for the partonic states produced by the event
generator, but it still contains the POWHEG Sudakov organization of the real
radiation.  We therefore performed a second validation in which the same
neutral-current NLO ingredients are evaluated as a fixed-order calculation.  The
calculation does not rely on the \Herwig\ event-generation machinery. We again use \POLDIS\
as the external reference against which the resulting distributions
are compared.

The physics setup is identical to that used in the neutral-current validation in
the main text.  We take $E_e=18~\mathrm{GeV}$ and $E_p=275~\mathrm{GeV}$,
impose $100~\mathrm{GeV}^2<Q^2\le2500~\mathrm{GeV}^2$ and $0.2\le y\le0.6$, and use
the unpolarized proton set \texttt{NNPDF40\_nlo\_pch\_as\_01180} and the
polarized-difference set \texttt{NNPDFpol20\_nlo\_as\_01180}.  The
electroweak input parameters, the $\gamma$, $Z$, and $\gamma Z$ neutral-current
coefficients, the polarized Born analysing powers, the PDF ratios, and the
running coupling with $\alpha_s(M_Z)=0.118$ are matched to the \Herwig\
configuration.  The central scale is $\mu^2=Q^2$. The uncertainty bands are
obtained by varying the scale to $\mu=Q/2$ and $\mu=2Q$, with the phase-space
points and cuts kept fixed.  This is the fixed-order scale convention used for
the \POLDIS\ comparison, and differs from the native POWHEG emission-scale
choice used by the no-shower \Herwig\ differential plots in the main text.

At each phase-space point we evaluate the Born contribution, the virtual
correction, the collinear finite terms, and the finite QCDC and BGF
contributions appearing in the \Herwig\ DIS matrix element.  The real-emission
part is treated at fixed order: the Born-projected contribution, the
two-parton QCDC or BGF configuration, and the corresponding subtraction
contribution are filled as separate signed contributions to the same
observables.  The real and subtraction terms therefore cancel in inclusive
quantities while contributing with their physical two-parton kinematics to
observables sensitive to resolved radiation.

For the polarized distributions, all four longitudinal helicity configurations
are evaluated on the same phase-space point and combined into
$\sigma=(\sigma_{++}+\sigma_{+-}+\sigma_{-+}+\sigma_{--})/4$ and
$\Delta\sigma_{LL}=(\sigma_{++}+\sigma_{--}-\sigma_{+-}-\sigma_{-+})/4$.
This common-point combination preserves the correlations required for stable
helicity-difference distributions.  The same Breit-frame reconstruction,
histogram definitions, and dijet selection are then applied to the independent
fixed-order result and to the \POLDIS\ reference.  No parton shower,
hadronization, decays, POWHEG Sudakov factor, or POWHEG hardest-emission
generation is included.

The distributions shown below mirror
Figs.~\ref{fig:validation-partonlevel-inclusive-unpol}-%
\ref{fig:validation-partonlevel-dijet-pol-kin}, replacing the no-shower POWHEG
comparison by the independent fixed-order result.  The agreement with \POLDIS\
in the inclusive, dijet-selected, unpolarized, and polarized distributions
validates the fixed-order components used in \Herwig.  Residual differences
with the main-text POWHEG plots therefore probe the organization of the hardest
emission, rather than the underlying fixed-order matrix elements.

\begin{figure}[!htbp]
  \centering
  \includegraphics[width=0.32\linewidth]{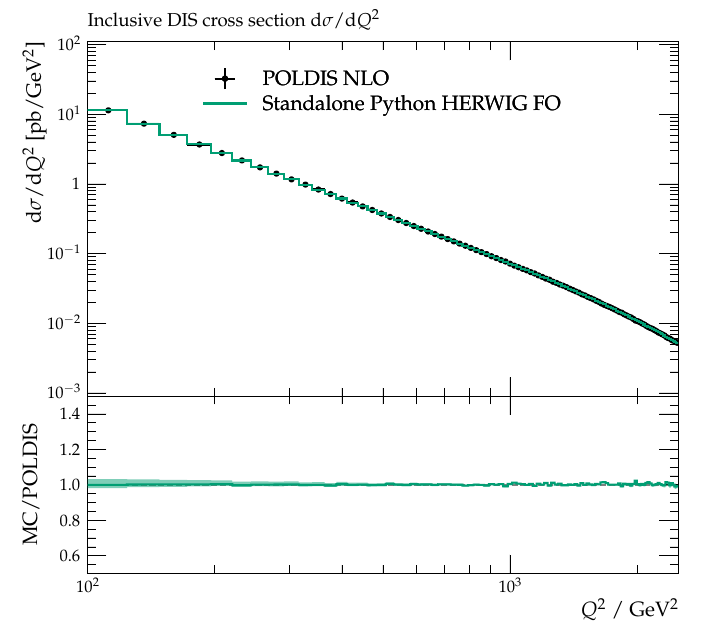}
  \includegraphics[width=0.32\linewidth]{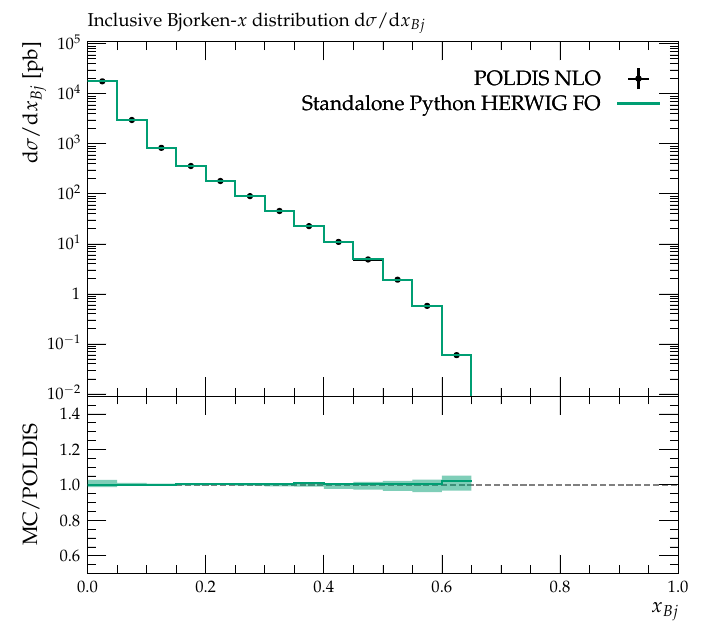}
  \includegraphics[width=0.32\linewidth]{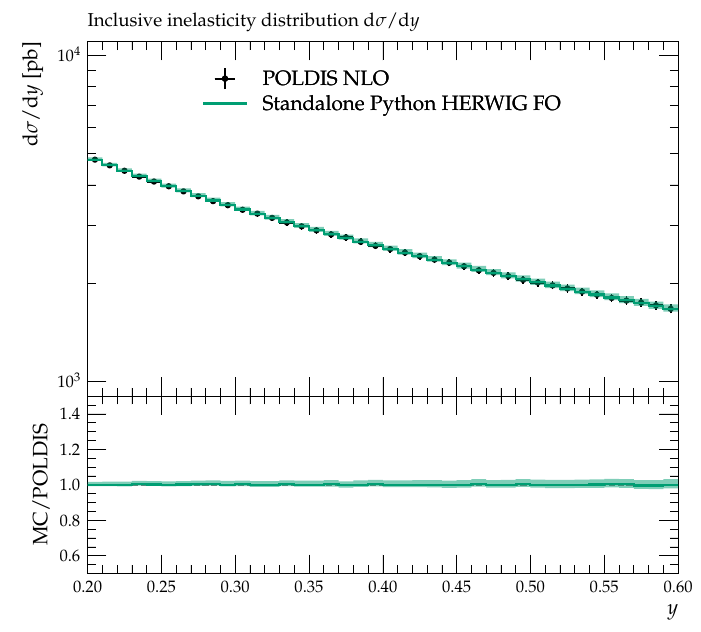}

  \includegraphics[width=0.32\linewidth]{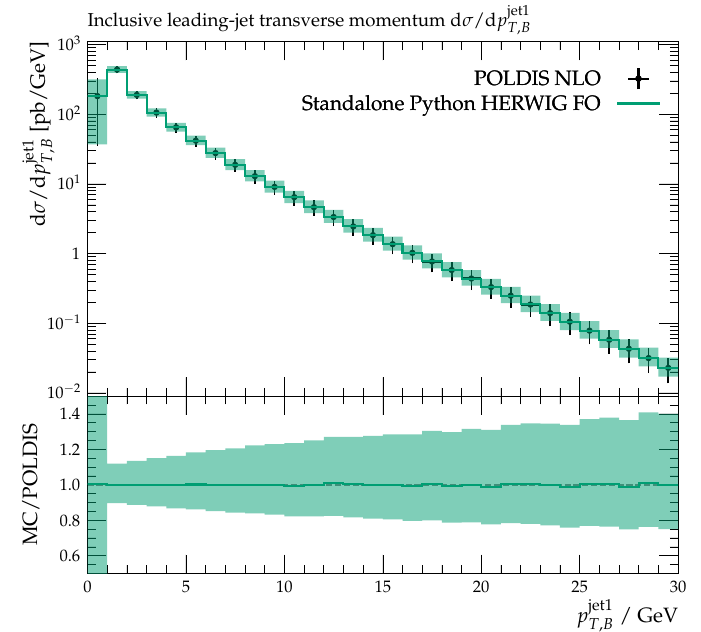}
  \caption{Independent fixed-order validation of inclusive unpolarized
  differential distributions before dijet cuts:
  $\mathrm{d}\sigma/\mathrm{d}Q^2$,
  $\mathrm{d}\sigma/\mathrm{d}x_{\mathrm{Bj}}$,
  $\mathrm{d}\sigma/\mathrm{d}y$, and
  $\mathrm{d}\sigma/\mathrm{d}p_{T,1}^{B}$.  The black points show the
  \POLDIS\ NLO reference, while the green curve and band show the independent
  fixed-order central prediction and scale variation.}
  \label{fig:app-standalone-fo-inclusive-unpol}
\end{figure}

\begin{figure}[!htbp]
  \centering
  \includegraphics[width=0.32\linewidth]{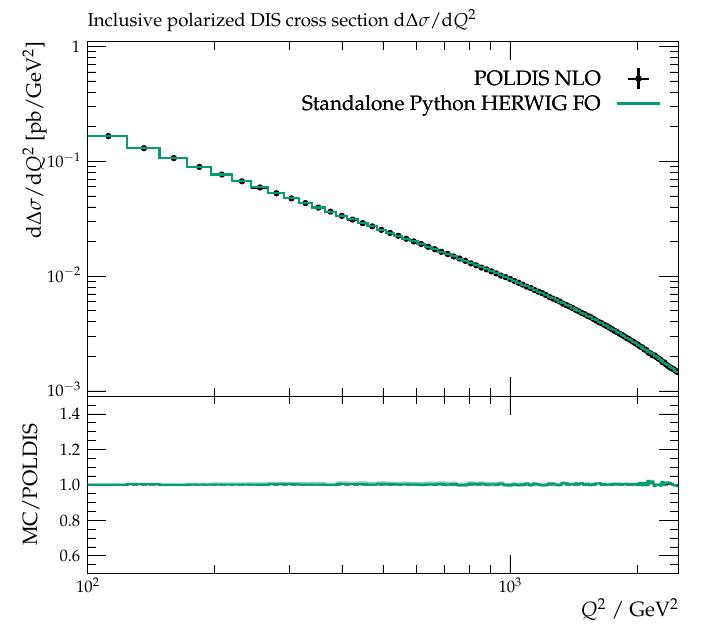}
  \includegraphics[width=0.32\linewidth]{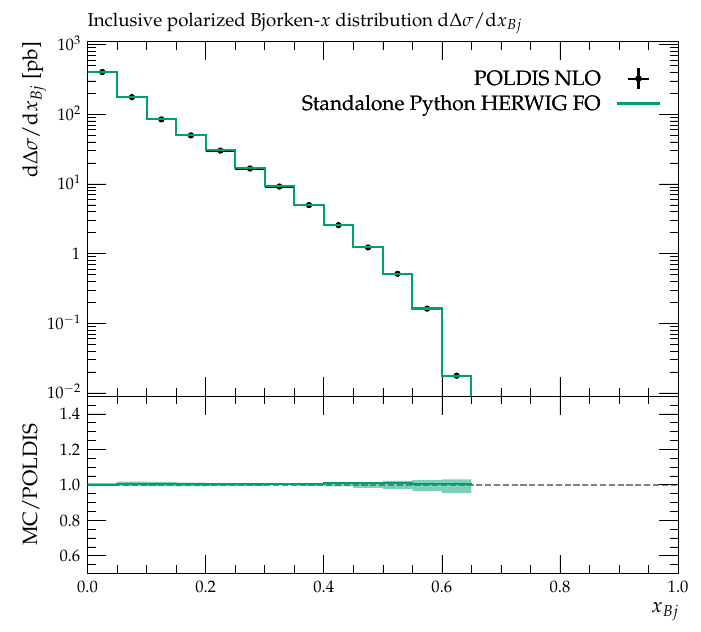}
  \includegraphics[width=0.32\linewidth]{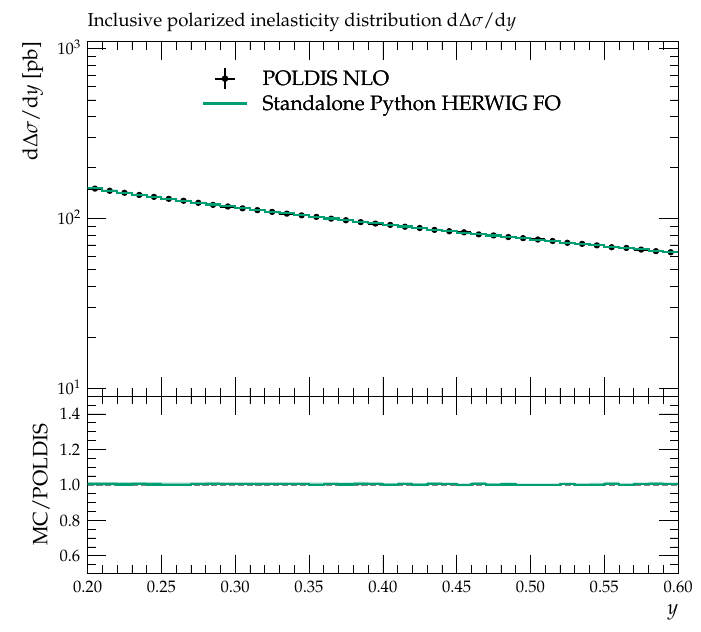}

  \includegraphics[width=0.32\linewidth]{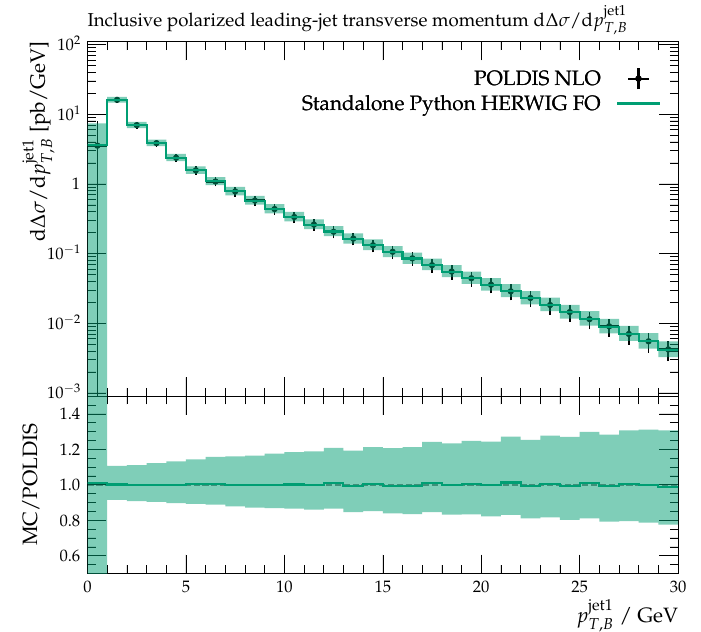}
  \caption{Independent fixed-order validation of inclusive polarized
  differential distributions before dijet cuts:
  $\mathrm{d}\Delta\sigma/\mathrm{d}Q^2$,
  $\mathrm{d}\Delta\sigma/\mathrm{d}x_{\mathrm{Bj}}$,
  $\mathrm{d}\Delta\sigma/\mathrm{d}y$, and
  $\mathrm{d}\Delta\sigma/\mathrm{d}p_{T,1}^{B}$. The black points show the
  \POLDIS\ NLO reference, while the green curve and band show the independent
  fixed-order central prediction and scale variation.}
  \label{fig:app-standalone-fo-inclusive-pol}
\end{figure}

\begin{figure}[!htbp]
  \centering
  \includegraphics[width=0.32\linewidth]{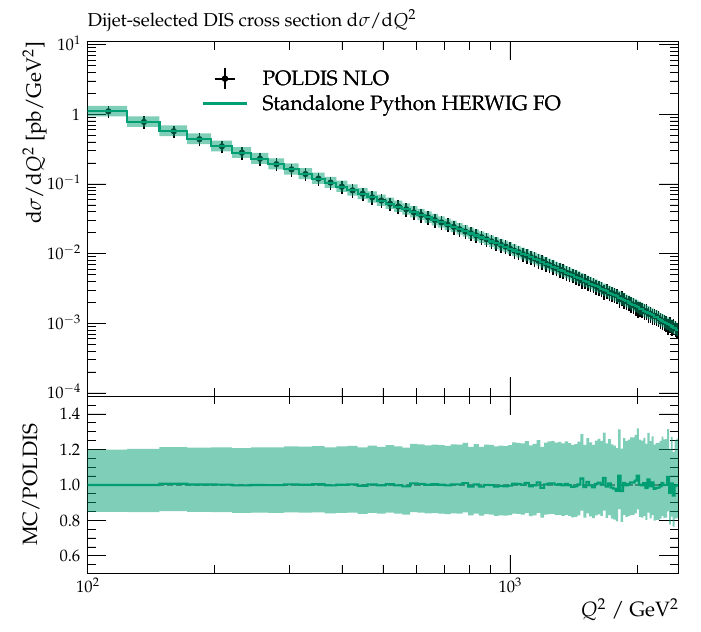}
  \includegraphics[width=0.32\linewidth]{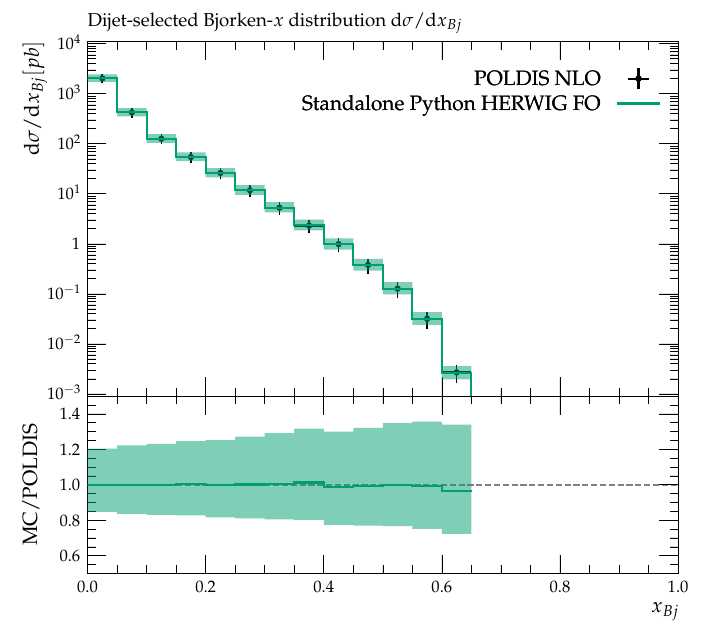}
  \includegraphics[width=0.32\linewidth]{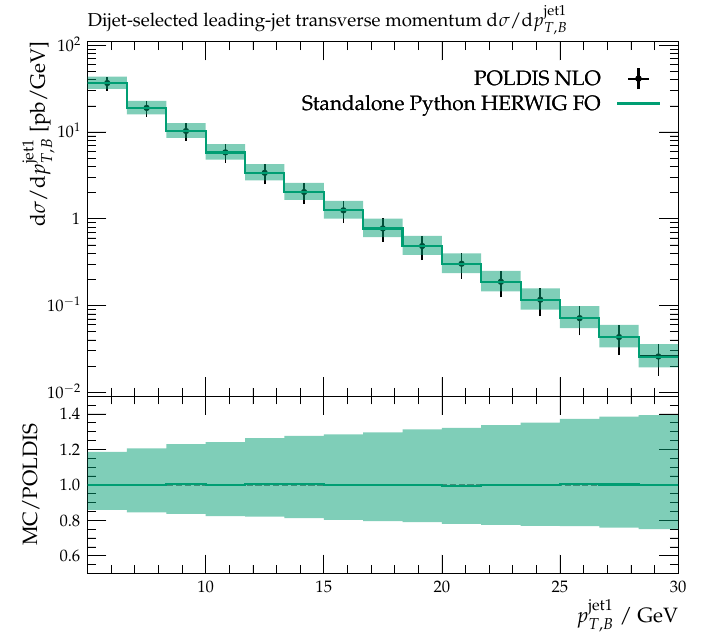}

  \includegraphics[width=0.32\linewidth]{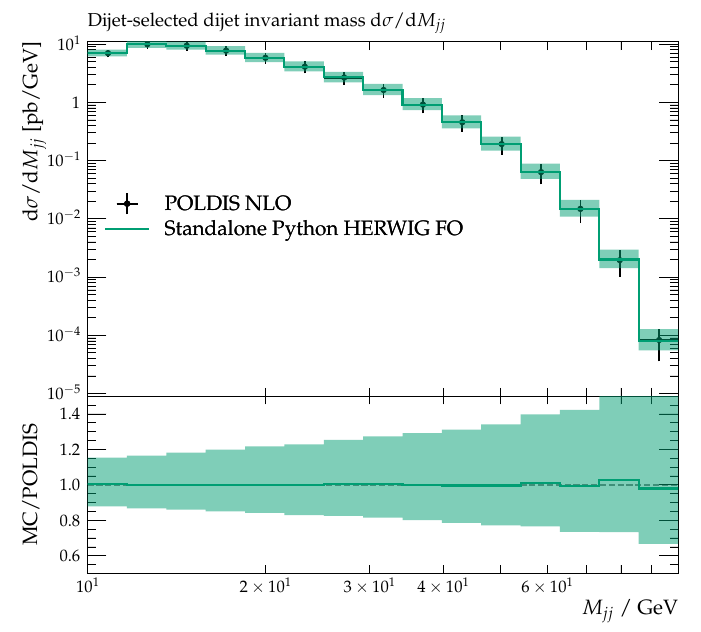}
  \includegraphics[width=0.32\linewidth]{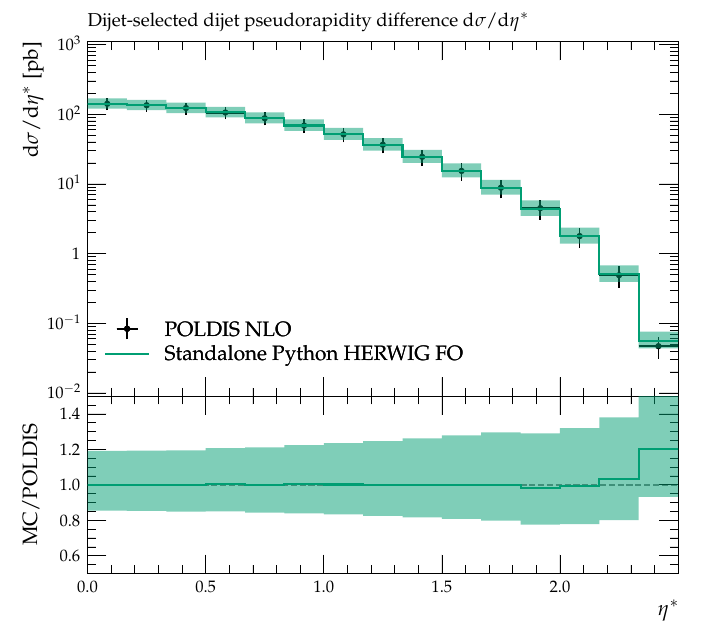}
  \includegraphics[width=0.32\linewidth]{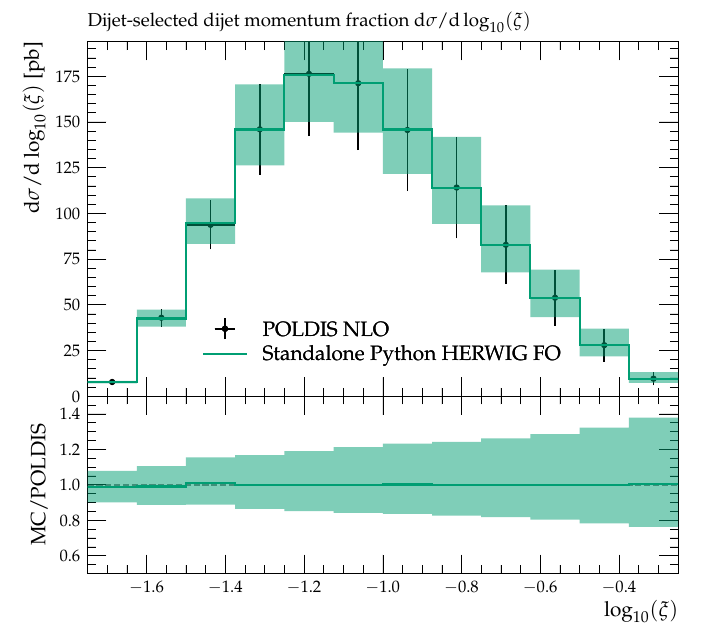}
  \caption{Independent fixed-order validation of dijet-selected unpolarized
  kinematic distributions after dijet cuts:
  $\mathrm{d}\sigma/\mathrm{d}Q^2$,
  $\mathrm{d}\sigma/\mathrm{d}x_{\mathrm{Bj}}$,
  $\mathrm{d}\sigma/\mathrm{d}p_{T,1}^{B}$,
  $\mathrm{d}\sigma/\mathrm{d}M_{jj}$,
  $\mathrm{d}\sigma/\mathrm{d}\eta^\ast$, and
  $\mathrm{d}\sigma/\mathrm{d}\log_{10}(\xi)$. The black points show the
  \POLDIS\ NLO reference, while the green curve and band show the independent
  fixed-order central prediction and scale variation.}
  \label{fig:app-standalone-fo-dijet-unpol}
\end{figure}

\begin{figure}[!htbp]
  \centering
  \includegraphics[width=0.32\linewidth]{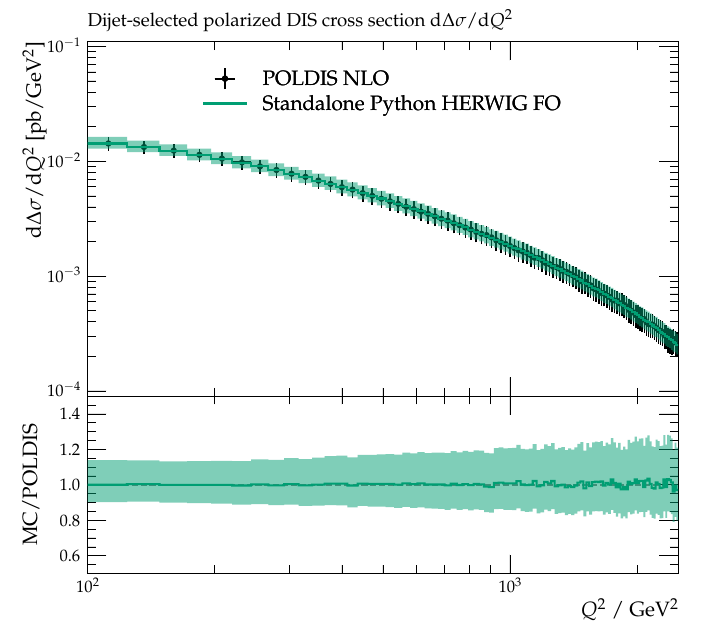}
  \includegraphics[width=0.32\linewidth]{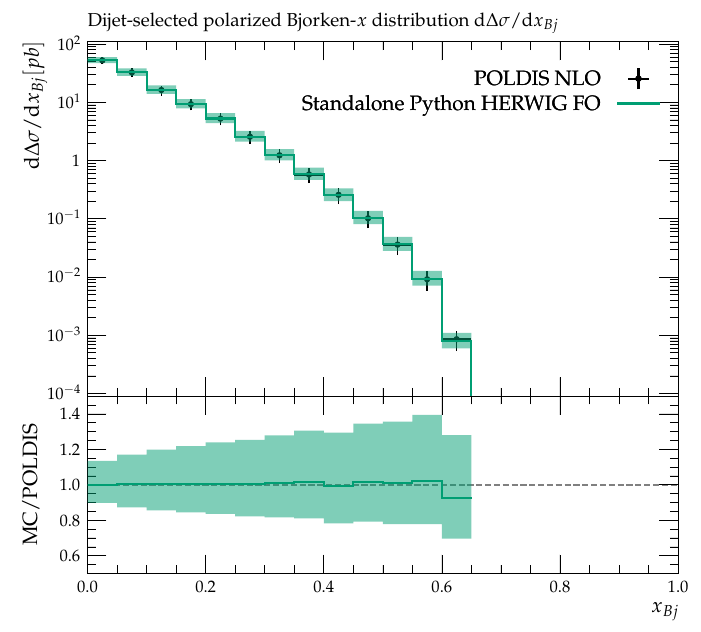}
  \includegraphics[width=0.32\linewidth]{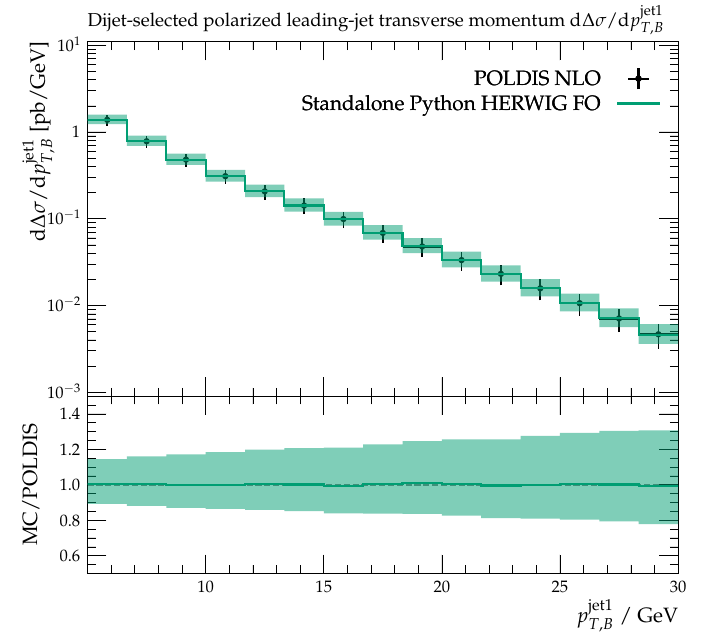}

  \includegraphics[width=0.32\linewidth]{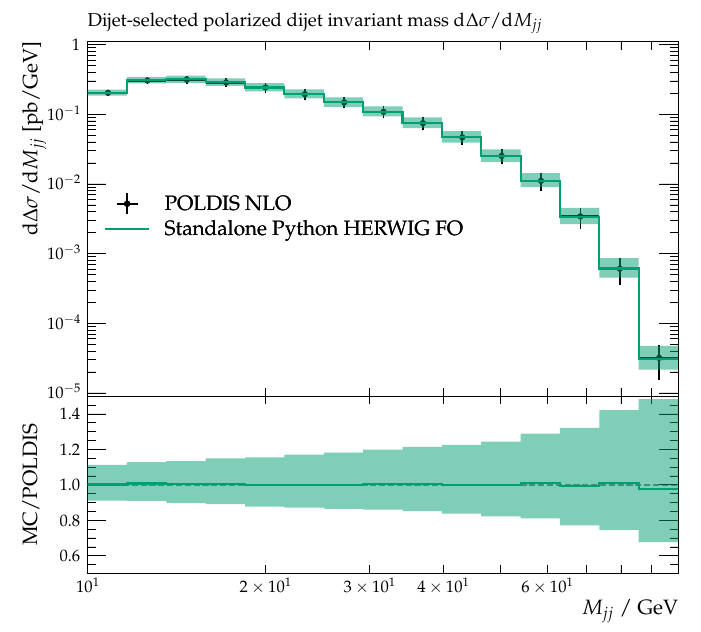}
  \includegraphics[width=0.32\linewidth]{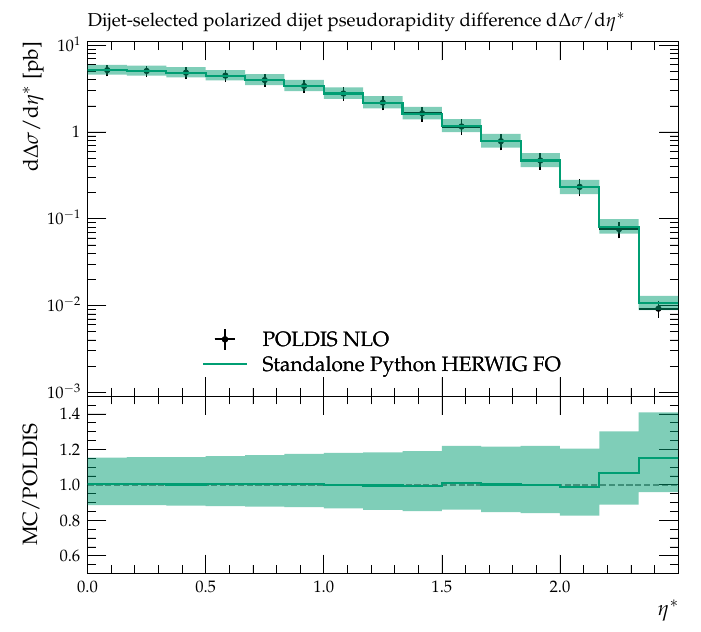}
  \includegraphics[width=0.32\linewidth]{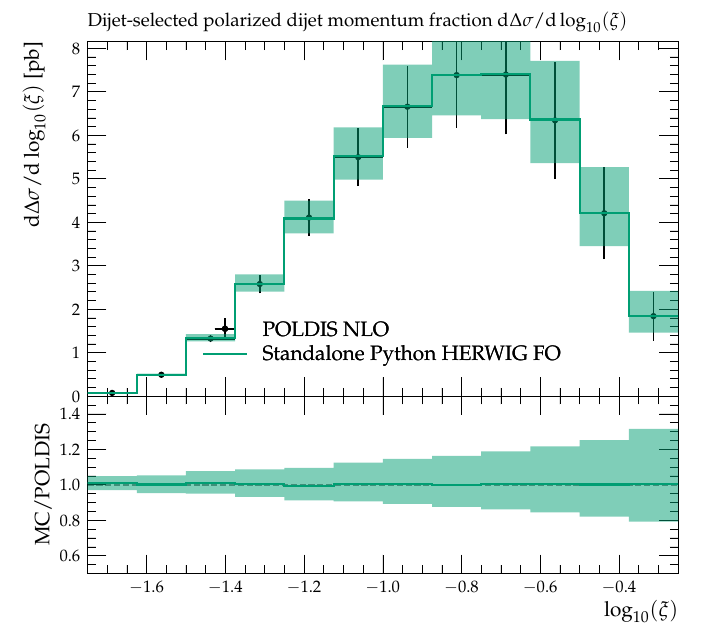}
  \caption{Independent fixed-order validation of dijet-selected polarized
  kinematic distributions after dijet cuts:
  $\mathrm{d}\Delta\sigma/\mathrm{d}Q^2$,
  $\mathrm{d}\Delta\sigma/\mathrm{d}x_{\mathrm{Bj}}$,
  $\mathrm{d}\Delta\sigma/\mathrm{d}p_{T,1}^{B}$,
  $\mathrm{d}\Delta\sigma/\mathrm{d}M_{jj}$,
  $\mathrm{d}\Delta\sigma/\mathrm{d}\eta^\ast$, and
  $\mathrm{d}\Delta\sigma/\mathrm{d}\log_{10}(\xi)$. The black points show the
  \POLDIS\ NLO reference, while the green curve and band show the independent
  fixed-order central prediction and scale variation.}
  \label{fig:app-standalone-fo-dijet-pol}
\end{figure}

\FloatBarrier

\section{Charged-Current Differential Validation}
\label{app:charged-current-differential-validation}

In the main text we present the neutral-current differential validation figures
and the charged-current integrated-rate comparison.  For completeness, we show
here the corresponding charged-current differential distributions.  The setup,
cuts, correlated-helicity construction, and no-shower parton-level analysis are
the same as in Section~\ref{sec:polarizeddis}. Only the exchanged boson is
changed to the charged-current $W$ channel.  The black points show the \POLDIS\
NLO reference, while the red curve and band show the \Herwig\ central prediction
and scale variation.  As in the neutral-current plots, the \Herwig\ band is the
native POWHEG emission-scale variation described in
Section~\ref{sec:polarizeddis}. The \POLDIS\ reference follows the fixed-order
$Q^2$-central scale convention.

\begin{figure}[!htbp]
  \centering
  \includegraphics[width=0.32\linewidth]{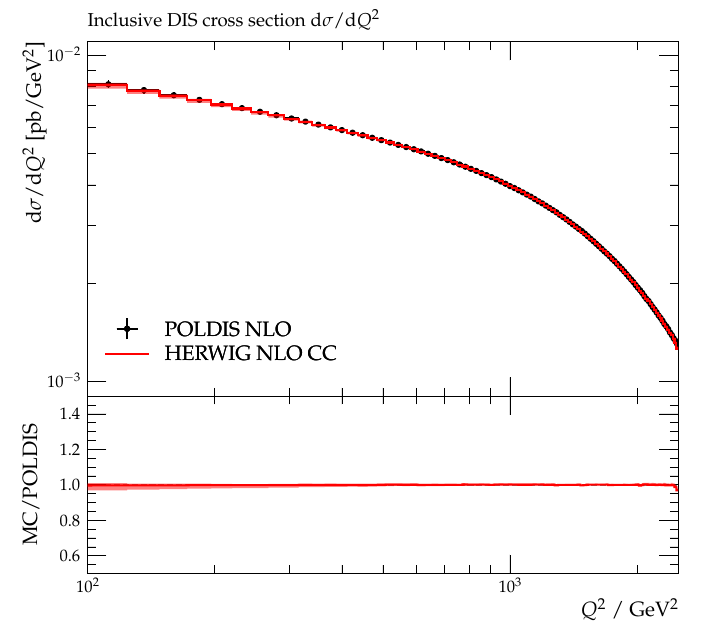}
  \includegraphics[width=0.32\linewidth]{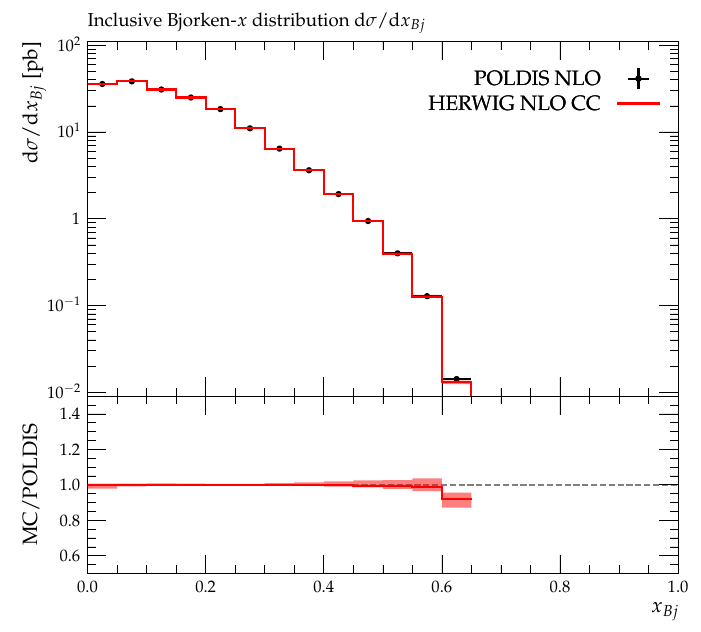}
  \includegraphics[width=0.32\linewidth]{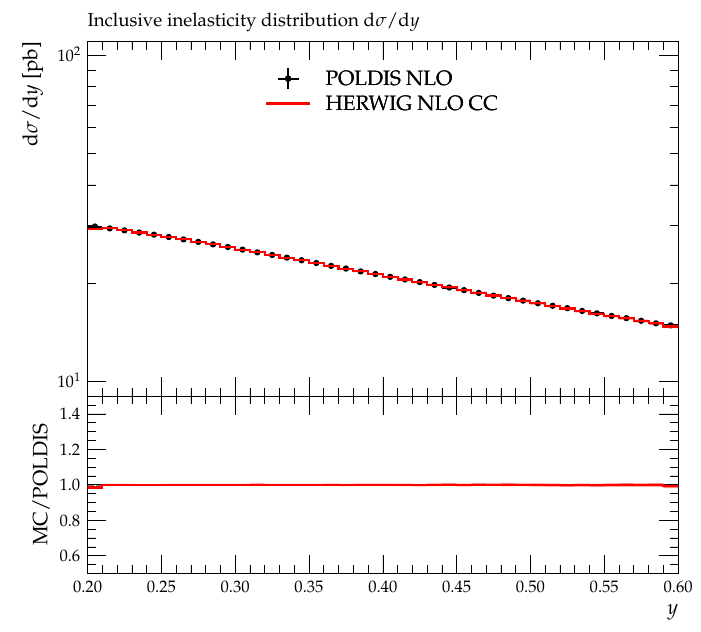}

  \includegraphics[width=0.32\linewidth]{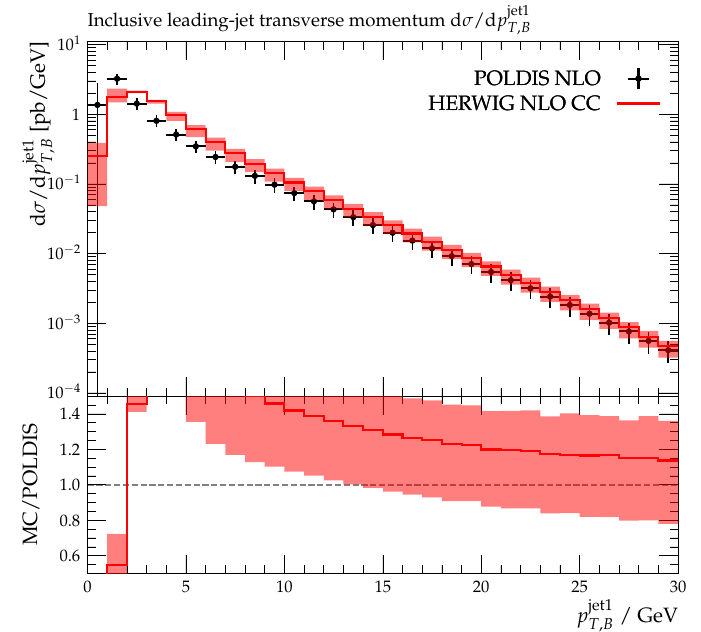}
  \caption{Charged-current parton-level validation without shower evolution of
  inclusive unpolarized differential distributions before dijet cuts:
  $\mathrm{d}\sigma/\mathrm{d}Q^2$,
  $\mathrm{d}\sigma/\mathrm{d}x_{\mathrm{Bj}}$,
  $\mathrm{d}\sigma/\mathrm{d}y$, and
  $\mathrm{d}\sigma/\mathrm{d}p_{T,1}^{B}$.}
  \label{fig:app-cc-inclusive-unpol}
\end{figure}

\begin{figure}[!htbp]
  \centering
  \includegraphics[width=0.32\linewidth]{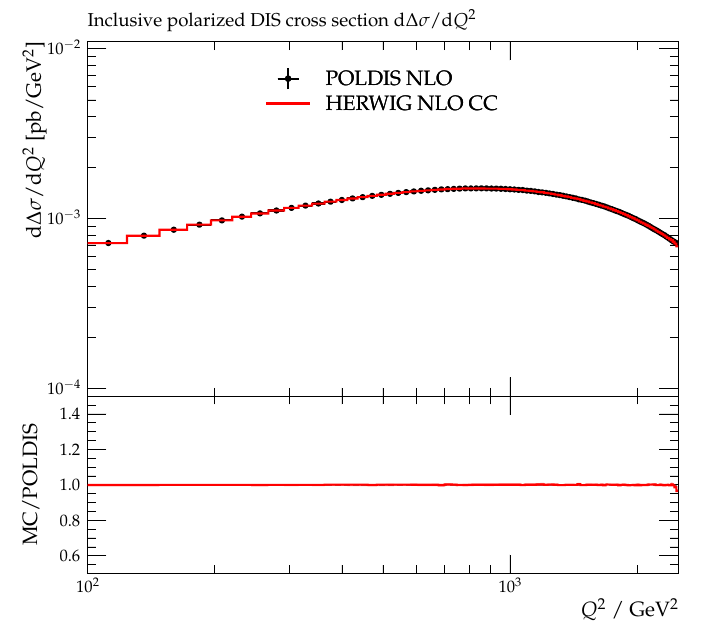}
  \includegraphics[width=0.32\linewidth]{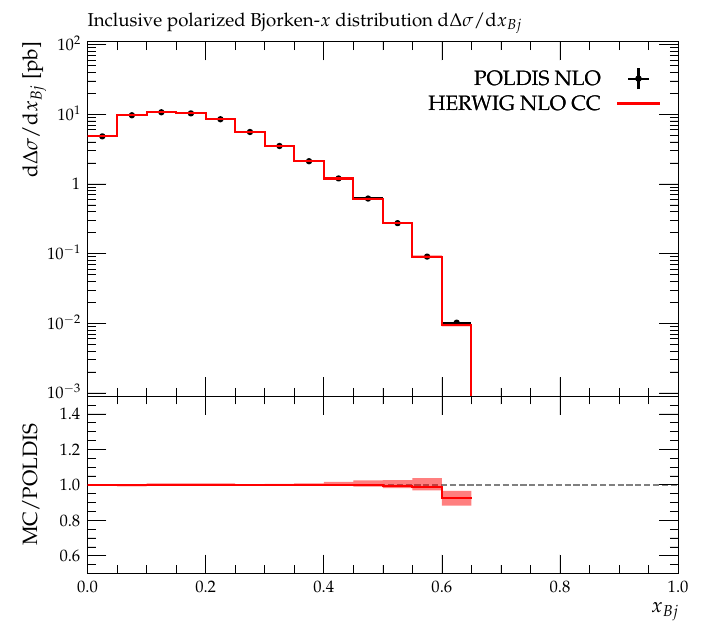}
  \includegraphics[width=0.32\linewidth]{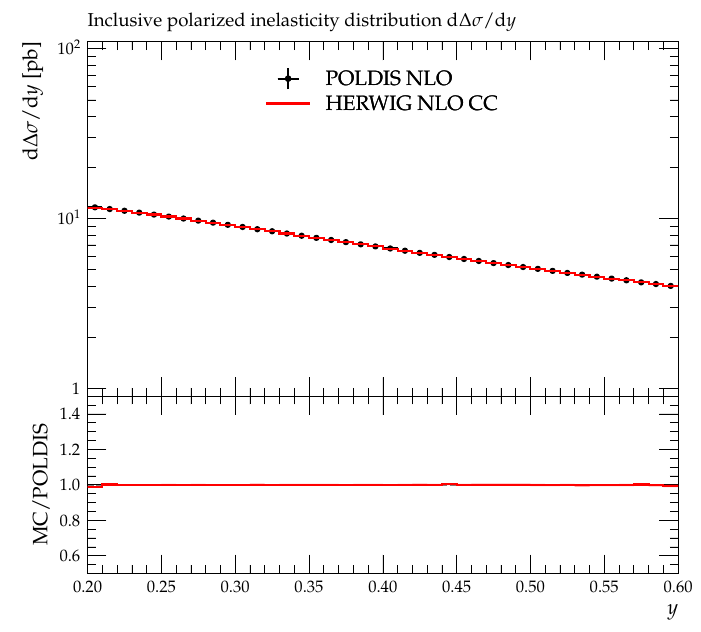}

  \includegraphics[width=0.32\linewidth]{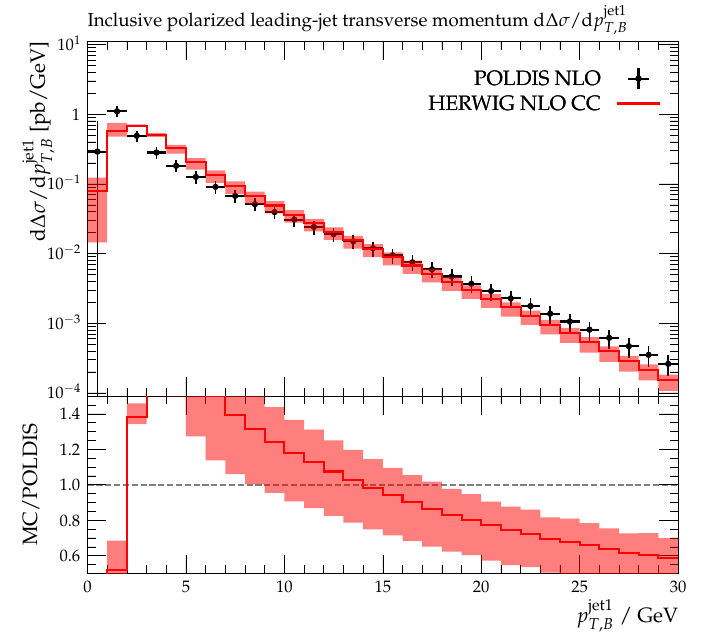}
  \caption{Charged-current parton-level validation without shower evolution of
  inclusive polarized differential distributions before dijet cuts:
  $\mathrm{d}\Delta\sigma/\mathrm{d}Q^2$,
  $\mathrm{d}\Delta\sigma/\mathrm{d}x_{\mathrm{Bj}}$,
  $\mathrm{d}\Delta\sigma/\mathrm{d}y$, and
  $\mathrm{d}\Delta\sigma/\mathrm{d}p_{T,1}^{B}$.}
  \label{fig:app-cc-inclusive-pol}
\end{figure}

\begin{figure}[!htbp]
  \centering
  \includegraphics[width=0.32\linewidth]{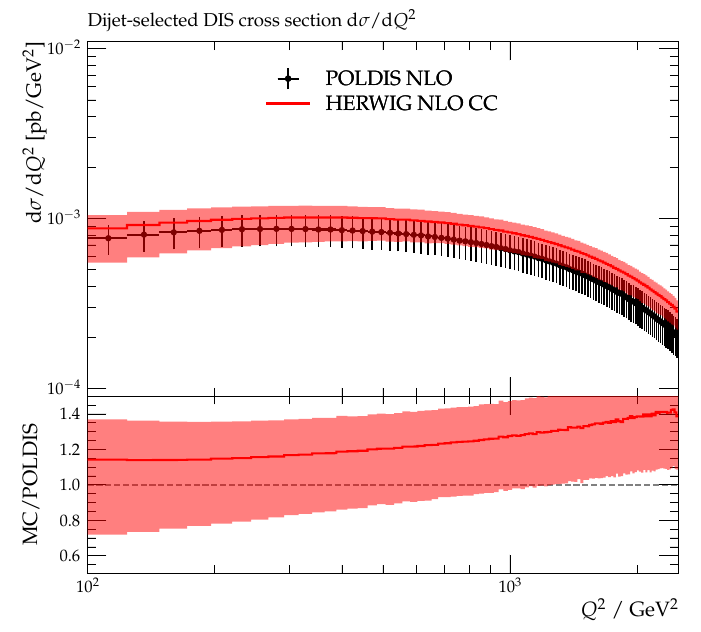}
  \includegraphics[width=0.32\linewidth]{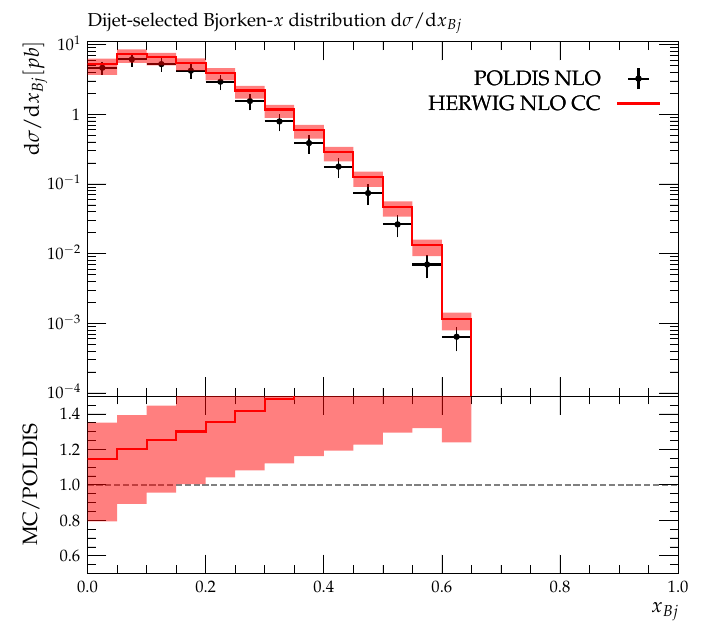}
  \includegraphics[width=0.32\linewidth]{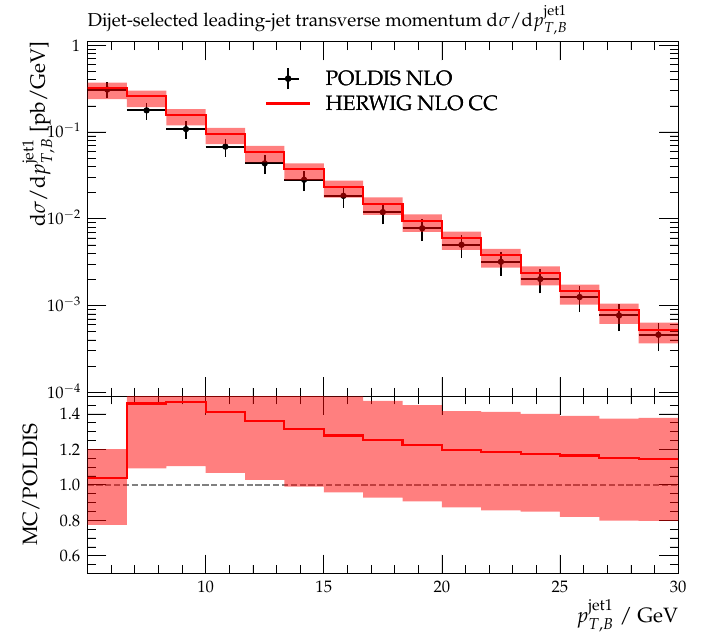}

  \includegraphics[width=0.32\linewidth]{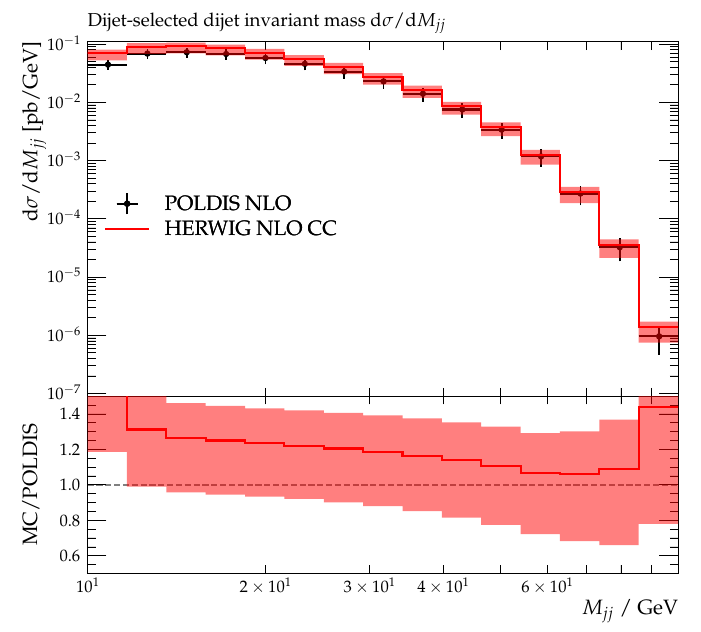}
  \includegraphics[width=0.32\linewidth]{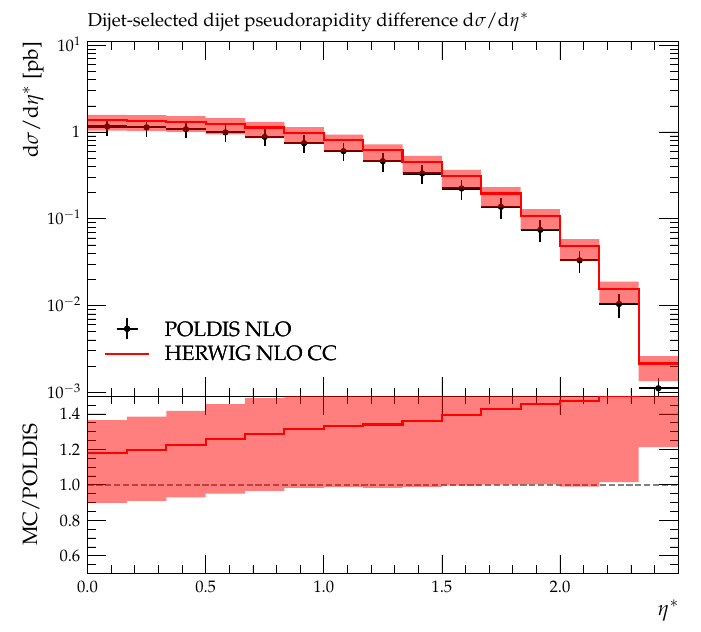}
  \includegraphics[width=0.32\linewidth]{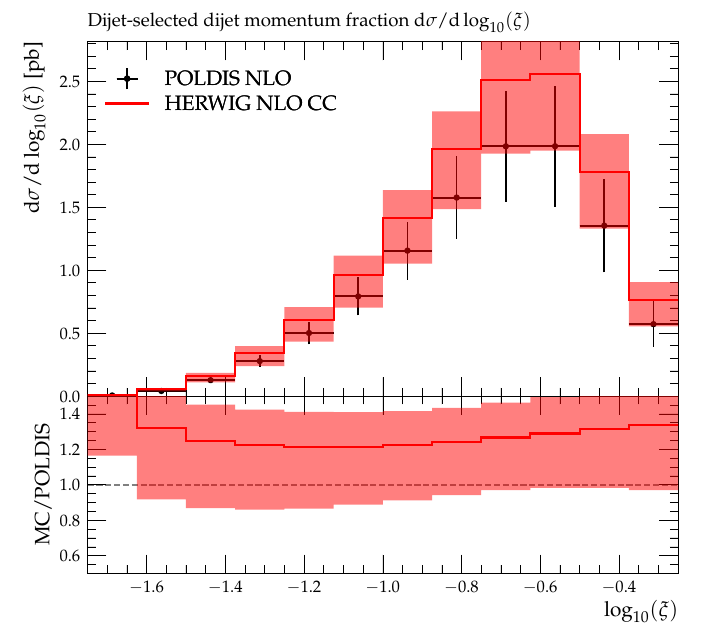}
  \caption{Charged-current parton-level validation without shower evolution of
  dijet-selected unpolarized kinematic distributions after dijet cuts:
  $\mathrm{d}\sigma/\mathrm{d}Q^2$,
  $\mathrm{d}\sigma/\mathrm{d}x_{\mathrm{Bj}}$,
  $\mathrm{d}\sigma/\mathrm{d}p_{T,1}^{B}$,
  $\mathrm{d}\sigma/\mathrm{d}M_{jj}$,
  $\mathrm{d}\sigma/\mathrm{d}\eta^\ast$, and
  $\mathrm{d}\sigma/\mathrm{d}\log_{10}(\xi)$.}
  \label{fig:app-cc-dijet-unpol}
\end{figure}

\begin{figure}[!htbp]
  \centering
  \includegraphics[width=0.32\linewidth]{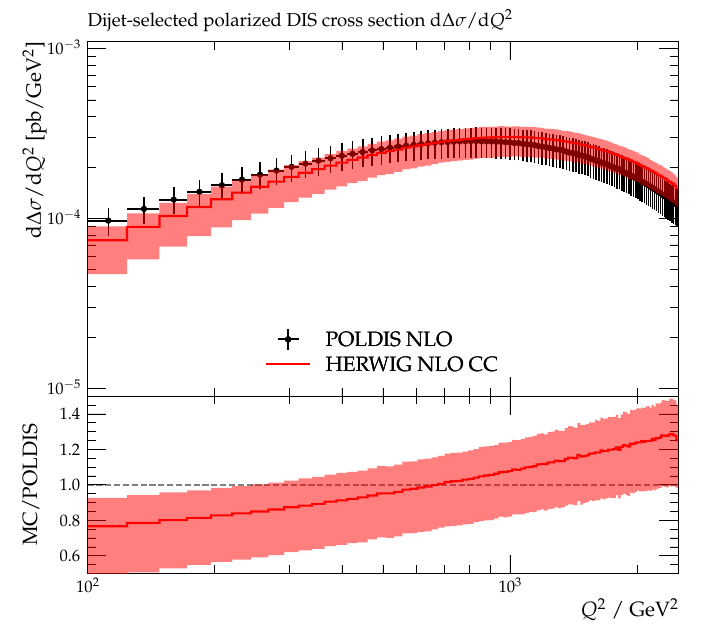}
  \includegraphics[width=0.32\linewidth]{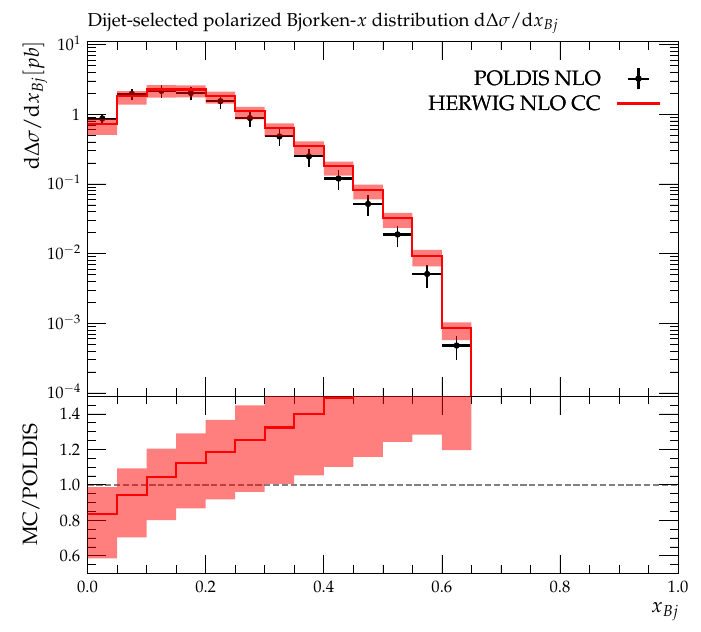}
  \includegraphics[width=0.32\linewidth]{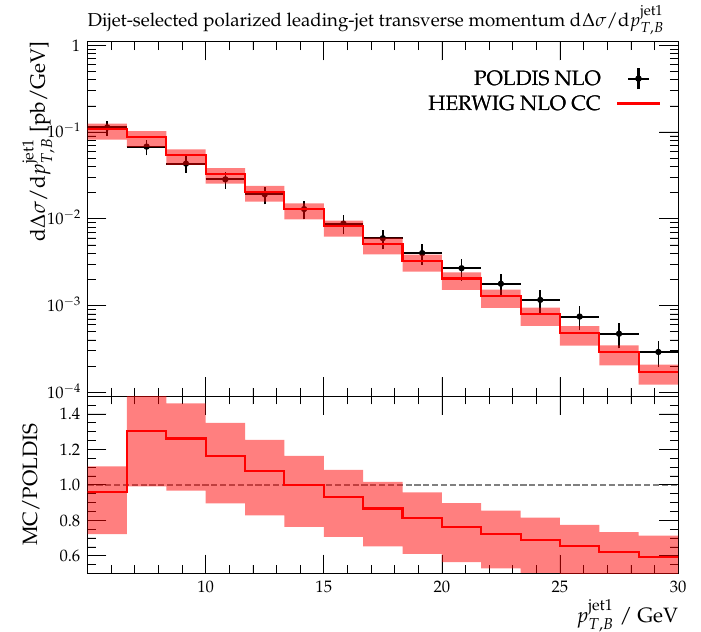}

  \includegraphics[width=0.32\linewidth]{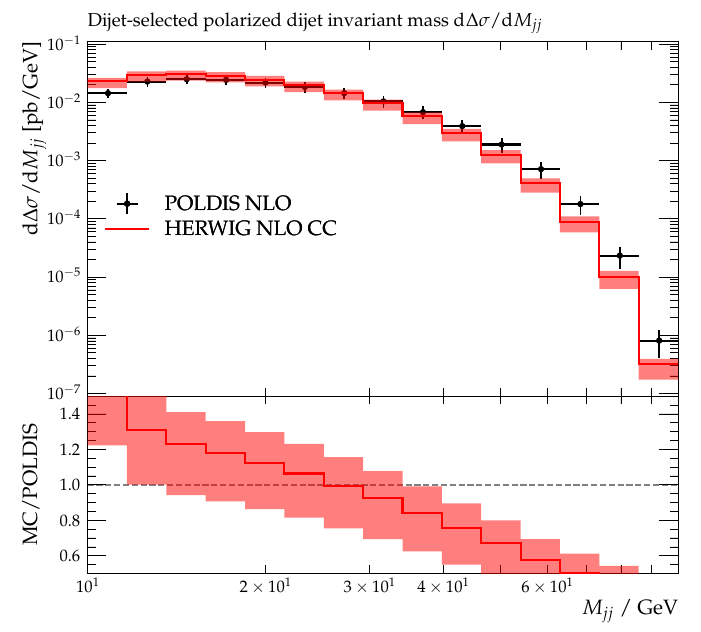}
  \includegraphics[width=0.32\linewidth]{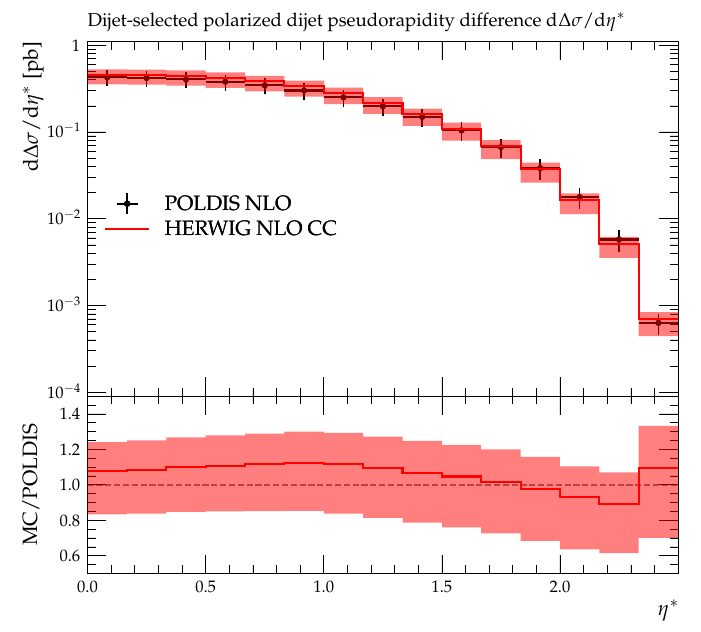}
  \includegraphics[width=0.32\linewidth]{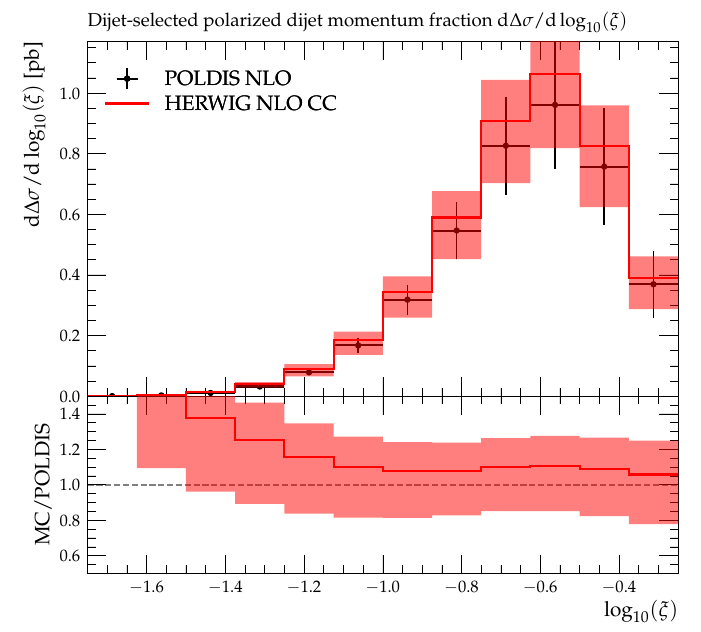}
  \caption{Charged-current parton-level validation without shower evolution of
  dijet-selected polarized kinematic distributions after dijet cuts:
  $\mathrm{d}\Delta\sigma/\mathrm{d}Q^2$,
  $\mathrm{d}\Delta\sigma/\mathrm{d}x_{\mathrm{Bj}}$,
  $\mathrm{d}\Delta\sigma/\mathrm{d}p_{T,1}^{B}$,
  $\mathrm{d}\Delta\sigma/\mathrm{d}M_{jj}$,
  $\mathrm{d}\Delta\sigma/\mathrm{d}\eta^\ast$, and
  $\mathrm{d}\Delta\sigma/\mathrm{d}\log_{10}(\xi)$.}
  \label{fig:app-cc-dijet-pol}
\end{figure}

\FloatBarrier

\bibliographystyle{JHEP}
\bibliography{biblio}

\end{document}